# IDENTIFICATION OF LEGGED LOCOMOTION VIA MODEL-BASED AND DATA-DRIVEN APPROACHES

A DISSERTATION SUBMITTED TO

THE GRADUATE SCHOOL OF ENGINEERING AND SCIENCE

OF BILKENT UNIVERSITY

IN PARTIAL FULFILLMENT OF THE REQUIREMENTS FOR

THE DEGREE OF

DOCTOR OF PHILOSOPHY

IN

ELECTRICAL AND ELECTRONICS ENGINEERING

By
İsmail Uyanık
May 2017

IDENTIFICATION OF LEGGED LOCOMOTION VIA MODEL-BASED AND DATA-DRIVEN APPROACHES
By İsmail Uyanık
May 2017

We certify that we have read this dissertation and that in our opinion it is fully adequate, in scope and in quality, as a dissertation for the degree of Doctor of Philosophy.

---
Ömer Morgül (Advisor)

---
Uluç Saranlı (Co-Advisor)

---
Hitay Özbay

---
Melih Çakmakcı

---
M. Kemal Leblebicioğlu

---
Mehmet Önder Efe

Approved for the Graduate School of Engineering and Science:

---
Ezhan Karaşan
Director of the Graduate School

# ABSTRACT

## IDENTIFICATION OF LEGGED LOCOMOTION VIA MODEL-BASED AND DATA-DRIVEN APPROACHES

İsmail Uyanık
Ph.D. in Electrical and Electronics Engineering
Advisor: Ömer Morgül
Co-Advisor: Uluç Saranlı
May 2017

Robotics is one of the core areas where the bioinspiration is frequently used to design various engineered morphologies and to develop novel behavioral controllers comparable to the humans and animals. Biopinspiration requires a solid understanding of the functions and concepts in nature and developing practical engineering applications. However, understanding these concepts, especially from a human or animal point of view, requires the significant use of mathematical modeling and system identification methods. In this thesis, we focus on developing new system identification methods for understanding legged locomotion models towards building better legged robot platforms that can locomote effectively as their animal counterparts do in nature.

In the first part of this thesis, we present our efforts on experimental validation of the predictive performance of mechanics-based mathematical models on a physical one-legged hopping robot platform. We extend upon a recently proposed approximate analytical solution developed for the lossy spring–mass models for a real robotic system and perform a parametric system identification to carefully identify the system parameters in the proposed model. We also present our assessments on the predictive performance of the proposed approximate analytical solution on our one-legged hopping robot data. Experiments with different leg springs and cross validation of results yield that our approximate analytical solutions provide a sufficiently accurate representation of the physical robot platform.

In the second part, we adopt a data-driven approach to obtain an input–output representation of legged locomotion models around a stable periodic orbit (a.k.a. limit cycle). To this end, we first linearize the hybrid dynamics of legged locomotor systems around a limit cycle to obtain a linear time periodic (LTP) system representation. Hence, we utilize the frequency domain analysis and identification methods for LTP systems towards the identification of input–output models (harmonic transfer functions) of legged locomotion. We propose simulation experiments on simple legged locomotion models to illustrate the prediction performance of the estimated input–output models.

Finally, the third part considers estimating state space models of legged locomotion using input–output data. To accomplish this, we first propose a state space identification method to estimate time periodic state and input matrices of a hybrid LTP system under full state measurement assumption. We then release this assumption and proceed with subspace identification methods to estimate LTP state space realizations for unknown stable LTP systems. We utilize bilinear (Tustin) transformation and frequency domain lifting methods to generalize our solutions to different LTP system models. Our results provide a basis towards identification of state space models for legged locomotion.

# ÖZET

# MODEL TABANLI VE VERİ GÜDÜMLÜ YÖNTEMLERLE BACAKLI HAREKETLİLİK İÇİN SİSTEM TANILAMASI

İsmail Uyanık
Elektrik ve Elektronik Mühendisliği Bölümü, Doktora
Tez Danışmanı: Ömer Morgül
Tez Eş Danışmanı: Uluç Saranlı
Mayıs 2017

Robotik, insan ve hayvan yapılarına uyumlu farklı yapısal morfolojilerin tasarlanması ve özgün davranışsal kontrolcülerin geliştirilmesi bakımından biyolojik sistemlerden esinlenmenin en sık kullanıldığı araştırma alanlarından birisidir. Biyolojik sistemlerden esinlenme doğadaki işlevleri ve kavramları kapsamlı bir şekilde anlamayı ve bunlar üzerinden pratik mühendislik uygulamaları geliştirmeyi gerektirir. Bununla birlikte, özellikle bir insan ya da hayvan için bazı kavramların anlaşılabilmesi, yüksek seviyede matematiksel modelleme ve sistem tanılama yöntemleri kullanımı gerektirmektedir. Bu tezde, doğadaki canlılar gibi başarılı bir şekilde hareket edebilen bacaklı robot platformlarının gerçeklenebilmesi amacıyla bacaklı hareketliliğin anlaşılabilmesi için yeni sistem tanılama yöntemlerinin geliştirilmesi üzerine odaklanılmıştır.

Bu tezin ilk bölümünde mekanik-tabanlı matematiksel modellerin fiziksel bir tek bacaklı zıplayan robot platformu üzerinde kestirim performansının deneysel doğrulamalarını içeren çalışmalarımızı sunmaktayız. Yakın bir zamanda kayıplı yay–kütle modeli için önerilen bir yakınsamalı analitik çözüm üzerine tarafımızca eklemeler yapılarak gerçek bir robot sistemine uyarlanmış ve önerdiğimiz yeni modeldeki sistem parametrelerini doğru kestirebilmek amacıyla bir parametrik sistem tanılama çalışması yapılmıştır. Aynı zamanda önerilen yakınsamalı analitik çözümün tek bacaklı zıplayan robotumuzun verileri üzerindeki kestirim performansına ait değerlendirmelerimiz de sunulmuştur. Farklı bacak yaylarıyla yapılan deneyler ve sonuçların çapraz doğrulamaları, önerdiğimiz yakınsamalı analitik çözümün fiziksel robot platformunu yeterince hassas bir şekilde tanılayabildiğini ortaya çıkarmıştır.

İkinci bölümde, kararlı bir periyodik yörünge (limit çevrimi) etrafında bacaklı hareketlilik için bir girdi–çıktı gösterimi elde edilmesini sağlayacak veri-güdümlü bir sistem tanılama yaklaşımı benimsenmiştir. Bu nedenle, doğrusal ve periyodik olarak zamanla değişen (DPZD) bir sistem gösterimi elde edebilmek amacıyla bacaklı hareketliliğin hibrit sistem dinamikleri bir limit çevrimi etrafında doğrusallaştırılmaktadır. Böylece, bacaklı hareketlilik için girdi–çıktı modellerinin tanılanabilmesi amacıyla DPZD sistemlerinin frekans düzleminde analizini ve tanılamasını yapan yöntemler kullanılmaktadır. Tanılaması yapılan girdi–çıktı modellerinin kestirim performansını gösterebilmek amacıyla basit bacaklı hareketlilik modelleri üzerinde benzetim ortamı deneyleri sunulmuştur.

Son olarak, üçüncü bölüm bacaklı hareketlilik için girdi–çıktı verisi kullanarak durum uzayı modellerinin veri-güdümlü tanılamasına odaklanmaktadır. Bunu başarabilmek için öncelikle tam durum ölçümü varsayımı altında bir hibrit DPZD sisteminin periyodik zamanlı durum ve giriş matrislerinin kestirilmesini sağlayan bir durum uzayı tanılama metodu sunulmuştur. Daha sonra, bu varsayımı kaldırarak bilinmeyen kararlı DPZD sistemleri için periyodik zamanlı durum uzayı modellerini kestirebilmek amacıyla altuzay tanılama yöntemleri kullanılmıştır. Çözümlerimizi farklı DPZD sistem modellerine genellemek amacıyla Tustin dönüşümü ve zamanla değişmeyen sistemlere yükseltme yöntemleri kullanılmıştır. Elde edilen sonuçlar bacaklı hareketlilik için durum uzayı modellerinin tanılamasına yönelik bir temel oluşturmaktadır.

*Anahtar sözcükler*: Sistem tanılama, bacaklı hareketlilik, matematiksel modelleme, yüklü-yay ters sarkaç (YYTS) modeli, doğrusal ve zamanla değişmeyen sistemler, harmonik transfer fonksiyonları, altuzay tanılama.

# Acknowledgement

The last six years was truly an amazing journey for me and reaching to a successful end would not have been possible without the inspiration and support of many great people.

Firstly, I owe my deepest gratitudes to my supervisors, Ömer Morgül and Uluç Saranlı for their guidance, encouragement and continous support. I can not find any proper words to convey my sincerest gratitudes and respect to them. They were always there to listen and to give advice when I needed. They encouraged me to become an independent researcher and helped me to discover and grow the creativity and enthusiasm that I didn't know I had.

I am hugely indebted to M. Mert Ankaralı and Noah J. Cowan for inspiring me to the world of system identification. The initial ideas of a great part of this thesis were formed with their invaluable contributions and stimulating discussions.

I would like to thank the distinguished members of my thesis jury Hitay Özbay, Melih Çakmakcı, M. Kemal Leblebicioğlu and M. Önder Efe for approving my work and guiding me all the way up to this point. I am also indebted to Orhan Arıkan for his support on my thesis.

Additionally, the members of our research group have contributed immensely to my personal and academic time at Bilkent. I am very thankful to Hasan Hamzaçebi, Ali Nail İnal, Deniz Kerimoğlu, Görkem Seçer, Bahadır Çatalbaş, Caner Odabaş, Eftun Orhon, Elvan Kuzucu, Mustafa Gül, Mansur Arısoy, Dilan Öztürk and Bengisu Özbay for our wonderful times at Bilkent.

Outside the laboratory, there are also some friends who directly or indirectly contributed to my completion of this thesis. I am grateful to my friends Serdar Öğüt, Necip Gürler, Veli Tayfun Kılıç, Serkan Sarıtaş, Ahmet D. Sezer, Furkan Çimen, Ali Alp Akyol, Merve B. Terzi, İmren Altepe, Saniye Vatansever and Selcen Deveci for always being there to listen and motivate.

I want to thank Mürüvet Parlakay and Aslı Tosuner for their helps on administrative works and thank Ergün Hırlakoğlu, Onur Bostancı and Ufuk Tufan for their technical support.

I am appreciative of the financial support from the Scientific and Technological Research Council of Turkey (TÜBİTAK). The work presented in this thesis was supported by TÜBİTAK through projects 215E050, 114E277 and 109E032. I also appreciate the PhD Fellowship of Aselsan A.Ş. and thank Nusrettin Güleç for guiding me as an industrial mentor.

Finally, but forever I owe my loving thanks to my wife Anıl Türel Uyanık for her unconditional love. I also would like to thank my family, Ayhan Uyanık, Meryem Uyanık, Ali Uyanık, Nurdan Uyanık, Serpil Tiryaki, Habib Tiryaki, Yakup Türel, Fahriye Türel and Işıl Türel for their support and encouragement.

# Contents

# List of Figures

# List of Tables

to my beloved wife Anıl Türel Uyanık

# Chapter 1

# Introduction

Legged locomotion emerges from a staggering diversity of animal morphologies in nature. However, despite the widespread use of legs by animals to achieve terrestrial locomotion [1, 2], the majority of mobile robots use wheels or tracks to move themselves. Unfortunately, this choice impairs mobility and performance on broken and unstable terrain [3], shifting attention to the use of legs in mobile and field robotics [4], despite significant challenges in the identification and control of legged robot platforms [5–7]. This thesis concerns the system identification problem of legged locomotion, since it still remains as a grand challenge in both biology and engineering [2, 8].

The primary objective in this thesis is to develop novel system identification tools that are applicable to legged locomotor systems. To this end, we utilize mechanics-based mathematical models, the harmonic transfer functions as well as the subspace identification theory. This thesis presents our efforts on utilizing these concepts to the system identification problem of legged locomotion. Our motivations from the results of existing studies and proposed methodology are explained in the following sections.

## 1.1 Mechanics-Based Mathematical Models of Legged Locomotion

A common approach to understanding and controlling robotic legged locomotion is the construction and analysis of simplified mathematical models that capture essential features of locomotor behaviors [9–20]. Running behaviors, in particular, are commonly represented by relatively simple spring–mass models such as the Spring-Loaded Inverted Pendulum (SLIP) model [1, 21]. However, modeling and analysis of even seemingly simple legged systems can be surprisingly complex due to the hybrid dynamics arising from intermittent foot contact as well as challenging nonlinearities in the equations of motion [9, 11, 22–25]. In this context, modeling of legged behaviors generally rely on a white-box approach, involving careful characterization of individual components in the system and the intended behavior together with informed (but possibly incorrect) "decisions" about what to neglect.

For instance, a common feature of such models is that their hybrid dynamics involve alternating flight and stance phases during locomotion. The Langrangian dynamics for these phases can be rather complex, with non-integrable equations of motion such as the case in stance phase [21, 26]. Given the utility of having accurate models and associated analytic solutions in constructing high performance controllers for nonlinear systems, substantial effort has been devoted to the construction of approximate analytical solutions to such non-integrable hybrid models [9–11, 13, 18, 27–29].

## 1.2 Estimating Input–Output Models of Legged Locomotion

The representational power of mechanics-based mathematical models is inevitably limited due to the nonlinear and complex nature of biological legged locomotor systems. Attempting to identify and explicitly incorporate these key nonlinearities into the model is daunting at best, increases complexity, and decreases the analytic utility of the resulting models. Despite our previous studies showing how accurate such models may be for simple spring–mass systems, there will always be unmodeled components in the physical system, resulting in discrepancies between the model and the experiments [22].

Consequently, we adopt a data-driven approach, with the goal of furnishing an input–output representation of a legged locomotor system, thereby eliminating the need to manually construct an explicit mathematical model for the system. Our main goal is to provide a system identification framework applicable to a useful (although not comprehensive) class of legged locomotion models [9], and possibly more complex robotic systems [30]. Our approach is based on considering legged locomotion as a hybrid nonlinear dynamical system with a stable periodic orbit (limit-cycle), corresponding to the locomotor behavior of interest. We introduce a formulation that addresses the input–output system identification problem in the frequency domain for a sub-class of hybrid legged locomotion models. More specifically, following certain assumptions on the hybrid dynamics of legged systems, we approximate their hybrid dynamics around the limit-cycle as a linear time-periodic system (LTP). Perturbing inputs to the locomotor system with small chirp signals yields input–output data necessary for the application of LTP system identification techniques, allowing us to estimate harmonic transfer functions (HTFs) associated with the local LTP approximation to the system dynamics around the limit cycle.

Existing studies on system identification of LTP systems focus on modeling these systems as multi-input single-output LTI systems [31]. This approach is based on the concept of harmonic transfer functions [32], which are infinite-dimensional operators that are analogous to frequency response functions for LTI systems. An identification strategy for such systems was developed in [33] using power spectral density and cross spectral density functions. A similar method was used in [34] considering the effects of noise in both input and output measurements. Different than these studies, local polynomial methods and lifting approaches were also used for the identification of harmonic transfer functions for multi-input single-output models of LTP systems [35]. Motivated by these studies, our main goal is to represent the dynamics of legged locomotion as a linear time periodic system, thereby enabling the use of the system identification method proposed in [33] for such systems.

## 1.3  Towards Identification of State Space Models of Legged Locomotion

Although all finite dimensional representations of a system will produce same input–output characteristics, state space models are accepted to be the natural and intuitive representation of a system. Therefore, in this section, we seek to develop novel system identification methods to estimate state space models of linear time periodic system towards application on legged locomotor systems.

A great majority of the state space identification methods that are available in the literature focus on linear time invariant (LTI) systems [36, 37]. However, as stated earlier, the dynamics of legged locomotor systems exhibit nonlinear characteristics, which yields a linear time periodic (LTP) behavior when linearized around a stable periodic orbit and under certain assumptions. Hence, we require novel tools for estimating time-periodic state space structures for these problems. To this end, we propose two different methods. The first one assumes full state measurement but considers hybrid linear time periodic systems, where each subsystem is also an LTP system with known periodic switching times. The second method considers a more general class of LTP systems by releasing the full state measurement assumption of the first method. We utilize frequency domain subspace identification methods to estimate LTP state space models for unknown stable LTP systems.

## 1.4 Organization of the Thesis

This thesis consists of three main parts, each of which are explained in detail in different chapters. The first part focuses on our efforts on mechanics-based mathematical models for legged locomotor systems. As stated earlier, even the simplest models, such as the Spring-Loaded Inverted Pendulum (SLIP) model, of legged locomotion includes non-integrable system dynamics. Thus, Chapter 2 extends upon a recently proposed analytical approximate solution for the SLIP dynamics and focuses on experimental validation of the predictive performance on a physical one-legged hopping robot platform.

In Chapter 3, we begin with linear time periodic (LTP) system modeling of legged locomotion around a stable periodic orbit. Hence, we utilize the frequency domain analysis methods for LTP systems towards obtaining data-driven models of legged locomotion. We illustrate the practicality of our approaches on different simulation models by estimating harmonic transfer functions (HTFs) of these models by just using input–output data without needing explicit mathematical modeling. We also show the predictive performance of estimated HTFs on a vertical hopping robot model.

Motivated by the successful results of Chapter 3, Chapter 4 focuses on estimating time-periodic state space structures from input–output data for LTP systems towards identification of state space models for legged locomotion. We explain two different novel methods for identifying state space models for LTP systems first under full state measurement assumption and then for a general class of LTP systems.

We finally conclude the thesis in Chapter 5 with some concluding remarks and possible extensions for future research.

## 1.5 Key Contributions

One of the first key contributions of this thesis is that we present an experimental validation study for an approximate analytical solution to the equations of motion of mechanics-based mathematical legged locomotion models. We provide systematic investigation of how mathematical models can present the state trajectories of a physical one-legged hopping robot with different initial conditions and control signals.

Another key contribution of this thesis is that we present a data-driven identification methodology for estimating frequency domain transfer functions of legged locomotor dynamics around a stable periodic orbit. We formulate the legged locomotion models as a linear time periodic (LTP) system around a limit cycle and show how data-driven identification methods for LTP systems can be utilized for system identification of legged locomotion models.

Our analysis on the identification of legged locomotion models with time delay yielded that the input–output identification method we consider in this thesis also allows estimation of transfer functions under input and measurement delay in the system. More importantly, our LTP formulation allows independent estimation of input and measurement delays which would otherwise be impossible to distinguish with an LTI system framework.

In addition, we provide a state space identification method for hybrid, piecewise smooth LTP systems under full state measurement assumption. Our formulation allows identification of switching time-periodic system and input matrices for an unknown stable LTP system. Besides, we extended our formulation for a general class of LTP systems by relaxing the full state measurement assumption and present a frequency domain subspace based state space identification methodology for LTP systems.

# Chapter 2

# Experimental Validation of a Feed-Forward Predictor for Legged Locomotion

Widely accepted utility of simple spring-mass models for running behaviors both as descriptive tools as well as literal control targets motivate accurate analytical approximations to their dynamics. Despite the availability of a number of such analytical predictors in the literature, their validation has been mostly done in simulation and it is yet unclear how well they perform when applied to physical platforms. In this study, we extend on one of the most recent approximations in the literature to ensure its accuracy and applicability to a physical monopedal platform. To this end, we present systematic experiments on a well-instrumented planar monopod robot, first to perform careful identification of system parameters and subsequently to assess predictor performance. The work presented in this chapter has also been reported and appeared in [22].

## 2.1 Introduction

Faced with an ever increasing need for mobile robotic platforms that can negotiate complex outdoor surfaces, it has become evident that traditional wheeled and tracked designs are approaching their morphological limits and the use of legs in various forms has to be explored [3]. Recent research and progress in both the theory [2] and practice [5–7, 30, 38] of building such machines provide ample evidence to support this observation. Nevertheless, numerous challenges remain before legged platforms can reach the level of autonomous performance already commonly observed in mobile wheeled and tracked robot platforms.

The ultimate promise of nimble locomotion on complex terrain led to both the construction of many legged morphologies as well as mathematical models to describe their underlying dynamics. Among the latter, the Spring-Loaded Inverted Pendulum (SLIP) model [21], an extended version of which is illustrated in Fig. 2.1, has become one of the most widely accepted and utilized model. The SLIP model is capable of accurately describing center-of-mass (COM) movements of running animals of widely varying sizes and morphologies [39, 40]. Originally motivated by biomechanical observations [41, 42], the SLIP model was adopted and refined by numerous robotics researchers in the last three decades [4], being established as an effective and appropriate dynamic abstraction for running behaviors [8].

The utility of this behavioral abstraction was also shown by its active embedding within more complex morphologies such as the RHex hexapod [43]. This provided further support to the idea pioneered by Raibert's robots [4] and other similar platforms [44–46], that the SLIP model could also act as the basis for hierarchical control strategies wherein the abstract running behavior would be regulated by SLIP controllers, unaware of the remaining redundancies in the complex morphology [14, 20]. This means that regardless of the complexity of the mechanical systems, the control problem can be solved by designing controllers for the mathematical representation [47].

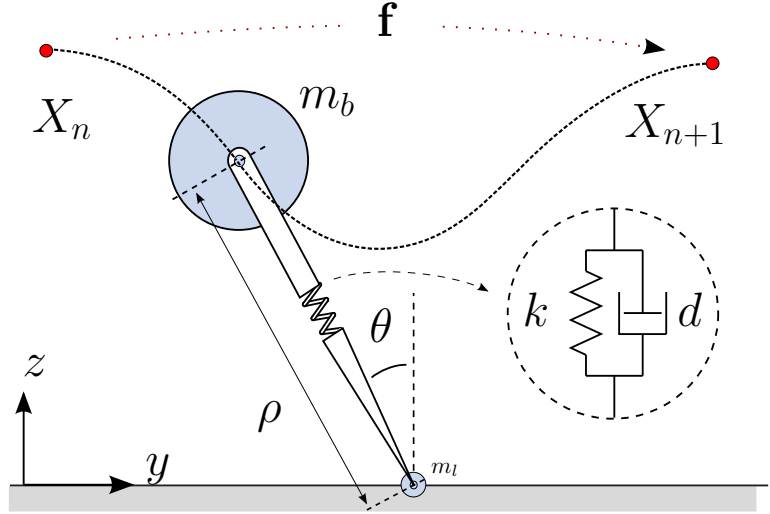

Figure 2.1: The Extended Spring-Loaded Inverted Pendulum (SLIP) model. Dashed curve illustrates a single stride from one apex event to the next, defining the return map $X_{n+1} = \mathbf{f}(X_n, \mathbf{u}_n)$.

The availability of analytic solutions to SLIP dynamics is crucial for formulating predictors for future steps as well as for the design of model-based controllers. Unfortunately, the non-integrable nature of SLIP stance phase dynamics necessitates approximate analytical solutions that can predict center of mass movements of legged locomotor systems under certain assumptions. A number of alternative approximate analytical solutions for the SLIP dynamics have been proposed in the literature. In this context, Schwind and Koditschek proposed an approximate analytical solution based on the iterative application of the mean value theorem, which converges to true SLIP dynamics after sufficient number of iterations [13]. Subsequently, Geyer formulated a simpler approximation based on certain assumptions on model parameters and trajectories such as small angular sweep and small leg compression [10]. Geyer's work was later extended with support for non-symmetric steps [18] and viscous damping in the leg [9]. Note that the extended SLIP model we use in this study considers the viscous damping in the leg as well as the effects of non-symmetric steps.

Experimental evidence for the relevance of the SLIP model to both biological and robotic running behaviors has also been established in a number of studies [12, 43]. However, the accuracy of approximate analytical solutions to the dynamics of this model have so far only been verified in simulation [9, 10], leaving their practical applicability an open question. The validity of approximate analytical predictors for SLIP trajectories strongly influences their usability in the design of model-based controllers [48]. The main goal of this part is to establish that even approximate analytical solutions to the SLIP model remain accurate for a physical one-legged hopping robot platform. Similar to the work presented in this study, Long et al. performed an experimental validation of approximate analytical solutions to the Simplest Parkour Model (SPM) on ParkourBot [49], a planar dynamic climbing robot with two compliant legs, exhibiting SLIP-like behavior. Unlike SPM, which relies on an instantaneous stance phase, we consider the full stance dynamics as proposed in [9].

Our primary contribution in this part is hence the experimental validation of a feed-forward predictor for SLIP trajectories. To this end, we also present the design of a well-instrumented monopod robot on which our validation experiments are performed. We also extend the solution presented in [9] to model the effects of non-negligible leg mass on system energy, an inescapable aspect of every legged platform, and viscous damping during the flight phase that can be used to model unexpected sources of energy losses. As a final step, we compare the prediction performance of our predictor with Geyer's approximation [10] as well as the numeric integration of SLIP model with and without damping in the leg to illustrate the practicality of the method proposed in [9].

## 2.2 The Extended SLIP Model

We begin our investigation by extending the ideal SLIP model to incorporate features necessary for its applicability to a physical monopod platform. First, we consider the effects of non-negligible leg mass, an inevitable component of all legged platforms effecting system dynamics both due to its moment of inertia and due to collision losses. Previous studies in this context focused on the effect of leg mass on gait stability considering its effects both throughout the entire stride [50] as well as just the touchdown collision [51]. Our extended model incorporates the latter, focusing on the energetic effects of the leg mass due to phase transitions with collision, since swing leg dynamics were found to have only a minor effect on locomotory dynamics [50]. In addition, integration of leg inertia to the system dynamics increases the complexity of the solutions for the stance dynamics. Therefore, we omit the effect of leg mass on system dynamics during the stance phase but only consider it for the impact collisions. This way, we can preserve approximate analytic nature of the solutions for the equations of motion of the stance phase. We will also show that the omission of leg dynamics during stance does not significantly impair the accuracy of our approximations for monopedal systems. The inclusion of this extension in our model substantially increases its applicability to physical legged platforms.

The second extension we consider is the presence of viscous damping during flight. Even though this is primarily useful to us for modeling mechanical properties of the central boom attachment for planarized robots such as our experimental platform, it generalizes the equations of motion in a way that allows modeling energy loss during flight for physical systems as well. This might be employed, for example, when leg retraction is found to effect flight dynamics or when air friction is found to be significant for fast running.

It would certainly have been desirable to integrate lateral dynamics or a torso in our model. However, it has been shown that the dynamics of steady-state running in three dimensions is largely determined by motion occurring in the sagittal plane, with negligible influence from the lateral plane [4]. Moreover, to the best of our knowledge, there are currently no analytical approximations to the dynamics of a 3D-SLIP with a torso and the feasibility of obtaining such approximations is not yet clear. Consequently, even though this is a problem that deserves and requires further theoretical and experimental investigation, we leave this inquiry outside the scope of the present study.

Note that the extensions we consider do not alter the analytical simplicity of the SLIP predictor and preserve the generality of our results. Both of our extensions can be adapted to different monopedal robot platforms by calibrating the leg mass and viscous damping during flight, whatever its source might be.

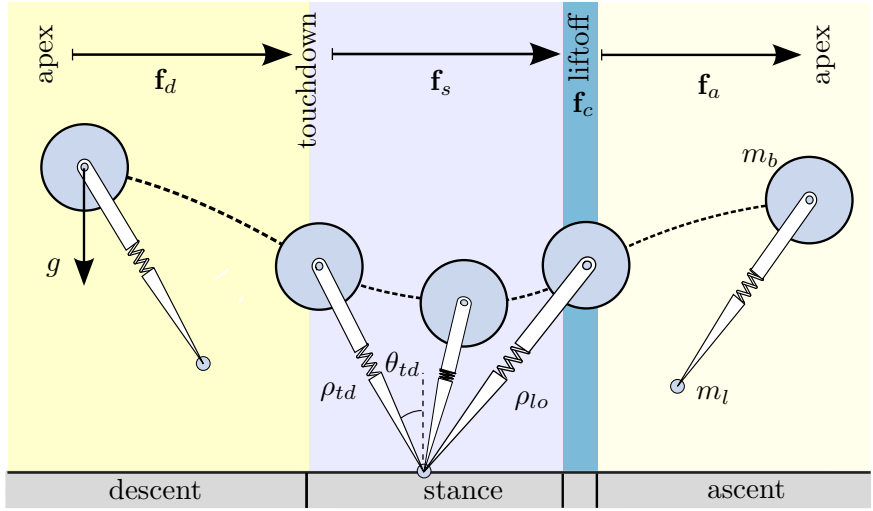

Figure 2.2: SLIP locomotion phases (shaded regions) and associated transition events (boundaries).

Also, our model reduces to the ideal SLIP model when the leg mass and flight damping are chosen to be zero, making our model applicable to a broad set of scenarios.

### 2.2.1 Model Structure and Definitions

The extended SLIP model we consider in this study consists of a point mass attached to a compliant leg with mass $m_l$ concentrated at the toe, stiffness $k$ and viscous damping $d$ as illustrated in Fig. 2.1. During locomotion, this model alternates between *stance* and *flight* phases as shown in Fig. 2.2, with the toe remaining stationary on the ground during stance. No torque is applied to the leg during stance and the body experiences gravitational acceleration with both vertical and horizontal viscous damping during flight. Table 2.1 details the notation we use throughout the chapter.

Touchdown and liftoff events mark transitions to and from the stance phase, respectively. Touchdown occurs when the toe comes into contact with the ground with the leg positioned at a fixed touchdown angle, $\theta_{td}$, during flight. We assume negligible toe dynamics during flight, with the toe mass positioned as necessary to achieve the desired touchdown leg angle and an uncompressed leg spring.

As usual, our study of this legged system relies on a Poincaré section defined at the "apex" point, which is indeed defined as the highest point on system trajectories during flight with $\dot{z} = 0$.

This leads to the definition of apex states as

$$X_n := [\, y_n,\ \dot{y}_n,\ z_n \,]^T,\qquad(2.1)$$

which is subsequently used to define the apex return map

$$X_{n+1} = \mathbf{f}(X_n, \mathbf{u}),\qquad(2.2)$$

with control inputs $\mathbf{u}$ appropriately defined as in [9]. For the current problem, the only control parameter we use is the leg angle at touchdown.

In contrast, liftoff occurs when the vertical component of the ground reaction force on the toe becomes negative. Unlike existing ideal SLIP models, our extended model incorporates a discrete change in the body velocity at liftoff due to the collision between the leg structure and a mechanical hard limiter on the leg length, typically included on almost all prismatic leg designs to prevent radial leg oscillations during flight. We model this discontinuity with an instantaneous liftoff map. Consequently, the apex return map can be decomposed as

$$X_{n+1} := (\mathbf{f}_a \circ \mathbf{f}_c \circ \mathbf{f}_s \circ \mathbf{f}_d)(X_n, \mathbf{u}),\qquad(2.3)$$

combining the descent map $\mathbf{f}_d$, the stance map $\mathbf{f}_s$, the instantaneous liftoff map $\mathbf{f}_c$ and the ascent map $\mathbf{f}_a$. Subsequent sections detail analytic derivations for each of these maps.

Table 2.1: Notation used throughout the chapter

| Extended SLIP Parameters | |
|---|---|
| $y, z, \dot{y}, \dot{z}$ | Body positions and velocities |
| $m_b, m_l$ | Body and leg mass of the robot |
| $k, d$ | Leg spring and damping constants |
| $d_h, d_v$ | Horizontal and vertical viscous damping during flight |
| $\rho, \theta$ | Leg length and angle |
| † Note that subscripts represent the system parameters at critical times such as $\rho_{td}$, $\rho_b$, and $\rho_{lo}$ represent the leg length at touchdown, bottom and liftoff times, respectively. | |

### 2.2.2 Descent and Ascent Maps

In contrast to the simple ballistic flight trajectories of [9], flight dynamics for the extended model have viscous damping in both horizontal and vertical directions. Hence, the associated equations of motion for the extended model take the form

$$\begin{bmatrix} \ddot{y} \\ \ddot{z} \end{bmatrix} = \begin{bmatrix} -d_h \dot{y} \\ -g - d_v \dot{z} \end{bmatrix}, \tag{2.4}$$

where $d_h$ and $d_v$ correspond to horizontal and vertical viscous damping during flight, respectively.

Analytic solutions to these equations are given by

$$y(t) = \frac{\dot{y}_0}{d_h}(1 - e^{-d_h t}) + y_0, \tag{2.5}$$

$$z(t) = \frac{g}{d_v^2}(1 - e^{-d_v t} - d_v t) + \frac{\dot{z}_0}{d_v}(1 - e^{-d_v t}) + z_0, \tag{2.6}$$

where $(y_0, z_0)$ and $(\dot{y}_0, \dot{z}_0)$ represent initial body positions and velocities, respectively. Velocity equations for the body can be obtained through differentiation as

$$\dot{y}(t) = \dot{y}_0 e^{-d_h t}, \tag{2.7}$$

$$\dot{z}(t) = \dot{z}_0 e^{-d_v t} - \frac{g}{d_v}(1 - e^{-d_v t}). \tag{2.8}$$

Using these solutions, time of touchdown can be found as the solution to the equation $z(t_{td}) = \rho \cos \theta_{td}$ whereas time of apex is the solution to the equation $\dot{z}(t_a) = 0$.

### 2.2.3 Approximate Analytical Solutions to Stance Trajectories of the SLIP Model

This section briefly summarizes the approximate analytical solutions proposed in [9] towards our experimental validation studies. A key point that needs to be noted for this section is that we utilize a non-dimensional coordinate system to generalize our results for legged locomotion models with different system parameters. Hence, using a non-dimensional formulation, we redefine time as $\bar{t} := t/\lambda$ with $\lambda := \sqrt{\rho_0/g}$ and scale all distances with the spring rest length $\rho_0$ to obtain equations of motion for stance in polar coordinates as

$$\ddot{\bar{\rho}} = \bar{\rho}\dot{\bar{\theta}}^2 - \kappa(\bar{\rho}-1) - c\dot{\bar{\rho}} - \cos(\bar{\theta}), \tag{2.9}$$

$$0 = \frac{d}{d\bar{t}}(\bar{\rho}^2\dot{\bar{\theta}}) - \bar{\rho}\sin\bar{\theta}. \tag{2.10}$$

Note that $(d/d\bar{t})^n = \lambda^n (d/dt)^n$, where all time derivatives are with respect to the newly defined, scaled time variable. Table 2.2 details descriptions and definitions of non-dimensional parameters used throughout the chapter.

Table 2.2: Notation for non-dimensional parameters

| Parameter | Definition | Description |
|---|---|---|
| $\bar{t}$ | $:= t/\lambda$ | Time (where $\lambda := \sqrt{\rho_0/g}$) |
| $[\bar{\rho}, \bar{\theta}]$ | $:= [\rho/\rho_0, \theta]$ | Leg length and leg angle |
| $\kappa$ | $:= k(\rho_0/(m_b g))$ | Leg spring stiffness |
| $c$ | $:= d(\rho_0/(\lambda m_b g))$ | Leg viscous damping |
| $\bar{p}_{\bar{\theta}}$ | $:= p_\theta/(\lambda/(m_b \rho_0^2))$ | Angular momentum |

We now define for the natural frequency $\hat{\omega}_0 := \sqrt{\kappa + 3\bar{p}_{\bar{\theta}}^2}$, the damping ratio $\xi := c/(2\hat{\omega}_0)$, the damped frequency $\omega_d := \hat{\omega}_0\sqrt{1-\xi^2}$ and the forcing term $F := -1 + \kappa + 4\bar{p}_{\bar{\theta}}^2$. Assuming the conservation of angular momentum and following approximations introduced in [9], approximate analytical solutions to stance trajectories can be computed as

$$\bar{\rho}(\bar{t}) = M e^{-\xi\hat{\omega}_0\bar{t}} \cos(\omega_d \bar{t} + \phi) + F/\hat{\omega}_0^2 \tag{2.11}$$

$$\dot{\bar{\rho}}(\bar{t}) = -M\hat{\omega}_0 e^{-\xi\hat{\omega}_0\bar{t}} \cos(\omega_d \bar{t} + \phi + \phi_2) \tag{2.12}$$

$$\bar{\theta}(\bar{t}) = \bar{\theta}_{td} + X\bar{t} + Y(e^{-\xi\hat{\omega}_0\bar{t}} \cos(\omega_d \bar{t} + \phi - \phi_2)$$
$$- \cos(\phi - \phi_2)) \tag{2.13}$$

$$\dot{\bar{\theta}}(\bar{t}) = 3\bar{p}_{\bar{\theta}} - 2\bar{p}_{\bar{\theta}}F/\hat{\omega}_0^2 - 2\bar{p}_{\bar{\theta}}Me^{-\xi\hat{\omega}_0\bar{t}} \cos(\omega_d \bar{t} + \phi), \tag{2.14}$$

where

$$M := \sqrt{A^2 + B^2} \qquad (2.15)$$
$$\phi := \arctan(-B/A) \qquad (2.16)$$
$$\phi_2 := \arctan(-\sqrt{1-\xi^2}/\xi) \qquad (2.17)$$
$$A := \rho_{td} - F/\hat{\omega}_0^2 \qquad (2.18)$$
$$B := (\dot{\rho}_{td} + \xi\hat{\omega}_0 A)/\omega_d. \qquad (2.19)$$

This approximate solution for stance trajectories allows us to find the time of occurrence for bottom and liftoff transitions. Bottom is reached when the leg is maximally compressed and can be found as the solution to the equation $\dot{\rho} = 0$. The liftoff event is more challenging since the presence of damping often results in the toe lifting off the ground prior to the spring reaching its rest length. Its time is computed as the minimum of these two conditions. Once these boundaries of the stance phase are found, the trajectories for an entire stride from an apex to the next can be computed.

Note that the derivations for the equations of motion explained above assumes constant angular momentum during each stride. However, this assumption is quickly violated for non-symmetric steps resulting in prediction errors for the center of mass trajectories. In order to overcome this issue, [18] introduces a correction for the effect of gravity on the angular momentum as a constant offset $\bar{p}_{\bar{\theta}}(\bar{t}_{td})$ added to the angular momentum at the time of touchdown. This correction term on the angular momentum increases the domain of validity of the approximate analytical solutions to the non-symmetric steps. Also, resolving this issue with a simple correction term to the angular momentum preserves the analytical simplicity of the solutions. This correction on the angular momentum is formulated as

$$\hat{p}_{\bar{\theta}} = \bar{p}_{\bar{\theta}}(\bar{t}_{td}) + \frac{\bar{t}_{lo}}{4}(\bar{\rho}(\bar{t}_{td})\sin\bar{\theta}(\bar{t}_{td}) + \bar{\rho}(\bar{t}_{lo})\sin\bar{\theta}(\bar{t}_{lo})) \ . \qquad (2.20)$$

### 2.2.4 Modeling the Liftoff Collision

The liftoff event marks the end of the stance phase. For the extended model, this is accompanied by an inelastic collision between the body and the leg structure, after which both masses end up moving with the same velocity together. This is captured in our model as an instantaneous liftoff map (collision map) $\mathbf{f}_c$, corresponding to a discontinuity in the body velocity with

$$\left[\, \dot{y}^+,\, \dot{z}^+ \,\right]^T := \frac{m_b}{m_b + m_l}\left[\, \dot{y}^-,\, \dot{z}^- \,\right]^T, \qquad (2.21)$$

where $m_b$ and $m_l$ are the body and leg masses while the $-$ and $+$ superscripts identify pre-collision and post-collision states, respectively. Even though the toe may have lifted off the ground prior to this collision (hence resulting in nonzero toe velocity prior to collision), its effect on the body through leg damping will also contribute to the decrease in the body velocity. We represent the entirety of this "liftoff phase" with the inelastic collision of (2.21), which has approximately the same energetic effect on system velocities since no external forces except gravity act on the system after liftoff and the leg mass is assumed to be small.

With all the maps in place, we now have an approximate analytical solution to the return map defined in (2.2). Subsequent sections use this approximate analytical solution for comparisons with experimental data collected for a wide range of initial conditions and parameters for the extended SLIP model.

## 2.3 Experimental Setup

Our focus in this study is the experimental evaluation of the predictive performance of our analytical approximations proposed in Section 2.2.3 to a SLIP-based physical robot trajectories within a *single stride*. To this end, we have designed and constructed a monopedal robot platform based on the SLIP morphology, instrumented to provide full state measurement while constraining robot motion to the sagittal plane. In this section, we first describe our experimental platform, and then conduct systematic experiments to identify various dynamic parameters for our setup.

### 2.3.1 Robot Platform

Our platform consists of the planarizer illustrated in Fig. 2.3 that constrains the motion of an end-plate to a cylindrical plane, approximating unconstrained motion in the sagittal plane while eliminating unmodeled lateral dynamics. Such designs are commonly used to investigate locomotion systems and their correspondence to sagittal plane models [4, 45, 52] while allowing sustained forward locomotion.

An important feature of our design is its ability to provide accurate and high-bandwidth positional measurements through optical encoders mounted on the central joint assembly. The main boom, a $5cm$ diameter, $1.67m$ long carbon-fiber tube, is connected to the central joint assembly which has incremental encoders with 8192 counts per revolution connected to each axis through 1 : 6 timing belts. This yields a resolution of $0.21mm$ in positional measurements of the robot attached to the end-plate.

The leg structure, also illustrated in Fig. 2.3, is affixed to the boom endplate, which is constrained to a fixed orientation in the sagittal plane. The rest length of the robot leg is $22cm$ and it is coupled to the boom plate through a DC motor. The hip motor is kept inactive during the stance phase but only used during the flight phase to maintain a fixed leg angle prior to touchdown and just after liftoff. The hip motor is a Maxon RE30-268215 60W brushed DC motor combined with a Maxon GP-32-C 1 : 18 planetary gear and is completely disabled during stance [53]. A three-channel Type L, MR encoder with 512 counts per revolution is used to measure the leg angle relative to the boom plate and hence the sagittal plane horizontal.

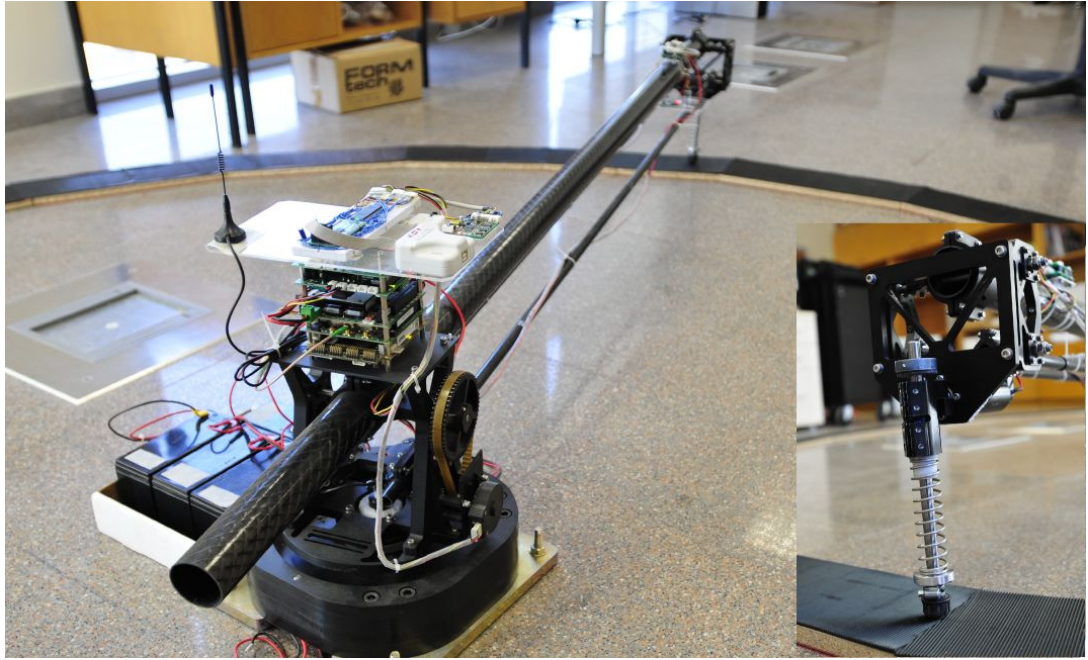

Figure 2.3: The hopper robot with an overall view of the planarizer and close view of the leg.

The robot is programmed with C/C++ programming language and all computations are performed at the center of the planarizer with a Cool LiteRunner-LX800 PC104 single-board computer. The central computer is mainly used for behavioral control of the robot and we utilized the Universal Robot Bus (URB) architecture for communication with the peripheral units such as the motor amplifiers and encoder interfaces [54].

### 2.3.2 Data Collection and Preprocessing

The planarized monopod platform we described in preceding sections is used for all the experiments presented in this study. To ensure general relevance of our results, we used four different helical springs, *hard*, *medium*, *soft* and *softer*, manufactured to have the same rest length but different stiffnesses and damping values. The identified compliance and damping values for each of these springs can be seen in Table 2.5.

Note that the stiffness range were chosen to be consistent with biomechanics literature. In particular, experiments on humans (with 80 kg mass and 1m leg length on average) running at different speeds (in the range $2.5 - 6.5 m/s$) reveal leg stiffnesses in the range $[12, 42]kN/m$ [55], which corresponds to the stiffness range $[15, 53]$ in non-dimensional coordinates. On the other hand, manual measurements of our leg springs yield a stiffness range $[16, 43]$ in non-dimensional units, which covers a large portion of the human stiffness range reported in [55].

Each experiment consisted of manually throwing the robot with different initial conditions, ensuring in each case that the vertical velocity was upwards to guarantee the occurrence of the first apex. Prior to this initial thrust, the leg was positioned at a desired angle (varied across different experiments), maintained throughout the initial flight phase using the hip motor without affecting flight dynamics. Upon touchdown, the hip motor was deactivated, letting natural SLIP stance dynamics govern the motion (see Remark 1). Immediately following liftoff, the hip motor was re-activated to maintain the liftoff leg angle until the second apex point was reached, following which it was positioned vertically to catch the robot and stop its motion. An example for such an experiment is illustrated in Fig. 2.5, with the corresponding approximate analytical solutions superimposed as dashed lines.

**Remark 1** *Note that in this study the only control parameter we use is the touchdown leg angle. However, there are also some approximate analytical solutions for the torque-actuated legged locomotion models in which a torque input is applied to inject additional energy to the system during the stance phase [11]. There are also some recent studies on experimental validation of the approximate analytical solutions to the torque-actuated legged locomotion models on physical robot platforms [56].*

All system states were recorded during the experiment at $500Hz$ using encoders mounted on the central assembly and the hip motor. Problematic experiments with foot slippage or other erroneous conditions were manually eliminated. Subsequently, positional data for clean experiments were filtered with a zero-phase fifth order Butterworth filter with a cutoff frequency of 50Hz to eliminate noise resulting from the oscillations and vibration of the boom. These positional encoder measurements were then numerically differentiated to obtain body velocity information. Following this filtering, key transition points along the trajectory, touchdown, bottom, liftoff and apex, were extracted based on their corresponding transition conditions and used for analysis and fitting.

### 2.3.3 Modeling of the Boom Dynamics

The center of mass of the boom–leg assembly is situated outside the sagittal plane of locomotion. However, since the SLIP model is formulated in this sagittal plane, we capture the inertial effect of the boom as an increased gravitational acceleration on the robot body. A simplified lateral model of the boom assembly is shown in Fig. 2.4, with the equations of motion taking the form

$$(I + Ml^2)\, \ddot{\phi} = -Mlg_0 \cos\phi - 0.5mlg_0 \cos\phi \,, \tag{2.22}$$

where $m$ and $I$ are the mass and moment of inertia for the boom and $M$ is the mass of the leg assembly. Assuming that $\phi$ stays small with $\cos\phi \approx 1$ and $\sin\phi \approx \phi$, we have

$$(I + Ml^2)\, \ddot{\phi} \approx -Mlg_0 - 0.5mlg_0 \,. \tag{2.23}$$

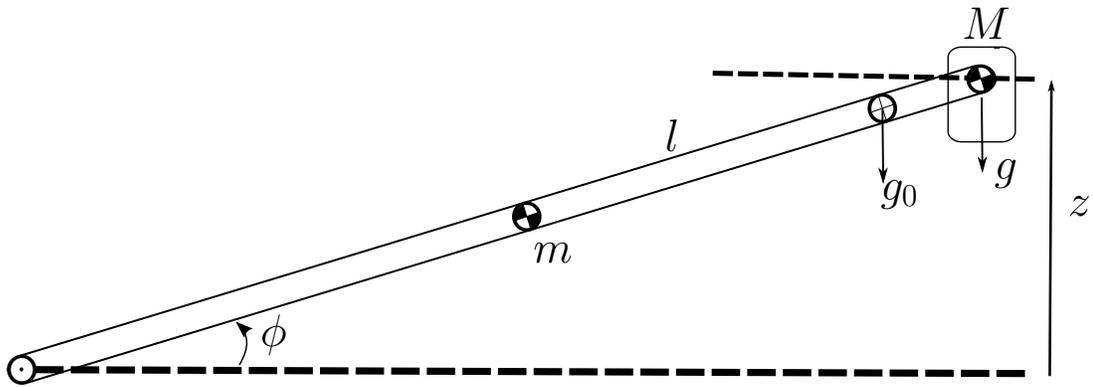

Figure 2.4: Simplified lateral model for the boom and the leg assembly during flight.

Vertical robot position depends on the boom angle through $z = l \sin \phi$. For this relation, our small angle approximation yields $z \approx l\phi$, whose second derivative $\ddot{z} \approx l\ddot{\phi}$ can be combined with (2.23) to yield

$$\ddot{z} \approx \frac{M + m/2}{M + m/3} g_0 ,  \qquad (2.24)$$

where we used $I = ml^2/3$ considering that the boom is a cylinder rotating around its tip. For our platform, we have $M = 3.4 kg$ and $m = 0.39 kg$, that yields the gravitational acceleration perceived in the body frame as $g = 9.99 m/s^2$.

## 2.4 Identification of the Experimental Platform

The two primary sources of inaccuracies in the predictive performance of our extended model are incorrect choices of model parameters, and inherent deficiencies in the model or associated approximations. In this study, we seek to isolate the latter to provide a fair assessment of our model and analytic approximations. Consequently, we use system identification methods to estimate dynamic model parameters which are difficult to measure. Similar parameter identification methods have been used in the literature to determine accurate models for complex legged platforms [57], but our focus is on the validation of our approximations.

### 2.4.1 Identification of Body and Leg Masses

We first focus our system identification efforts on the body and leg mass parameters, $m_b$ and $m_l$ respectively, for the extended SLIP model since their influence on system dynamics, particularly energy losses due to the liftoff collision are substantial. To this end, we first use vertical hopping experiments with the leg kept vertical by the hip motor. For the flight phase, (2.6) and (2.8) remain valid and yield the vertical position and velocity. In contrast, stance trajectories, (2.11) to (2.14), take a much simpler form when we constrain the motion to vertical dimension.

Using the data collection and filtering procedure described in Section 2.3.2, we ran 50 vertical experiments for each one of all four springs for a total of 200 experiments with $\theta_{td} = 0$ and $\dot{y}_0 = 0$. Analytic solutions for these vertical trajectories have three common parameters: the body mass $m_b$, the leg mass $m_l$ and the vertical flight damping $d_v$ in addition to the spring specific compliance $k$ and damping $d$ parameters. In order to find these parameters, we construct a nonlinear least-squares error problem with the cost function defined as the percentage difference between measured and predicted apex and bottom positions, taking the form

$$C_v := \frac{||\,[z_a, z_b] - [\hat{z}_a, \hat{z}_b]\,||_2}{||\,[z_a, z_b]\,||_2}. \tag{2.25}$$

We use Matlab's `lsqnonlin` function to find solutions for $m_b$, $m_l$ and $d_v$ common to all 200 experiments. Our results are shown in Table 2.3, while stiffness and damping parameters for all four springs are detailed in Table 2.5. We will use these parameters throughout the chapter during the experimental validation process of our approximate analytical solution.

Table 2.3: Estimates of mass and vertical damping parameters based on vertical experiments.

| $m_b(kg)$ | $m_l(kg)$ | $d_v(Ns/m)$ |
|---|---|---|
| 3.21 | 0.19 | 0.06 |

## 2.4.2 Identification of Horizontal Flight Damping

Vertically constrained experiments do not exercise horizontal degrees of freedom in our boom assembly. Consequently, we use our entire set of single-stride experiments to identify the horizontal damping coefficient during the flight phase.

We begin by introducing a first order approximation to horizontal flight dynamics, which normally have exponential decay terms in their solution, making linear fitting methods inapplicable. In particular, we will assume that the horizontal velocity during the descent phase can be approximated as

$$\dot{y}(t) \approx \dot{y}_0 - d_h t \;, \tag{2.26}$$

while relaxing the initial condition $\dot{y}_0$ to possibly be different than the measured initial condition $\dot{y}(0)$ to increase the accuracy of the approximation. Recall that the parameter of interest in this fitting procedure is the damping coefficient $d_h$, which justifies this relaxation in the fitting.

Having identified the touchdown states through the preprocessing steps described in Section 2.3.2, we can now formulate a linear set of equations $Ax = b$, by equating multiple predicted and measured state points along each trajectory as

$$\begin{bmatrix} 1 & 0 & \cdots & 0 & t_1^1 \\ \vdots & & & & \vdots \\ 1 & 0 & \cdots & 0 & t_{n_1}^1 \\ \vdots & & & & \vdots \\ 0 & 0 & \cdots & 1 & t_1^m \\ \vdots & & & & \vdots \\ 0 & 0 & \cdots & 1 & t_{n_m}^m \end{bmatrix} \begin{bmatrix} \dot{y}_0^1 \\ \dot{y}_0^2 \\ \vdots \\ \dot{y}_0^m \\ -d_h \end{bmatrix} = \begin{bmatrix} \dot{y}^1(t_1^1) \\ \vdots \\ \dot{y}^1(t_{n_1}^1) \\ \vdots \\ \dot{y}^m(t_1^m) \\ \vdots \\ \dot{y}^m(t_{n_m}^m) \end{bmatrix} \;, \tag{2.27}$$

where $t_i^j$ is the $i^{th}$ data point for the $j^{th}$ experiment, with the corresponding horizontal velocity $\dot{y}^j(t_i^j)$. The best fit to this set of data points is given by the regressor

$$x = (A^T A)^{-1} A^T b. \tag{2.28}$$

Using this procedure, our experiments result in the horizontal flight damping coefficient common to all experiments identified as $d_h = 0.3 \; Ns/m$.

## 2.5 Experimental Validation of Approximate Analytic Solutions

Having identified fixed mass and flight damping parameters for the leg assembly and the planarizing boom, we now continue with the evaluation of the predictive performance of our analytic approximations to the extended SLIP model together with the identification of spring compliance and damping coefficients for four different leg springs. In order to ensure the validity of our evaluation, we ran experiments with a wide range of initial conditions and touchdown leg angles as described in Section 2.3.2. In particular, 181, 208, 267 and 174 valid experiments were completed for the softer, soft, medium and hard springs, respectively, for a total of 830 experiments. The initial conditions for single stride experiments were chosen in the ranges $\dot{y} \in [0.3, 2.5](m/s)$ and $z \in [0.24, 0.48](m)$.

### 2.5.1 Performance Criteria

As a common basis for our cost function for system identification as well as the evaluation of the predictive performance for our approximations, we first define percentage apex position, velocity and time error measures for each stride as

$$E_{ap} := 100 \frac{|| [y_a, z_a] - [\hat{y}_a, \hat{z}_a] ||_2}{|| [y_a, z_a] ||_2} \qquad (2.29)$$

$$E_{av} := 100 \frac{|| \dot{y}_a - \hat{\dot{y}}_a ||_2}{|| \dot{y}_a ||_2} \qquad (2.30)$$

$$E_{at} := 100 \frac{|| t_a - \hat{t}_a ||_2}{|| t_a ||_2} , \qquad (2.31)$$

where variables with hats denote our predictions. These definitions mirror similar measures defined in [9]. In order to improve convergence for the system identification, we also define a position error for the bottom transition as

$$E_{bp} := 100 \frac{|| [y_b, z_b] - [\hat{y}_b, \hat{z}_b] ||_2}{|| [y_b, z_b] ||_2} . \qquad (2.32)$$

The cost function we define for system identification is composed of four components corresponding to the error measures defined above, taking the form

$$C := \sqrt{C_{ap}{}^2 + C_{av}{}^2 + C_{at}{}^2 + C_{bp}{}^2}, \qquad (2.33)$$

where individual cost functions $C_{ap}$, $C_{av}$, $C_{at}$ and $C_{bp}$ correspond to arithmetic mean of corresponding errors.

## 2.5.2 Predictive Performance with Cross-Validation

In this section, we present a comprehensive evaluation of the predictive performance of our analytic approximations (AAS) to the extended SLIP model, first identifying the stiffness and damping coefficients for the compliant leg, then using the error measures defined in Section 2.5.1 to quantify the accuracy of the approximations. In addition to AAS, we also evaluate the prediction performance of Geyer's approximation [10] as well as the numeric integration of the original stance dynamics in (2.9) and (2.10) both with (SLIPD) and without (SLIP) viscous damping in the leg.

For statistical validity, we used a cross-validation approach by dividing experiments into disjoint subsets for training (estimating leg compliance and damping) and testing (evaluating predictive performance). In this context, we considered 5-fold, 10-fold, 30-fold and leave-one-out options and observed their results separately. Consistent with observations described in [58], we confirmed that using higher number of folds yields low deviations in training results but high deviations in test results. Consequently, we use 30-fold cross-validation for this study, ensuring that test results represent the worst case performance figures for our approximations.

For the estimation of leg compliance and damping from training data, we use the `lsqnonlin` method of MATLAB, which uses the *trust-region-reflective* optimization algorithm [59]. We use the compliance and damping parameters in Table 2.5 obtained from vertical experiments to initialize the optimization, further refining resulting parameters through repeated runs of the optimization.

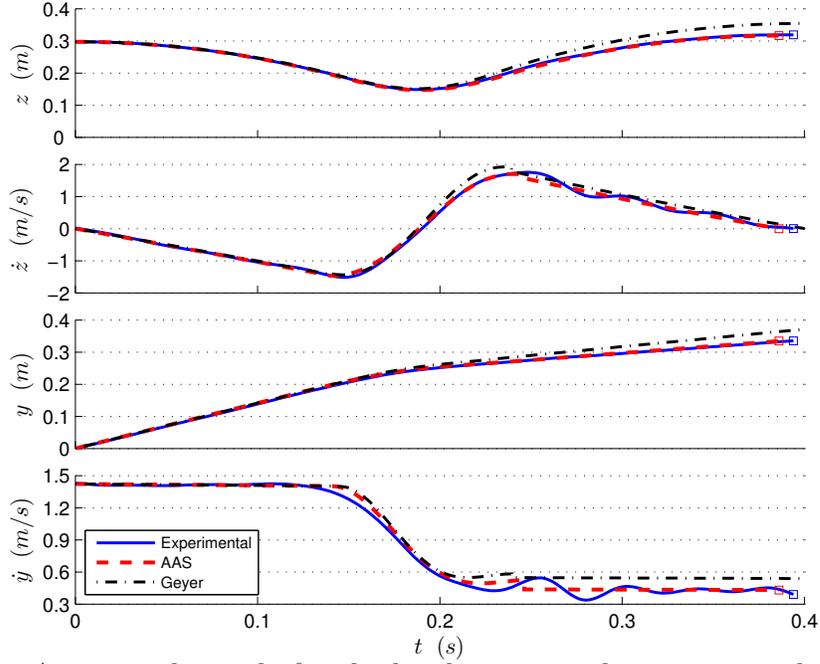

Figure 2.5: An example stride for the hard spring with experimental data (solid blue) and analytic predictions of AAS (dashed red) and Geyer's method (dashed black) for position and velocity trajectories are shown together.

Fig. 2.5 illustrates the results of our system identification for one of the experiments, showing filtered system states superimposed with the predictions of our analytic approximation and Geyer's predictor. Initial conditions for the analytic solutions were chosen to be the same as the experiment, except the initial horizontal velocity which uses the estimate obtained from (2.28). Velocity oscillations in the experimental data right after $t = 0.235s$ are due to vibrations of the boom assembly following the liftoff collision (also visible as a discontinuity in horizontal and vertical velocities at around t = 0.235s), but dissipate long before the end of the stride and hence do not effect the predictive performance of the return map. Apart from this unmodeled effect, the extended SLIP model and our approximations show an accurate performance in capturing the behavior of the experimental platform as compared to Geyer's predictor.

Table 2.4 details average percentage prediction errors for apex position, velocity and time as well as parameter estimations and their standard deviations across all experiments including training as well as test sets. Overall, our results show that prediction errors in positional, velocity and time variables are 2%, 7% and 1.85% on average, respectively. The standard deviations are also well below 0.1% and 2.5% for training and test data, respectively as a result of 30-fold cross validation. Nevertheless, these experimentally validated single-stride prediction errors are sufficiently low to be compensated by using adaptive controllers such as in [48] when additional feedback can be introduced. Similarly, reactive control algorithms, which are robust against model and measurement uncertainty, can be used to compensate for such errors [60].

In contrast, numeric integration of SLIP model and Geyer's prediction show prediction errors around 10% on average for positional variables. The main reason for this significant error is the unmodeled but inescapable damping loss in experimental robot platforms. Note that AAS and Geyer's predictor are approximations to SLIPD and SLIP dynamics, respectively. This is why numeric integrations perform better than their corresponding analytic predictors. Consequently, since the numeric SLIP predictor represents an upper bound for the accuracy of all methods that approximate the trajectories of lossless SLIP models and still performs worse than our method, we have not included results from any other approximations in our comparative study.

We can also observe that prediction errors of AAS decrease with increasing spring stiffness, which is expected since stiffer springs compress less, with trajectories coming closer to satisfying the assumptions underlying our approximations [9]. In the case of hard spring, the average percentage position prediction error is 1.53%, which corresponds to approximately $0.75cm$ for our robot running at a maximum height of 50 cm.

It is interesting to note that average prediction errors for AAS with respect to SLIPD were 0.75% and 1.40% for position and velocity coordinates, respectively [9]. However, the relative prediction performance of AAS with respect to SLIPD has drastically decreased to 0.1% on average in our experimental study. This is due to our fitting procedure, which allows AAS and SLIPD to choose different leg compliance and damping parameters in order to minimize their prediction errors.

Table 2.4: Percentage prediction errors and parameter estimates resulting from 30-fold cross validation experiments.

| | Method | Test Runs | | | Training Runs | | | Leg Parameters | |
|---|---|---|---|---|---|---|---|---|---|
| | | $E_{ap}$ | $E_{av}$ | $E_{at}$ | $E_{ap}$ | $E_{av}$ | $E_{at}$ | $k$ (N/m) | $d$ (Ns/m) |
| **Hard** | *SLIPD* | $1.41 \pm 0.51$ | $3.98 \pm 1.04$ | $1.69 \pm 0.62$ | $1.36 \pm 0.01$ | $3.75 \pm 0.04$ | $1.54 \pm 0.02$ | $6560 \pm 3.09$ | $12.3 \pm 0.07$ |
| | ***AAS*** | $\mathbf{1.53 \pm 0.40}$ | $\mathbf{4.23 \pm 1.12}$ | $\mathbf{1.74 \pm 0.54}$ | $1.52 \pm 0.02$ | $4.21 \pm 0.04$ | $1.74 \pm 0.02$ | $6605 \pm 6.14$ | $12.1 \pm 0.06$ |
| | *SLIP* | $8.13 \pm 0.63$ | $4.32 \pm 1.22$ | $3.17 \pm 0.67$ | $7.9 \pm 0.01$ | $4.18 \pm 0.04$ | $3.06 \pm 0.03$ | $8136 \pm 9.06$ | – |
| | *Geyer* | $10.46 \pm 0.43$ | $8.82 \pm 2.28$ | $2.74 \pm 0.99$ | $10.45 \pm 0.03$ | $8.68 \pm 0.07$ | $2.75 \pm 0.04$ | $11569 \pm 93.97$ | – |
| **Medium** | *SLIPD* | $1.69 \pm 0.46$ | $5.41 \pm 1.34$ | $1.24 \pm 0.42$ | $1.54 \pm 0.02$ | $5.26 \pm 0.06$ | $1.19 \pm 0.02$ | $4931 \pm 7.00$ | $11.3 \pm 0.05$ |
| | ***AAS*** | $\mathbf{1.89 \pm 0.40}$ | $\mathbf{6.31 \pm 1.20}$ | $\mathbf{1.27 \pm 0.35}$ | $1.88 \pm 0.02$ | $6.29 \pm 0.04$ | $1.26 \pm 0.01$ | $4828 \pm 6.83$ | $11.9 \pm 0.05$ |
| | *SLIP* | $7.43 \pm 0.56$ | $6.24 \pm 1.35$ | $3.44 \pm 0.32$ | $6.92 \pm 0.01$ | $5.34 \pm 0.06$ | $3.47 \pm 0.04$ | $5921 \pm 9.47$ | – |
| | *Geyer* | $9.49 \pm 0.35$ | $9.11 \pm 2.54$ | $2.09 \pm 0.43$ | $9.49 \pm 0.01$ | $9.05 \pm 0.09$ | $2.10 \pm 0.02$ | $8308 \pm 30.50$ | – |
| **Soft** | *SLIPD* | $1.88 \pm 0.72$ | $5.56 \pm 2.34$ | $1.98 \pm 0.64$ | $1.83 \pm 0.02$ | $5.34 \pm 0.06$ | $1.72 \pm 0.02$ | $3570 \pm 3.88$ | $9.94 \pm 0.05$ |
| | ***AAS*** | $\mathbf{2.07 \pm 0.66}$ | $\mathbf{7.54 \pm 2.45}$ | $\mathbf{2.68 \pm 0.72}$ | $2.06 \pm 0.02$ | $7.54 \pm 0.08$ | $2.68 \pm 0.03$ | $3529 \pm 3.05$ | $9.84 \pm 0.04$ |
| | *SLIP* | $8.37 \pm 0.92$ | $8.12 \pm 2.55$ | $2.48 \pm 0.41$ | $7.93 \pm 0.02$ | $7.73 \pm 0.06$ | $2.13 \pm 0.02$ | $4645 \pm 2.34$ | – |
| | *Geyer* | $12.29 \pm 0.68$ | $15.25 \pm 4.12$ | $3.47 \pm 0.91$ | $12.28 \pm 0.03$ | $15.23 \pm 0.13$ | $3.48 \pm 0.04$ | $7602 \pm 25.65$ | – |
| **Softer** | *SLIPD* | $2.03 \pm 0.40$ | $8.23 \pm 1.74$ | $1.67 \pm 0.62$ | $2.02 \pm 0.02$ | $8.22 \pm 0.09$ | $1.65 \pm 0.02$ | $2598 \pm 5.79$ | $5.33 \pm 0.05$ |
| | ***AAS*** | $\mathbf{2.21 \pm 0.47}$ | $\mathbf{7.80 \pm 1.84}$ | $\mathbf{1.68 \pm 0.49}$ | $2.19 \pm 0.03$ | $7.74 \pm 0.06$ | $1.67 \pm 0.02$ | $2572 \pm 4.18$ | $6.49 \pm 0.03$ |
| | *SLIP* | $7.14 \pm 0.86$ | $11.57 \pm 2.36$ | $3.06 \pm 0.61$ | $5.95 \pm 0.01$ | $9.19 \pm 0.11$ | $2.79 \pm 0.03$ | $3128 \pm 8.67$ | – |
| | *Geyer* | $9.49 \pm 1.15$ | $19.97 \pm 5.85$ | $1.37 \pm 0.40$ | $9.47 \pm 0.06$ | $19.78 \pm 0.19$ | $1.38 \pm 0.02$ | $4117 \pm 29.88$ | – |

Our system identification process also reveals leg compliance and damping parameters for all four springs as listed in Table 2.5. Due to our adoption of the 30-fold cross-validation approach, we obtain 30 different values for these parameters, whose mean and standard deviation figures are summarized in Table 2.4. Note that the estimated leg compliance and damping parameters through AAS and SLIPD are very close to those revealed by vertical experiments. On the contrary, estimation through Geyer's predictor and SLIP results in unrealistic leg compliance values, since they assume zero damping in the leg.

Table 2.5: Estimated Leg compliance and damping parameters.

| Spring: | Softer | | Soft | | Medium | | Hard | |
|---|---|---|---|---|---|---|---|---|
| | $k$ | $d$ | $k$ | $d$ | $k$ | $d$ | $k$ | $d$ |
| SLIPD | 2598 | 5.3 | 3570 | 9.9 | 4931 | 11.3 | 6560 | 12.3 |
| **AAS** | **2572** | **6.5** | **3529** | **9.8** | **4828** | **11.9** | **6605** | **12.1** |
| SLIP | 3128 | – | 4645 | – | 5921 | – | 8136 | – |
| Geyer | 4117 | – | 7602 | – | 8308 | – | 11569 | – |
| Vertical | 2600 | 6.7 | 3536 | 9.9 | 4972 | 12.4 | 6630 | 12.7 |
| Manual | 2322 | – | 2915 | – | 4298 | – | 6282 | – |

Finally, we have also investigated the dependence of prediction performance on the asymmetry of the stride trajectory with respect to the gravitational vector. The concept of a *neutral touchdown angle* plays an important role in the characterization of equilibrium gaits for the ideal SLIP model [21]. Moreover, analytic approximations to SLIP trajectories preceding our contributions relied on the assumption of symmetric gaits, decreasing their efficacy for transient, asymmetric steps. Consequently, an evaluation of how prediction performance degrades as the touchdown leg angle deviates from its neutral choice was investigated in [9], revealing that the gravity correction featured in approximations substantially improves prediction performance. We present a similar evaluation on our experimental platform in the remainder of this section.

We begin by defining the *relative angle*, $\theta_{td,rel}$ as

$$\theta_{td,rel} := \theta_{td} - \theta_{td,n}, \tag{2.34}$$

to represent the deviation from the neutral angle $\theta_{td,n}$. An important difference from the corresponding definition in [9] is the fact that stance trajectories are never symmetric for our lossy SLIP model or the experimental platform. Consequently, our definition of a neutral angle focuses on forward velocity as the solution to the equation

$$\theta_{td,n} = \operatorname*{argmin}_{\theta}((\dot{y}_n - \dot{y}_{n+1}(\theta))^2) \tag{2.35}$$

which we use to compute the relative angle value corresponding to the initial condition associated with each experiment.

Fig. 2.6 shows our results for each of the four different spring stiffnesses, where marked data points represent different bins for the relative angle and the vertical axes represent mean and standard deviation values for the average positional prediction errors all experiments grouped in each bin. Continuous graphs show quadratic fits to the mean errors in each bin to reveal the dependence of the errors on the relative angle and coincide very well with mean data. Our results are consistent with those obtained from pure simulation studies, confirming that the gravity correction introduced by our approximations substantially improves the degradation in prediction performance away from symmetric gaits with positional errors remaining below 5% even when considerable asymmetry is present in the stride.

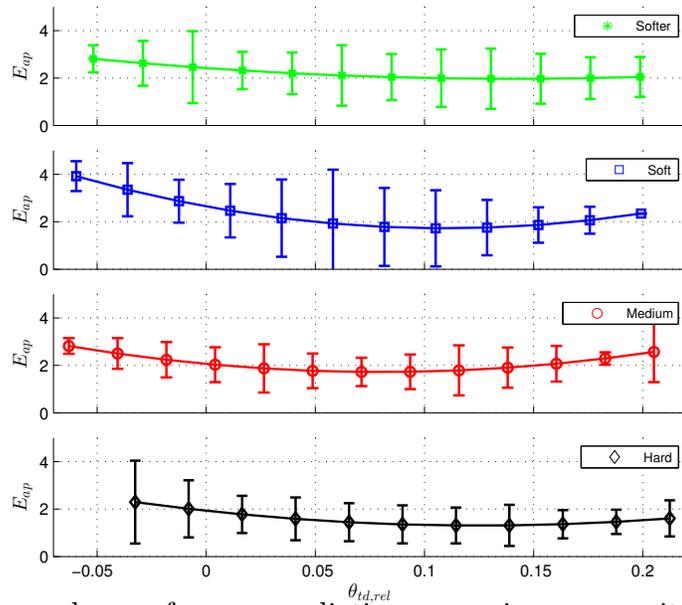

Figure 2.6: Dependence of mean prediction errors in apex position to the deviation from the neutral touchdown angle (relative angle) for all leg springs.

## 2.6 Conclusion

In this chapter, we presented the experimental validation of an approximate but accurate feed-forward predictor we recently introduced for the well-known Spring-Loaded Inverted Pendulum template. Our verification method first identifies unknown system parameters for our SLIP-based experimental platform, then evaluates prediction performance of the proposed predictor on the experimental data. We also compare the prediction performance of our predictor with Geyer's approximation as well as the numeric integration of SLIP dynamics with and without damping in the leg. Key extensions to the basic SLIP model, including viscous damping during flight and leg collision at liftoff, were also introduced to improve model performance in comparison to the experimental platform.

Our validation experiments include systematic tests by using four different leg springs, covering a large range of initial conditions and control inputs to show that the proposed map can provide accurate estimates for all trajectories of the experimental platform. Our method not only provides an experimental validation strategy for the SLIP predictors but also reveals insight into the effects of mechanical parameters of the physical robot platform. In particular, we observed that harder springs yield better prediction performance, confirming theoretical observations based on the nature of our approximations. Overall, our approximations can predict positional and velocity trajectories with mean 2% and 7%, respectively, well within the range of errors that can be tolerated by adaptive [48] or reactive [60] control strategies.

In addition to this performance characterization, we also investigated the performance of the predictions with respect to the relative touchdown angle, which is defined as the deviation from the touchdown angle that would yield symmetric trajectory. Hence, we also validate the practicality of the gravity correction incorporated by our approximations to the angular momentum. Our approximations preserve the accuracy even for non-symmetric trajectories where the angular momentum around the toe is no longer conserved.

# Chapter 3

# Input–Output Models of Legged Locomotion via Harmonic Transfer Functions

In this chapter, we adopt a data-driven approach, with the goal of furnishing an input–output representation of legged locomotor systems. Under certain assumptions, we can approximate hybrid dynamics of such systems around their limit cycle as a piecewise smooth linear time periodic (LTP) system. In this study, we first use a simple one-dimensional hybrid model in which a limit-cycle is induced through the actions of a linear actuator to illustrate the details of our method. We derive theoretical harmonic transfer functions (HTFs) of our example model. Then, we excite the model with small chirp signals to introduce perturbations around its limit-cycle and present systematic identification tests to estimate the HTFs for this model. Comparison between the data-driven HTFs model and its theoretical prediction illustrates the potential effectiveness of such empirical identification methods in legged locomotion. Besides, we present how these methods can be used to estimate input–output delays and to investigate stability characteristics of legged locomotion models. Finally, we test the practical usability of our approach on a more realistic legged locomotion model. The work presented in this chapter has also been reported and appeared in [61–65].

## 3.1 Introduction

Legged morphologies admit a wide range of locomotor behaviors, for which a variety of mathematical models have been proposed. For example, simple spring-mass models, including the Spring-Loaded Inverted Pendulum (SLIP) model [21], have been very successful in representing running and walking behaviors [1, 39]. The hybrid structure of these models alternates between flight and stance phases, each possessing smooth dynamics with continuous flows; transitions are punctuated with discrete, state-based transitions. Despite the seemingly simple nature of these models, however, their dynamics during stance include non-integrable parts [26], which prevent the derivation of exact closed-form solutions. Various approximate solutions were developed in the literature to address this problem [9–11, 13], some of which have also been verified experimentally [22] as detailed in Chapter 2.

It has also been shown that these models and associated solutions can support the design of hierarchical controllers for more complex platforms and morphologies [20, 66, 67]. Moreover, [48] showed that the structure and efficiency of these analytic solutions can also be exploited to yield effective solutions for parameter identification and adaptive control. Nevertheless, such explicit modeling efforts will always suffer from inaccuracies resulting from unmodeled aspects of physical platforms [22]. While one can certainly introduce more complexity along the template–anchors continuum [8], beyond a certain point these extended models begin to lose the analytic tractability in order to gain improvements in accuracy.

Here, we propose an alternative approach that steers away from explicit mechanical modeling towards data-driven system identification. Rather than introducing more and more specific, detailed mechanistic features of increasing complexity to mathematical models, such a data-driven approach treats the system as a black box, focusing on the adequacy of available data and the identification method to increase accuracy. The goal with such empirical models is to complement, not to replace, mechanistic models which have the benefit of trying to explicitly connect system behavior to physical design details and controller parameters.

More specifically, our approach in this study is to use a Linear Time Periodic (LTP) system structure to approximately model locomotor behaviors around their limit cycles, using associated system identification techniques to obtain a linearized input–output representation for the system. To this end, we first approximate state dependent hybrid transitions of these systems as time dependent transition functions at steady state. We then linearize these approximate dynamics around the limit cycle, yielding a piecewise smooth LTP system.

An important and powerful tool for the analysis and data-driven identification of LTP systems is the concept of Harmonic Transfer Functions (HTF), which are analogous to traditional transfer functions for Linear Time Invariant (LTI) systems [32]. Unlike LTI systems, an input signal with a specific frequency supplied to an LTP system produces output components spread across different harmonics of the periodic system frequency and components of HTF structure defines the coupling between different harmonics. Within this framework, [33] developed an identification strategy to estimate the individual HTFs for an LTP system by exciting the system using specially designed chirp signals and using modified "power spectral density" and "cross spectral density" functions as in the case of LTI systems.

In this work, we adopt this technique to estimate the harmonic transfer functions for a simplified vertical hopping robot model. Alternative identification strategies for LTP systems were also proposed by [34, 68, 69]. In the present study, we rely on chirp signals with their well-defined frequency range and predetermined power spectral density [70], rather than the sum of sines inputs used by [68]. [34] uses single sine inputs in their work, but their method requires multiple experiments to cover the frequency range that we are interested in this study. For these reasons, our identification method is based on [33].

Prior to our work, a system-identification method for smooth rhythmic dynamical systems was developed by [71] using continuous-time HTFs. Later, [69] developed a new identification method for hybrid (non-smooth) dynamical systems based on discrete-time HTFs. In this study, we show that identification methods based on continuous-time HTFs can be applied to a clock-driven hybrid dynamical model of locomotion.

In this context, we extend and apply the HTF based system identification method described in [33], which we briefly review in Section 3.3.3, to a clock-driven hybrid vertical hopping robot model for which the analytic derivation for the HTF representation appears to be challenging (or infeasible). In so doing, we show that identification methods normally designed for continuous systems can be applied to systems with hybrid components that are inevitable for legged locomotion, while also establishing the accuracy of the identification process through systematic simulation studies. In addition, our investigations revealed the fact that the LTP modeling of legged locomotor systems can also be used to identify input and measurement delays in the system separately which would otherwise be impossible with an LTI system model.

## 3.2 Legged Locomotion Models as a Linear Time Periodic (LTP) System

Our goal in this section is to represent the legged locomotion models as a hybrid dynamical system using linear time periodic (LTP) system models. As mentioned before, our ultimate goal is to provide a system identification framework for a class of models related to legged locomotion. For the current study, we limit ourselves to "clock-driven" locomotion models as described in Section 3.2.1, representative of controllers used by a wide variety of existing robots [30, 72], with open-loop central pattern generators (CPG) coordinating control actions to achieve time periodic behavior. This will allow us to directly use time periodicity in our LTP analysis, while eliminating a variety of complications associated with estimating the phase [73].

### 3.2.1 Smooth Clock-driven Oscillators

In general, the dynamics of smooth, clock-driven oscillators with external inputs can be written as

$$\begin{aligned}
\dot{q} &= f(q, \phi, u), \\
\dot{\phi} &= 1 \\
f &: \mathbb{R}^n \times S^1 \times \mathbb{R}^p \mapsto \mathbb{R}^n \\
(q, \phi) &\in \mathbb{R}^n \times S^1, \\
u &\in \mathbb{R}^q
\end{aligned} \quad (3.1)$$

where $(q, \phi)$ and $\mathbb{R}^n \times S^1$ denote the state vector and the state space of the oscillator, respectively. The circle component $S^1 = \mathrm{mod}(\mathbb{R}^+, T)$ enforces the periodicity of the dynamics, while the external input $u(t)$ represents small external perturbations which we will use for system identification.

In this study, we focus on oscillators of the form (3.1) with asymptotically stable, isolated periodic orbits (limit cycle) $\bar{q}(t) = \bar{q}(t - T)$ when $u(t) = 0$. For such systems, if we let $q(t) = \bar{q}(t) + x(t)$ and linearize the dynamics in (3.1) around the limit cycle $\bar{q}(t)$, and $u(t) = 0$ we get

$$\begin{aligned}
\dot{x}(t) &= A(t)x(t) + B(t)u(t) \\
y(t) &= C(t)x(t) + D(t)u(t)
\end{aligned} \quad (3.2)$$

where

$$A(t) = \left[\frac{\partial f}{\partial q}\right]_{\substack{q(t) = \bar{q}(t) \\ u(t) = 0}}, \quad (3.3)$$

$$B(t) = \left[\frac{\partial f}{\partial u}\right]_{\substack{q(t) = \bar{q}(t) \\ u(t) = 0}}. \quad (3.4)$$

This corresponds to a linear time periodic (LTP) system, with all system matrices sharing a common period, $T$.

## 3.2.2 Modeling Framework for Hybrid Systems

Legged systems are often modeled using hybrid dynamics due to intermittent foot contact with the ground, which cannot be represented with a single, smooth dynamical flow. In the broadest sense, a hybrid dynamical system is a set of smooth flows together with discrete transitions (and associated transformations) between these flows triggered by intersections of system trajectories with sub-manifolds of the continuous state space [74]. These flows are called *charts*, indexed with unique labels $\mathcal{I} := \{0, \cdots, d\}$ each with possibly different equations of motion. Along its trajectories, a hybrid system transitions from one chart to another, with each transition defined by the zero crossing of a *threshold function*. For each source chart $\alpha \in \mathcal{I}$ and destination chart $\beta \in \mathcal{I}$, the threshold function $h_\alpha^\beta$ defines the transition from chart $\alpha$ to chart $\beta$. An example transition graph for a hybrid dynamical system is illustrated in Fig. 3.1.

Since we are interested in the local behavior around the limit cycle, we assume that there is only one transition function associated with each chart.[1] We further assume that the hybrid dynamical system we consider has an isolated periodic orbit ensuring that chart transitions within the limit cycle are also periodic and consistent.

It is important to note that these assumptions are generally satisfied by models of common locomotory behaviors such as running and walking [9, 75] as well as a wide range of legged robots for which leg masses are negligible compared to the dynamics of a larger body [30, 72]. Consequently, the system identification methods we introduce will remain applicable to systems other than the simplified examples we will present in this chapter.

---

[1]This approach does not apply to gaits such as pronking that nominally involve multiple legs making contact with the ground at the same time on the limit cycle, because small deviations from the limit cycle can lead to different touchdown order between legs, violating our assumption.

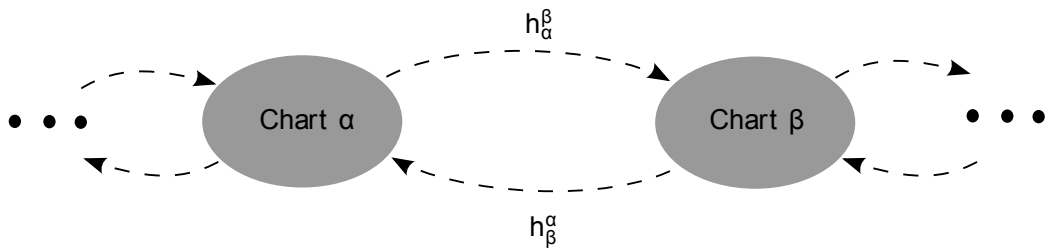

Figure 3.1: A simple state transition graph for a hybrid dynamical system.

### 3.2.3 Modeling Legged Locomotion as a Linear Time Periodic (LTP) System

For clarity, we limit our focus in this section to an example hybrid dynamical system with only two charts, $\mathcal{I} = \{0, 1\}$, designed to capture stance and flight phases of simple spring-mass models of locomotion. Based on a clock driven assumption, for each $i \in \mathcal{I}$ the continuous dynamics can be represented with

$$\dot{\phi} = 1$$
$$\dot{q}_i = f_i(q, \phi, u) , \qquad (3.5)$$
$$q_i \in \mathbb{R}^n$$

and let the associated threshold function be $h_i^{\mathrm{mod}(i+1,2)}(q)$. The transition map associated with each hybrid event is simply the identity map, $q_i \mapsto q_i$, due to the continuity assumption. Our linearization of these hybrid dynamics towards an LTP approximation assumes that these transition times, $\hat{t}$, zero crossings of $h_0^1(q)$ and $h_1^0(q)$, maintain their periodicity and offsets within the period in close proximity of the limit cycle, resulting in the following form of the nonlinear dynamics

$$\dot{\phi} = 1 \qquad (3.6)$$

$$\dot{q} \approx \begin{cases} f_0(q, \phi, u) , & \text{if } \mathrm{mod}(t, T) \in [0, \hat{t}) \\ f_1(q, \phi, u) , & \text{if } \mathrm{mod}(t, T) \in [\hat{t}, T) \end{cases} . \qquad (3.7)$$

Assuming that the system given above has a limit cycle $\bar{q}(t)$ with a period $T$, linearization around $\bar{q}(t)$ yields the piecewise smooth LTP system

$$\dot{x}(t) = \begin{cases} A_0(t)x(t) + B_0(t)u(t), & \text{if } \mathrm{mod}(t, T) \in [0, \hat{t}) \\ A_1(t)x(t) + B_1(t)u(t), & \text{if } \mathrm{mod}(t, T) \in [\hat{t}, T) \end{cases}$$

where

$$A_0(t) := \left[\frac{\partial f_0}{\partial q}\right]_{\substack{q(t) = \bar{q}(t) \\ u(t) = 0}}, B_0(t) := \left[\frac{\partial f_0}{\partial u}\right]_{\substack{q(t) = \bar{q}(t) \\ u(t) = 0}},$$

$$A_1(t) := \left[\frac{\partial f_1}{\partial q}\right]_{\substack{q(t) = \bar{q}(t) \\ u(t) = 0}}, B_1(t) := \left[\frac{\partial f_1}{\partial u}\right]_{\substack{q(t) = \bar{q}(t) \\ u(t) = 0}}.$$

It is natural to assume that direct measurement of all $x(t)$ may not be available or we may only measure a subset of $x(t)$. Consequently, we also define a time-periodic output equation as in the form (3.9).

Since system matrices $A_i(t)$, $B_i(t)$, $C_i(t)$ and $D_i(t)$ with $i \in \{0, 1\}$ are time parametrized functions, the system has infinite parametric degrees of freedom, making parametric system identification challenging even when Harmonic Transfer Functions (HTFs) are used. At this point, we hypothesize that for hybrid systems, the variability within a chart is small compared to the change between charts and we approximate the LTP dynamics using a piecewise LTI approximation that preserves the LTP structure of the system. The LTP equations of motion then take the form

$$\dot{x}(t) \approx \begin{cases} A_0 x(t) + B_0 u(t), & \text{if } \mathrm{mod}(t, T) \in [0, \hat{t}) \\ A_1 x(t) + B_1 u(t), & \text{if } \mathrm{mod}(t, T) \in [\hat{t}, T) \end{cases} \quad (3.8)$$

$$y(t) \approx \begin{cases} C_0 x(t) + D_0 u(t), & \text{if } \mathrm{mod}(t, T) \in [0, \hat{t}) \\ C_1 x(t) + D_1 u(t), & \text{if } \mathrm{mod}(t, T) \in [\hat{t}, T) \end{cases} \quad (3.9)$$

The formulation above constitutes the basis of our framework for analyzing and identifying clock-driven legged locomotion models.

## 3.3 Harmonic Transfer Functions (HTFs)

Many finite-dimensional Linear Time Periodic systems can be described by a state space model of the form

$$\begin{aligned} \dot{\mathbf{x}}(t) &= \mathbf{A}(t)\mathbf{x}(t) + \mathbf{B}(t)\mathbf{u}(t) \\ \mathbf{y}(t) &= \mathbf{C}(t)\mathbf{x}(t) + \mathbf{D}(t)\mathbf{u}(t), \end{aligned} \quad (3.10)$$

where $\mathbf{A}(t)$, $\mathbf{B}(t)$, $\mathbf{C}(t)$, and $\mathbf{D}(t)$ are all periodic with a common period $T$. Despite its linear nature, the time dependence of matrices in this representation make it impossible to directly apply analysis and identification techniques developed for LTI systems.

System identification methods for asymptotically stable LTI systems are well established, thanks in large measure to the one-to-one mapping between frequency response characteristics of input and output signals at steady state. This allows one to obtain empirical frequency response functions (i.e. "Bode plots") describing the magnitude and phase changes in the input signal at each specific frequency for the desired system. Due to the time dependence of matrices in (3.10), however, the response to a sinusoidal input with a specific frequency passing through an LTP system may include multiple (possibly infinite) harmonics, with different magnitudes and phases. In general, neglecting all higher order harmonics of the system [76] to obtain a one-to-one mapping between the input and output signals in frequency domain may lead to unacceptable inaccuracies. Consequently, a different approach is required for data-driven identification of such systems.

Wereley proposed a solution to this problem by transforming the input and output signals to exponentially modulated periodic (EMP) signals [32]. In this domain, it is possible to obtain a one-to-one mapping between the Fourier coefficients of the input and output EMP signals. In the resulting structure input–output representation is determined by multiple (possibly infinite) parallel LTI subsystems whose inputs are multiplied by complex periodic signals. Fig. 3.2 illustrates the resulting HTF structure. These LTI subsystems are called Harmonic Transfer Functions (HTFs) and they characterize frequency response characteristics of an LTP system. A detailed frequency domain analysis of linear time periodic systems based on harmonic transfer functions has also been investigated by [77]. Subsequent subsections in this section review theoretical derivations behind the HTF framework as proposed by [32] and [78].

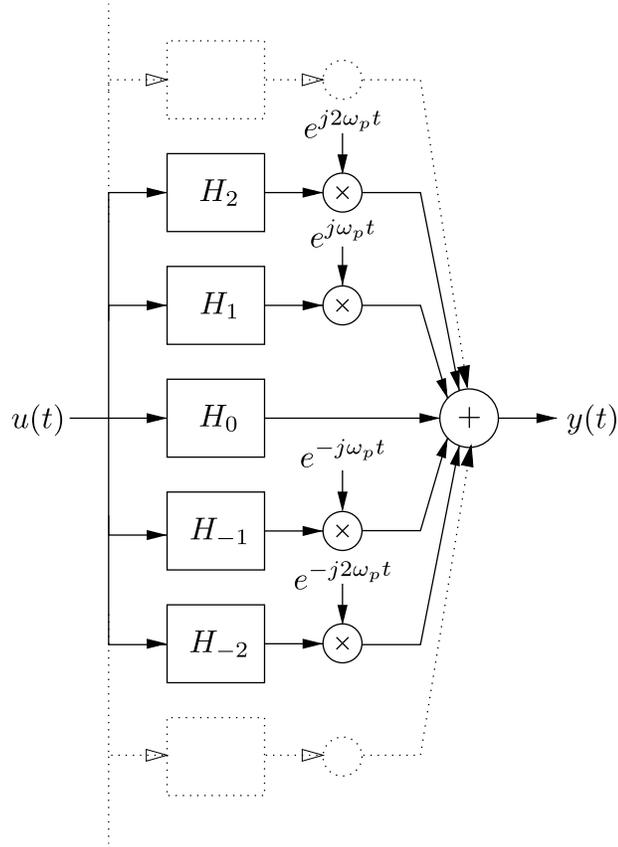

Figure 3.2: Illustration of HTF structure. The input–output relation of an LTP system can be expressed by multiple (possibly infinite) parallel LTI sub-systems

### 3.3.1 Derivation of HTFs via Harmonic Balance

In this section, we overview the derivation of harmonic transfer functions (HTFs) as presented in [32, 79] using the principle of harmonic balance starting from the state space representation (3.10). Note that system matrices in (3.10) are all T-periodic. Consequently, they can be exactly represented by an infinite Fourier series expansion with pumping frequency $\omega_p = 2\pi/T$, where $T$ is the common period of the system matrices. That is to say, we have

$$A(t) = \sum_{n=-\infty}^{\infty} A_n e^{j\omega_p n t} . \qquad (3.11)$$

where the matrices $B(t)$, $C(t)$ and $D(t)$ can be similarly decomposed.

Since both the input and output vectors are EMP signals [32], we can also expand the state and output vectors as

$$x(t) = \sum_{n=-\infty}^{\infty} x_n e^{s_n t} \tag{3.12}$$

$$\dot{x}(t) = \sum_{n=-\infty}^{\infty} s_n x_n e^{s_n t} \tag{3.13}$$

$$y(t) = \sum_{n=-\infty}^{\infty} y_n e^{s_n t}, \tag{3.14}$$

where we have $s_n = s + j\omega_p n$.

Substituting these expansions into (3.10) and grouping similar terms, one obtains

$$\begin{aligned}
0 &= \sum_{n=-\infty}^{\infty} (s_n x_n - \sum_{m=-\infty}^{\infty} A_{n-m} x_m - \sum_{m=-\infty}^{\infty} B_{n-m} u_m) e^{s_n t} \\
0 &= \sum_{n=-\infty}^{\infty} (y_n - \sum_{m=-\infty}^{\infty} C_{n-m} x_m - \sum_{m=-\infty}^{\infty} D_{n-m} u_m) e^{s_n t}
\end{aligned} \tag{3.15}$$

Multiplying each equation by $e^{-st}$, one then gets

$$\begin{aligned}
0 &= \sum_{n=-\infty}^{\infty} (s_n x_n - \sum_{m=-\infty}^{\infty} A_{n-m} x_m - \sum_{m=-\infty}^{\infty} B_{n-m} u_m) e^{j\omega_p n t} \\
0 &= \sum_{n=-\infty}^{\infty} (y_n - \sum_{m=-\infty}^{\infty} C_{n-m} x_m - \sum_{m=-\infty}^{\infty} D_{n-m} u_m) e^{j\omega_p n t}.
\end{aligned}$$

The set of exponentials $e^{j\omega_p n t}$ create an orthonormal basis on $L_2[0, T]$ and as such each term enclosed by braces must individually be equal to zero. This procedure is also known as the principle of harmonic balance. After applying this procedure, we have $\forall n \in \mathbb{Z}$:

$$\begin{aligned}
s_n x_n &= \sum_{m=-\infty}^{\infty} A_{n-m} x_m + \sum_{m=-\infty}^{\infty} B_{n-m} u_m, \\
y_n &= \sum_{m=-\infty}^{\infty} C_{n-m} x_m + \sum_{m=-\infty}^{\infty} D_{n-m} u_m.
\end{aligned} \tag{3.16}$$

Even though the above equations are a concise representation of the input-output relationship between the Fourier series coefficients of the input and output signals, performing operations with infinite sums is tedious. Consequently, we adopt a Toeplitz form notation to expand the infinite summations with matrix operations. The system of equations in (3.16) can hence be expressed as the doubly infinite matrix equation,

$$\begin{aligned} s\mathcal{X} &= (\mathcal{A} - \mathcal{N})\mathcal{X} + \mathcal{B}\mathcal{U} \\ \mathcal{Y} &= \mathcal{C}\mathcal{X} + \mathcal{D}\mathcal{U}, \end{aligned} \tag{3.17}$$

where the doubly infinite vectors representing the harmonics of the state, control, and output signals are

$$\begin{aligned} \mathcal{X}^T &= [\cdots, x_{-2}^T, x_{-1}^T, x_0^T, x_1^T, x_2^T, \cdots], \\ \mathcal{U}^T &= [\cdots, u_{-2}^T, u_{-1}^T, u_0^T, u_1^T, u_2^T, \cdots], \\ \mathcal{Y}^T &= [\cdots, y_{-2}^T, y_{-1}^T, y_0^T, y_1^T, y_2^T, \cdots] \end{aligned} \tag{3.18}$$

and the doubly infinite block diagonal matrix containing all harmonics of the pumping frequency is

$$\mathcal{N} = \text{blockdiag}\{jn\omega_p I\} \quad \forall n \in Z. \tag{3.19}$$

The T-periodic dynamics matrix, $A(t)$, is expressed in terms of its complex Fourier series coefficients, $\{A_n | n \in \mathbb{Z}\}$, as a doubly infinite block Toeplitz matrix as

$$\mathcal{A} = \begin{bmatrix} \ddots & \vdots & \vdots & \vdots & \vdots & \vdots \\ \cdots & A_0 & A_{-1} & A_{-2} & A_{-3} & A_{-4} & \cdots \\ \cdots & A_1 & A_0 & A_{-1} & A_{-2} & A_{-3} & \cdots \\ \cdots & A_2 & A_1 & A_0 & A_{-1} & A_{-2} & \cdots \\ \cdots & A_3 & A_2 & A_1 & A_0 & A_{-1} & \cdots \\ \cdots & A_4 & A_3 & A_2 & A_1 & A_0 & \cdots \\ & \vdots & \vdots & \vdots & \vdots & \vdots & \ddots \end{bmatrix} \tag{3.20}$$

with a similar definition for B in terms of its Fourier coefficients represented by $\{B_n | n \in \mathbb{Z}\}$, C in terms of $\{C_n | n \in \mathbb{Z}\}$, and D in terms of $\{D_n | n \in \mathbb{Z}\}$.

This doubly infinite matrix representation (3.17) is called the harmonic state space model of the system given in (3.10). However, it will also be useful to determine an explicit input-output functional relationship between the Fourier series coefficients, or the harmonics of the input, $\{u_n | n \in \mathbb{Z}\}$, and those of the output, $\{y_n | n \in \mathbb{Z}\}$.

This relationship is represented by the harmonic transfer functions, $G(s)$, which is an infinite dimensional matrix of Fourier series coefficients describing the relationship between the harmonics of the input signal, and those of the output signal, such that

$$\mathcal{Y} = G\mathcal{U}, \tag{3.21}$$

where

$$G = \mathcal{C}[sI - (\mathcal{A} - \mathcal{N})]^{-1}\mathcal{B} + \mathcal{D} \tag{3.22}$$

as long as the inverse within this equation exists.

There are, however, two problems associated with the harmonic transfer function as stated above. First, it is not clear whether the inverse of the doubly infinite matrix in the definition of the harmonic transfer function will always exist. This problem will be dealt with by application of the Floquet Theorem, where it is simpler to check the conditions for the existence of the inverse operation. Second, the harmonic transfer functions is a doubly infinite matrix operator, which cannot practically be implemented on a computer. This second problem will be mitigated by truncating the HTF in order to implement analysis on computer.

### 3.3.2 Derivation of HTFs via Time Periodic Impulse Response

Different than Section 3.3.1, in this section, we summarize the derivation of HTFs using the time periodic impulse response representation of an LTP system as explained in [78] to establish some notation regarding this chapter. The output of an LTP system, such as those in the form (3.10), can also be represented by using its time periodic impulse response functions as

$$y(t) = \int_0^t H(t, \tau) u(\tau) d\tau, \tag{3.23}$$

where $H(t, \tau) = H(t - T, \tau - T)$ and $T$ is the period of the system. In other words, all time-varying impulse response functions of the system are periodic in both arguments. Letting $\tau = t - r$, we have $H(t, \tau) = H(t, t - r)$ which is $T$ periodic in $t$. This periodicity allows us to expand $H(t, t-r)$ through an infinite Fourier series expansion with pumping frequency, $\omega_p = 2\pi/T$, yielding

$$H(t, t-r) = \sum_{k=-\infty}^{\infty} H_k(r) e^{jk\omega_p t},$$
$$H_k(r) := \frac{1}{T} \int_0^T e^{-jk\omega_p t} H(t, t-r) dt. \tag{3.24}$$

Switching back to $\tau$ through $r = t - \tau$ gives

$$H(t, \tau) = \sum_{k=-\infty}^{\infty} H_k(t - \tau) e^{jk\omega_p t}. \tag{3.25}$$

Substituting (3.25) into (3.2), we get

$$\begin{aligned} y(t) &= \sum_{k=-\infty}^{\infty} \int_0^T H_k(t - \tau) e^{jk\omega_p t} u(\tau) d\tau \\ &= \sum_{k=-\infty}^{\infty} \int_0^T H_k(t - \tau) e^{jk\omega_p t - \tau} u(\tau) e^{jk\omega_p \tau} d\tau \\ &= \sum_{k=-\infty}^{\infty} (H_k(t) e^{jk\omega_p t}) * (u(t) e^{jk\omega_p t}), \end{aligned} \tag{3.26}$$

where $*$ denotes the convolution operator. Finally, taking the Laplace transform of both sides yields

$$Y(s) = \sum_{k=-\infty}^{\infty} H_k(s - jk\omega_p) U(s - jk\omega_p). \tag{3.27}$$

Now, let us define $G_k(s) := H_k(s - jk\omega_p)$ as the elements of the HTF structure.

### 3.3.3 Data-Driven Identification of Harmonic Transfer Functions

When an explicit representation of a system is given, either in state space form (3.10) or through an impulse response function as in (3.23), the derivations of Section 3.3 can be used to obtain the corresponding harmonic transfer functions. However, manual construction of such explicit models is often impractical beyond a certain level of complexity. Consequently, the estimation of harmonic transfer functions without the need for such explicit models is of great practical interest. In this section, we review the data-driven system identification method introduced by [33] for LTP systems, together with our extensions to support its application for clock–driven legged locomotion models.

The HTF structure of (3.27) includes an infinite number of harmonics, which is problematic for practical applications. Consequently, these harmonic components are often truncated beyond a certain order to enable effective computational implementations. Similarly, some LTP system identification methods also focus on a preselected number of harmonics. For clarity, our review of the method given in [33] considers only three harmonic transfer functions, $\hat{\mathbf{G}}_0$, $\hat{\mathbf{G}}_{-1}$ and $\hat{\mathbf{G}}_1$, leading to a representation of the system output in frequency domain as

$$Y(j\omega) \approx \hat{Y}(j\omega) := \hat{\mathbf{G}}_0(j\omega)U(j\omega) \tag{3.28}$$
$$+ \hat{\mathbf{G}}_{-1}(j\omega)U(j\omega + j\omega_p) \tag{3.29}$$
$$+ \hat{\mathbf{G}}_1(j\omega)U(j\omega - j\omega_p), \tag{3.30}$$

where variables annotated with a hat denote estimated versions of their system counterparts. Based on this definition, the data-driven system identification problem can be reduced to the problem of estimating the quantities $\hat{\mathbf{G}}_0(j\omega)$, $\hat{\mathbf{G}}_{-1}(j\omega)$ and $\hat{\mathbf{G}}_1(j\omega)$ at each specific frequency, $\omega$, such that the difference between the measured and estimated output vectors is minimized.

The correct choice of input signals plays a crucial role in the system identification process. Input signals must be designed to expose as much dynamic behavior in the system as possible. To this end, chirp signal inputs as shown in

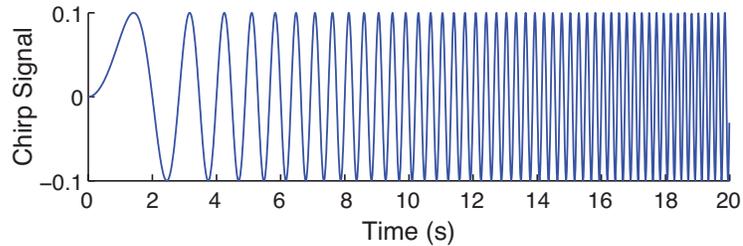

Figure 3.3: Chirp signal used to perturb the system for system identification, consisting of a sinusoid with amplitude 0.1 and frequency increasing linearly in time within the range $(0, 5]\,\text{Hz}$ in $20s$.

Fig. 3.3, can be used to to cover a sufficiently wide frequency spectrum. The accuracy of the identification critically depends on particular aspects of these chirp inputs, such as its duration, frequency range and sweep rate. Moreover, the phase timing of the input signal relative to the LTP system also effects the activation of different dynamic components within the system. [33] addresses these problems by designing a single input sequence, incorporating phase-shifted replicas of an original chirp signal spanning a sufficiently wide range of frequencies. Phase shifts ensure that the system is excited evenly throughout the system's period, while the wide frequency range explores different modes in the system towards a complete characterization of the effects of internal system dynamics on system output.

Using the resulting input–output pairs, one can then compute "extended power spectral density" and "extended cross spectral density" functions $\Phi_{UU}(\omega)$ and $\Phi_{UY}(\omega)$, respectively. Even though $\Phi_{UU}(\omega)$ and $\Phi_{UY}(\omega)$ are analogous to power spectral density and cross spectral density functions for LTI systems, they differ in their representation as matrix operators. Detailed definitions of these functions can be found in [33] and are omitted here for space considerations.

Now, the three harmonic transfer function components of the LTP system can

be estimated as

$$\hat{\mathbf{G}}(j\omega) := \begin{bmatrix} \hat{\mathbf{G}}_1 \\ \hat{\mathbf{G}}_0 \\ \hat{\mathbf{G}}_{-1} \end{bmatrix} = (\Phi_{UU})^{-1}\Phi_{UY} . \tag{3.31}$$

An important problem with this formulation, however, is that it embodies an under-determined fitting problem. A single pair of input and output vectors may not be sufficient to accurately estimate all three harmonic transfer functions even though (3.30) represents a single-input single-output LTP system. In order to address this problem, [33] observes continuity properties of physical transfer functions and introduces additional constraints to penalize high curvature within each harmonic transfer function towards better output prediction performance. More formally, having modeled the system with three harmonic transfer functions, its output response due to an input can be expressed as

$$Y(j\omega) = \sum_{k=-1}^{1} U(j\omega - kj\omega_p)\hat{\mathbf{G}}_k + E(j\omega) \tag{3.32}$$

$$= \hat{Y}(j\omega) + E(j\omega). \tag{3.33}$$

The error term captures the difference between the measured system response and the predictions of the estimated harmonic transfer functions. The cost function adopted by [33] for the minimization problem penalizes the quadratic output prediction error and the curvature of the harmonic transfer functions, taking the form

$$\mathbf{J} := [(\mathbf{Y} - \mathbf{U}^T\hat{\mathbf{G}})^2 + \alpha(\mathbf{D^2}\hat{\mathbf{G}})^2], \tag{3.34}$$

where $\mathbf{D}^2$ is the second order difference operator, $\alpha$ is a scalar weight to tune the smoothness of resulting transfer functions and $\mathbf{Y}$, $\mathbf{U}$ and $\hat{\mathbf{G}}$ are defined as in [33]. Differentiating $\mathbf{J}$ with respect to $\hat{\mathbf{G}}$ and equating to zero yields the estimated harmonic transfer functions as

$$\hat{\mathbf{G}}(j\omega) = (\Phi_{UU} + \alpha\mathbf{D}^4)^{-1}\Phi_{UY} . \tag{3.35}$$

## 3.4 Application to a Simplified Legged Locomotion Model with Hybrid System Dynamics

In this section, we describe a simple, vertically constrained spring-mass-damper system that possesses hybrid structural properties similar to the extensively studied Spring-Loaded Inverted Pendulum (SLIP) model for running behaviors. This will provide a simple example to illustrate the application of the proposed system identification method to such systems.

### 3.4.1 System Dynamics

Fig. 3.4 illustrates the vertical leg model we focus on in this section. It consists of a mass attached to a leg with a spring-damper mechanism as well as a force transducer. Unlike the SLIP model, we assume that the toe is permanently affixed on the ground. Nevertheless, we recover the hybrid nature of locomotory gaits by assuming that the damper is turned on during a "stance phase" (lossy) and off during a "flight phase" (lossless). This construction recovers the hybrid nature of the dynamics, while allowing active input throughout the entire trajectory to support the generation of system identification data, as well as admitting theoretical computation of its harmonic transfer functions for a comparative investigation.

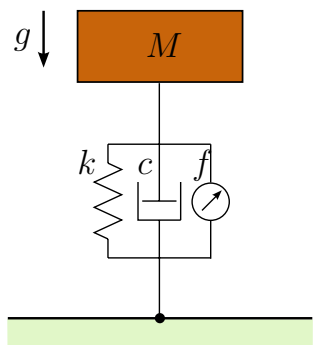

Figure 3.4: A Simplified leg model as a spring-mass-damper mechanism with linear force transducer.

We use the force transducer $f$ in this system for two purposes. Firstly, active energy input to the system must be provided to maintain the limit cycle and compensate for energy losses due to the presence of damping. Second, it will be used as an exogenous input to the system to support the system identification process. Many physical legged platforms include similar active components in their legs to regulate their mechanical energy [80, 81]. Notwithstanding differences in how these actuators are incorporated into the system, they can all be used as the necessary exogenous inputs to perform system identification. A similar model was also investigated in [35] but using an additional nonlinear spring for energy regulation.

The equations of motion for this simplified legged locomotion model are given by
$$m\ddot{x} = \begin{cases} -mg - c\dot{x} - k(x - x_0) + \omega_p, & \text{if } \dot{x} > 0 \\ -mg - k(x - x_0) + \omega_p, & \text{otherwise.} \end{cases} \quad (3.36)$$
The lossy and lossless dynamics in (3.36) correspond to different charts in Fig. 3.1 and zero crossings of $\dot{x}$ represent threshold functions for both phases.

Our illustrative examples use the parameters $g = 9.81$, $k = 200$, $c = 2$, $m = 1$ and $x_0 = 0.2$, chosen to be similar to the parameters of a vertical hopper platform in our laboratory [82]. As noted above, we choose the linear actuator input $\omega_p = f_0(t) + u(t)$, consisting of a forcing term $f_0(t)$ to compensate for energy losses, and a chirp signal $u(t)$ to introduce small periodic perturbations for system identification.

### 3.4.2 Theoretical Computation of Harmonic Transfer Functions

The goal of this section is to compute harmonic transfer functions for our model around its limit cycle as outlined in Section 3.3.1.

We first assume that the forcing input $f_0(t)$ is appropriately chosen to induce an asymptotically stable limit cycle for this system. For example, our simple leg model achieves a stable limit cycle with $f_0(t) = \cos(2\pi t)$. At this point, changing into error coordinates away from the limit cycle with $\xi = x(t) - \bar{x}(t)$, and substituting into (3.36), the equations of motion take the form

$$\ddot{\xi} = \begin{cases} -c\dot{\xi} - k\xi, & \text{if } \dot{\xi} + \dot{\bar{x}}(t) > 0 \\ -k\xi, & \text{otherwise} \end{cases} \qquad (3.37)$$

Due to the simplicity of the dynamics, this corresponds to a piecewise LTI system without necessitating any additional approximations, taking the form

$$\begin{bmatrix} \dot{\xi}_1 \\ \dot{\xi}_2 \end{bmatrix} = \begin{bmatrix} 0 & 1 \\ -k & -cs(\dot{\xi}, t) \end{bmatrix} \begin{bmatrix} \xi_1 \\ \xi_2 \end{bmatrix} + \begin{bmatrix} 0 \\ 1 \end{bmatrix} u(t), \qquad (3.38)$$

where the hybrid nature of the system is captured by the flag $s(\dot{\xi}, t)$, with $s = 1$, when $\dot{\xi} + \dot{\bar{x}}(t) > 0$ and $s = 0$ otherwise.

We now need to represent this piecewise LTI system as a linear time periodic system. However, even though the binary valued function $s(\dot{\xi}, t)$ can be considered time-periodic on the limit cycle itself, this is not the case for trajectories away from the limit cycle. To proceed, we hence assume that input induced perturbations are small, and that the binary valued function $s(\dot{\xi}, t)$ maintains its period and becomes strictly time dependent rather than state dependent, taking the form $s(\dot{\xi}, t) \approx s(t)$. We can now perform a Fourier series expansion on $s(t)$ by treating it as a square wave with an offset to obtain a linear time periodic system in the form

$$\begin{bmatrix} \dot{\xi}_1 \\ \dot{\xi}_2 \end{bmatrix} = \begin{bmatrix} 0 & 1 \\ -k & -cs(t) \end{bmatrix} \begin{bmatrix} \xi_1 \\ \xi_2 \end{bmatrix} + \begin{bmatrix} 0 \\ 1 \end{bmatrix} u(t), \qquad (3.39)$$

$$y = \begin{bmatrix} 1 & 0 \end{bmatrix} \begin{bmatrix} \xi_1 \\ \xi_2 \end{bmatrix}.$$

Plugging these equations into the HTF framework described in Section 3.3.1 yields analytic solutions to the harmonic transfer functions. We use the resulting analytic solutions for the harmonic transfer functions up to $n_h = 10$ to evaluate the output of our system identification method.

### 3.4.3 Estimation of Harmonic Transfer Functions using Input–Output Data

In this section, we obtain harmonic transfer functions corresponding to the linearized dynamics of (3.39) by using input–output data without assuming prior knowledge of the state space model. Using $f_0(t) = \cos(2\pi t)$ and $u(t) = 0$ for 30 cycles without a perturbation, our example system stabilized to a limit cycle $\bar{x}(t)$ with a period $T = 1s$. We use the $30^{th}$ period as the numerical limit cycle of the nonlinear system and subtract it from the trajectories of subsequent experiments to obtain the error function $\xi_1$.

In order to obtain input–output data for system identification, we apply an input signal consisting of nine subsequent $30s$ long chirp signals, each with a linearly increasing frequency in the range $(0, 7]$ Hz over its duration but with a different starting phase evenly distributed across the system's period, $T = 1s$. Each chirp signal has an amplitude of 0.004, chosen to be large enough to perturb system dynamics but small enough to keep the system close to the periodic orbit. A sample chirp signal with zero phase can be generated by

$$u(t) = 0.004 \sin(14\pi t^2/30). \tag{3.40}$$

The resulting output is then subtracted from the numerically measured limit cycle to obtain error trajectories $\xi_1$ for vertical position. The input signal and $\xi_1$ are then used as in Section 3.3.3 to estimate harmonic transfer functions for our system. Since our theoretical computations showed that responses beyond the third harmonic were very small, we only consider the fundamental harmonic and three harmonics on both sides for our experiments.

Fig. 3.5 illustrates the estimation performance of our algorithm for the magnitude and phase of the fundamental harmonic. Both graphs show that the application of the identification algorithm in [33] works well even for nonlinear periodic systems with hybrid dynamics.

We also show our identification results for three harmonics in both the negative and positive sides in Fig. 3.6. Even though magnitudes for the harmonic transfer functions are small compared to the fundamental, the identification algorithm can provide accurate estimates for these transfer functions except in some narrow regions of $G_{-2}$ and $G_2$. The identification algorithm could not correctly estimate these two harmonics around $12-15$ (rad/s). One possible reason for this discrepancy is the presence of strong responses in all harmonics around the same frequency except $G_{-2}$ and $G_2$, resulting in the inability of the identification algorithm to distinguish between the contributions from each harmonic absent knowledge of the internal system dynamics. Alternatively, these discrepancies may also be a result of the fact that hybrid transitions are not strictly time periodic (rather, they are state-dependent) which likely has effects on different frequencies and harmonics. We plan on investigating these issues further in the future.

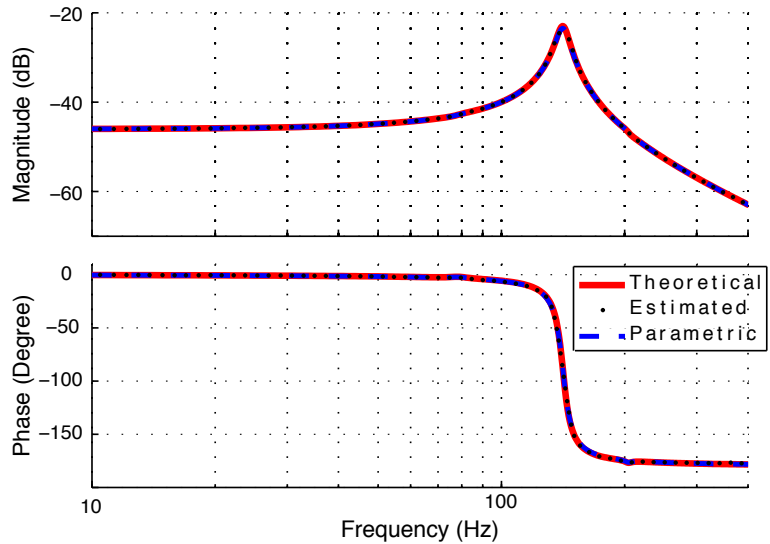

Figure 3.5: Estimation results for the fundamental harmonic. The upper figure illustrates the magnitude plots obtained via theoretical derivation and data-driven identification of HTFs as well as parametric system identification. The lower plot illustrates the phase responses.

For a comparative analysis, we also present results from a parametric identification in order to show that further corrections on estimation results from a non-parametric method are possible. To this end, we fit the system parameters $k$ and $c$ in (3.39) by comparing root mean square error between theoretically computed and estimated harmonic transfer functions $G_0$, $G_{-1}$ and $G_1$. We truncate the system response after the first harmonic in order to discard erroneous regions in higher harmonics. The resulting estimates were $\hat{k} = 200$ for the spring constant and $\hat{c} = 2.12$ for the damping coefficient, which closely coincide with the parameters used to generate the input–output data. As such, harmonic transfer functions obtained from parametric identification were found to closely match those obtained from theoretical computations as seen in Fig. 3.6.

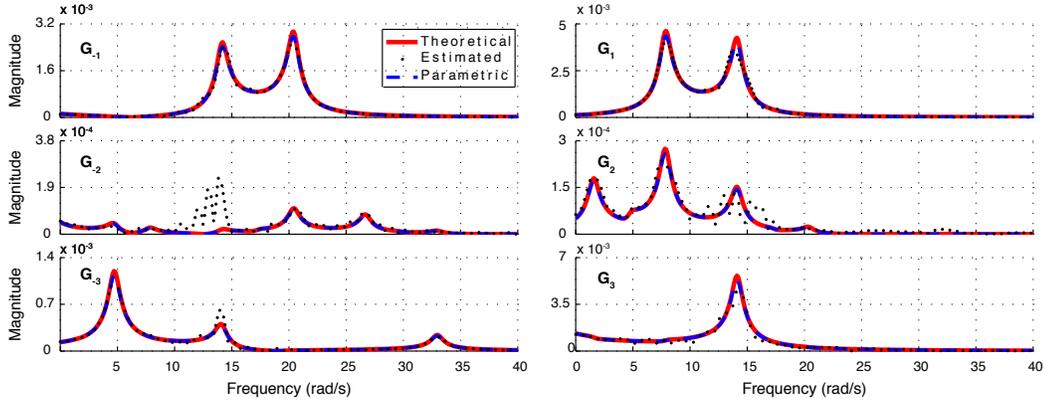

Figure 3.6: Estimation results for the higher harmonics. The magnitude plots for the first three harmonics are illustrated as a comparison of theoretical derivation, data-driven identification and parametric identification.

## 3.5 Identifying Stability Properties via Harmonic Transfer Functions

In this section, our goal is to develop a method for the identification of the stability properties of legged locomotor dynamics using input–output data based system identification approaches. The proposed method does not require downsampling the raw data and does not presume full state measurement as in the case of model-based identification methods. Previous sections show that an input–output linear time periodic (LTP) system structure can be used to represent rhythmic locomotor behaviors around their limit cycles. In this section, we show that our data-driven system identification technique can be utilized to characterize the stability properties of limit cycles for clock-driven legged locomotion models. To this end, we again use the concept of harmonic transfer functions (HTFs), first to obtain a non-parametric system model based on input–output data of the system and then to identify associated, explicitly parametrized model to estimate the eigenvalues of a suitably defined Poincaré map. We utilize the simplified legged locomotion model described in Section 3.4.1 to present our results on simulation studies.

### 3.5.1 Estimating the Linearized Return Map

Let $\Sigma_0$ and $\Sigma_1$ be two Poincaré sections associated with the two distinct hybrid transitions for our model. This results in a nonlinear map

$$P_{0\to1} : \Sigma_0 \to \Sigma_1 \tag{3.41}$$

which is defined by continuous forward trajectories of the system from $\Sigma_0$ until their intersection with $\Sigma_1$. Likewise, $P_{1\to0} : \Sigma_1 \to \Sigma_0$, resulting in an overall, single-stride return map on $\Sigma_0$ constructed as

$$P_{0\to0} := P_{1\to0} \circ P_{0\to1} \tag{3.42}$$

Approximating the hybrid equations of motion around a limit cycle as a piecewise LTI system, linearized versions of $P_{0\to1}$ and $P_{1\to0}$ correspond to transition states between the two linear systems $\dot{x}(t) = A_0 x(t)$, $x(t_0) = x_0$, at time $t = t_0 + \hat{t}$ and $\dot{x}(t) = A_1 x(t)$, $x(t_1) = x_1$, at time $t = t_1 + (T - \hat{t})$, respectively. The overall linearized return map on $\Sigma_0$ is then computed as

$$DP_{0\to0} = e^{A_1(T-\hat{t})} e^{A_0 \hat{t}}. \tag{3.43}$$

In this study, we both analytically derive and parametrically estimate $A_0$ and $A_1$ matrices for the hybrid model of Section 3.4.1. However, one should note that even the analytic version of (3.43) is an approximation to the "true" linearized return map, since we approximate hybrid transitions as being dependent on time rather than state in close proximity to the limit cycle. For this reason, we use a numerically computed Jacobian to the return map for the hybrid model as a ground truth against which we compare the eigenvalues estimated with both the parametric LTP model as well as our analytic approximations.

## 3.5.2 Stability characteristics of the simplified legged locomotion model

For the analytic approximation, we explicitly derive the system matrices in (3.39) using the simulated system parameters. For the data-driven identification step, we first estimate the non-parametric HTFs using the input–output method detailed in Section 3.3.3. Subsequently, we perform a parametric fitting to estimate the values for $k$ and $c$ that minimize the error between non-parametric HTFs and analytically derived HTFs of the explicitly parameterized LTP system structure in (3.39) (See Section 3.4.3 for details). Having a piecewise LTI representation with explicit estimates of system parameters, we obtain the eigenvalues of the LTP system as described in Section 3.5.1 for both analytic and the identified model.

We repeated the above steps for different values of spring constants as illustrated in Fig. 3.7. Note that our model is a two-dimensional system due to its clock-driven structure and we observe two complex conjugate eigenvalues except around $k = 160$, where we obtain two distinct real eigenvalues. Our results show that the LTP framework yields accurate estimates for the true eigenvalues of the system. We believe that the small error (see Fig. 3.7) originates primarily from our approximating of state-dependent hybrid transitions as time dependent away from steady state.

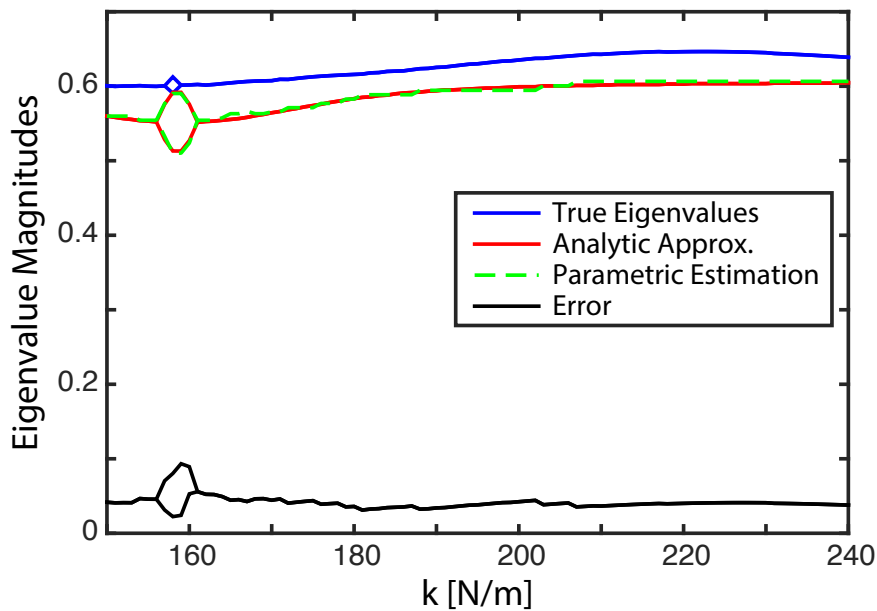

Figure 3.7: Eigenvalues of the linearized return map for the dynamics around the limit cycle, computed using three different methods as a function of the spring stiffness.

## 3.6 System Identification of LTP Systems under Input and Measurement Delays

In the previous sections, we showed input–output characterization for LTP systems both via analytical derivation of harmonic transfer functions for known LTP systems as well as data driven estimation of harmonic transfer functions for unknown stable LTP systems. Although there are some different tools to solve the input–output characterization problem of LTP systems, few if any model-based or data-driven identification methods for time-periodic systems address the problem of input and measurement delays in the system. In this section, we focus on data-driven system identification for a simple mechanical system (shown in Section 3.4.1) and analyze its dynamics in the presence of input and measurement delays using HTFs. By exploiting the way input delays are modulated by the periodic dynamics, our results enable the separate, independent estimation of input and measurement delays, which would be indistinguishable were the system linear and time invariant.

### 3.6.1 Background

The finite dimensional, piecewise LTI representation given in Section 3.2.2 cannot capture time delays (input, measurement, or internal/transmission), which constitute an inevitable aspect of both biological and artificial locomotor systems with significant impact on behavior. For instance, sensor latency and delays can significantly limit neural control performance [83–85]. In the context of robotics, delays can be introduced by different sources, including communication during teleoperation [86] or between multi-agent systems [87] and latency arising from the computational complexity and filtering associated with processing sensory information such as visual and LiDAR data [88]. Even though these phenomena can often be approximated as pure delays in the system [89], their impact on the stability and performance of the closed loop system can be rather significant and should be carefully taken into account in all stages of the analysis including system identification and the design of controllers.

In this section, our main contribution is the extension of our prior work on the identification and analysis of robotic legged locomotion [61] to explicitly consider input and measurement delays, leaving the modeling of internal system delays as future work. There is a long history of modeling and analyzing delays in both biological [84, 90] and robotic [86, 89] control systems. Most of this previous work, however, use linear time-invariant (LTI) models to approximate system dynamics and their nominal trajectories. As we have shown before, such LTI representations can be inadequate in capturing time-varying characteristics of locomotor behaviors where nominal trajectories are large limit-cycles with distinct hybrid phases [61, 62, 69, 71]. We now show that linear analysis can still be applied in this context, using the LTP framework to relax the time-invariance assumption, allowing us to identify and analyze input and measurement delays in rhythmic legged locomotor behaviors.

## 3.6.2 System Model with Input and Measurement Delays

Even though there are many different forms in which delay can be observed in practical systems, our model for input and measurement delays in this study takes the form of constant, frequency independent time shifts $\tau_u$ and $\tau_y$ in the action of the input $u(t)$ and the observation of the system output $y(t)$, respectively. If we define two intermediate variables $\bar{u}(t)$ and $\bar{y}(t)$, where

$$\begin{aligned} \bar{u}(t) &= u(t - \tau_u) \\ y(t) &= \bar{y}(t - \tau_y) \, , \end{aligned}$$

and write the system dynamics using these new variables, we get

$$\begin{aligned} \dot{x}(t) &= \begin{cases} A_0 x(t) + B_0 \bar{u}(t), & \text{if } \operatorname{mod}(t, T) \in [0, \hat{t}) \\ A_1 x(t) + B_1 \bar{u}(t), & \text{if } \operatorname{mod}(t, T) \in [\hat{t}, T) \end{cases} \\ \bar{y}(t) &= \begin{cases} C_0 x(t) + D_0 \bar{u}(t), & \text{if } \operatorname{mod}(t, T) \in [0, \hat{t}) \\ C_1 x(t) + D_1 \bar{u}(t), & \text{if } \operatorname{mod}(t, T) \in [\hat{t}, T) \end{cases} \end{aligned} \quad (3.44)$$

We intentionally represent in a form that reveals the input–output dynamics between $\bar{u}(t)$ and $\bar{y}(t)$, since the same structure is valid for $u(t)$ and $y(t)$.

The most important benefit of this representation arises from the difficulty of trying to explicitly model delay in the harmonic transfer function framework, which would have required substantial modifications to its derivation as well as the associated system identification method. We will instead adopt a two stage approach. First we will perform system identification on input-output pairs using only the magnitude plots of the non-parametric HTFs and assume that neither the input, nor the output signals are delayed. Then, we will analyze the resulting model in the frequency domain using the phase plots of the non-parametric HTFs to estimate the delays.

### 3.6.3 The Effects of Delays on Harmonic Transfer Functions

Since the delayed LTP dynamics of (3.44) have the same structure with delayless dynamics, the associated input–output relation between $\bar{u}(t)$ and $\bar{y}(t)$ has the same form as (3.27) in the frequency domain. However, we need the relationship between $u(t)$ and $y(t)$, which are the actual input and output signals of the system. Fortunately, our assumption of fixed time linear delays allows us to express actual input and output signals as a function of their undelayed counterparts in (3.44) in the frequency domain, satisfying $\bar{U}(w) = U(w)e^{-j\omega\tau_u}$ and $Y(w) = \bar{Y}(w)e^{-j\omega\tau_y}$. Substitution into (3.27) yields

$$Y(j\omega) = \sum_{m=-\infty}^{\infty} G_m(j\omega) e^{-j[(\omega - m\omega_p)\tau_u + \omega\tau_y]} U(j(\omega - m\omega_p)) \qquad (3.45)$$

where the terms

$$H_m(j\omega) := G_m(j\omega) e^{-j[(\omega - m\omega_p)\tau_u + \omega\tau_y]}$$

correspond to the harmonic transfer functions between the actual input $U(w)$, and the actual measurement $Y(w)$ for the delayed system. Comparing the HTFs for the zero-delay input–output representation, $G_m(j\omega)$, with their delayed counterparts, $H_m(j\omega)$, we will show that we can separately identify input and measurement delays in the system. This will be the main contribution of this section.

We begin by noting that the harmonic transfer functions both with and without delay have the same magnitudes. This is easily shown through the definition of $H_m(j\omega)$, with

$$\begin{aligned} |H_m(j\omega)| &= |G_m(j\omega) e^{-j[(\omega - m\omega_p)\tau_u + \omega\tau_y]}|, \\ &= |G_m(j\omega)|. \end{aligned} \qquad (3.46)$$

On the other hand, phase responses with and without delays can be different. More specifically, we have

$$\angle H_m(j\omega) = \angle G_m(j\omega) - [(\omega - m\omega_p)\tau_u + \omega\tau_y]. \qquad (3.47)$$

Note that for $m = 0$, these derivations are analogous to LTI systems, where input and measurement delays cannot be distinguished since the fundamental harmonic is phase-shifted according to their sum. More importantly, however, when additional harmonics with $m \neq 0$ are considered, the frequency dependence of contributions from input and measurement delays to the HTF phase shift will be different. This property of harmonic transfer functions allows us to independently estimate the input and measurement delays in an LTP system, which are otherwise indistinguishable for LTI systems.

We incorporate both of these observations in our approach to estimate system delays. We begin by using input–output pairs $u(t)$ and $y(t)$ from the original, delayed system to obtain the harmonic transfer functions $H_m(j\omega)$. Since the magnitudes of these HTF components are identical to their counterparts for the undelayed version of the system, we can use our previously proposed method to estimate unknown parameters for an explicitly constructed, undelayed system model based on $|H_m(j\omega)|$ alone [61]. This parametric model gives us the phase responses of the undelayed system, $\angle G_m(j\omega)$, as a reference against which the phase characteristics of the delayed system, $\angle H_m(j\omega)$ can be compared. This comparison allows independent and robust computation of the input and measurement delays, $\tau_u$ and $\tau_y$, when multiple harmonics $m \in \mathbb{Z}$ are considered to as closely satisfy (3.47) as possible.

## 3.6.4 Application to Hybrid, Vertical, Spring–Mass System Example

In this section, we illustrate the application of the proposed method to estimate the input and measurement delays for a hybrid, vertical, spring–mass system example with system dynamics defined as in Section 3.4.1 and harmonic transfer functions computed as in Section 3.4.2.

Similar to previous sections, we begin to our analysis by estimating the harmonic transfer functions of the linearized LTP system dynamics by using input–output data, but this time under input and measurement delay, without assuming any prior knowledge of the state space model. We use $f_0(t) = \cos(2\pi t)$ and $u(t) = 0$ to achieve an asymptotically stable limit cycle and record steady state data.

Subsequently, we start perturbing the limit cycle with an input signal, constructed as the concatenation of nine consecutive, $30s$ long chirp signals. Each chirp signal is designed to linearly increase in the range $(0, 7]$ Hz over its duration, taking the form

$$u_i(t) = 0.004 \sin(7\pi t^2/30). \tag{3.48}$$

In contrast, the starting phases of consecutive copies of the chirp signal are chosen to be evenly separated throughout system's period, $T = 1s$ as explained in Section 3.3.3.

In order to evaluate the performance of our identification method, we present 9 separate experiments with different combinations input and measurement delays, listed in Table 3.1. We feed the sequence of chirp signals described above with an input delay, $\tau_u$ and simulate the system with the input $\bar{u}(t) = u(t - \tau_u)$. We then impose an output delay with $y(t) = \bar{y}(t - \tau_y)$ to simulate the effect of measurement delays on system response. This yields input–output data we use to estimate harmonic transfer functions as described in Section 3.3.3. At this stage, as noted before, the identification process will be unaware of the amount of input and measurement delays that are present in the system.

Table 3.1: Estimation results for input and measurement delay via different harmonic transfer functions.

| Exp. | Actual Delays | | $G_0$ Estimates | | $G_{-1}$ Estimates | | | | $G_1$ Estimates | | | |
|---|---|---|---|---|---|---|---|---|---|---|---|---|
| | $\tau_u$ [ms] | $\tau_y$ [ms] | $\hat{\tau}_{u+y}$ [ms] | PE$_{u+y}$ | $\hat{\tau}_u$ [ms] | PE$_u$ | $\hat{\tau}_y$ [ms] | PE$_y$ | $\hat{\tau}_u$ [ms] | PE$_u$ | $\hat{\tau}_y$ [ms] | PE$_y$ |
| #1 | 0 | 0 | 0 | - | 0 | - | 1.6 | - | 0 | - | 0 | - |
| #2 | 0 | 50 | 49.9 | 0.20 % | 0 | - | 51.6 | 3.2 % | 0 | - | 47.5 | 5.0 % |
| #3 | 0 | 100 | 99.9 | 0.10 % | 0 | - | 101.6 | 1.6 % | 0 | - | 97.5 | 2.5 % |
| #4 | 40 | 0 | 40.0 | 0 % | 41.1 | 2.75 % | 0 | - | 38 | 5.0 % | 0 | - |
| #5 | 40 | 50 | 90.1 | 0.10 % | 41.7 | 4.25 % | 49.7 | 0.6 % | 41.2 | 3.0 % | 47.6 | 4.8 % |
| #6 | 40 | 100 | 140.1 | 0.07 % | 41.9 | 4.75 % | 99.5 | 0.5 % | 41.2 | 3.0 % | 95.1 | 4.9 % |
| #7 | 80 | 0 | 80.1 | 0.12 % | 81.7 | 2.12 % | 0.5 | - | 76.7 | 4.1 % | 2.9 | - |
| #8 | 80 | 50 | 130.1 | 0.08 % | 81.8 | 2.25 % | 50.4 | 0.8 % | 76.8 | 4.0 % | 52.4 | 4.8 % |
| #9 | 80 | 100 | 180.1 | 0.55 % | 82.1 | 2.62 % | 100.2 | 0.2 % | 76.9 | 3.9 % | 102.7 | 2.7 % |

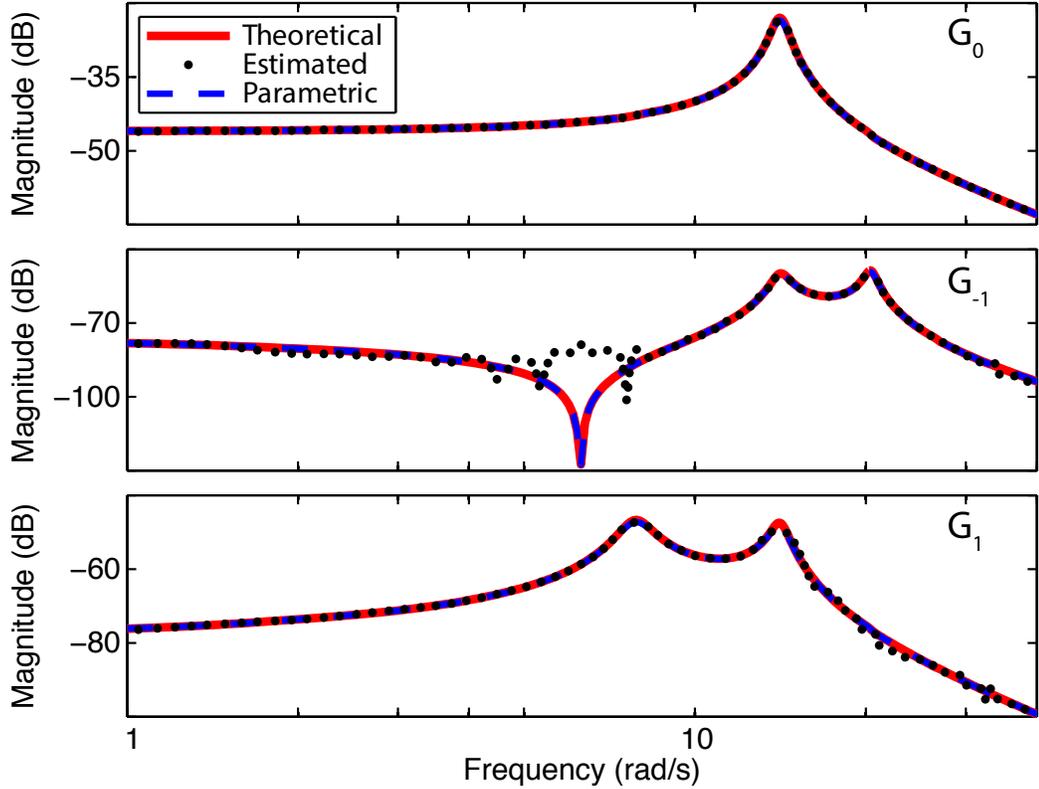

Figure 3.8: Magnitudes of HTF components for theoretically computed (solid red), estimated (dotted black) and parametrically fitted (dashed blue) models.

Fig. 3.8 illustrates the magnitudes of harmonic transfer functions obtained through the theoretical derivations of Section 3.4.2 (solid red) and the data-driven estimates of Section 3.3.3 (dotted black). However, the theoretical transfer functions rely on the knowledge of dynamic systems parameters which may normally not be available for a physical system. Consequently, before we proceed with the estimation of delays using the phase characteristics, we first estimate these unknown parameters $k$ and $c$ in (3.39) using the identification strategy presented in [61]. Equation (3.46) shows that the input and measurement delay does not effect the magnitude of the harmonic transfer functions. We can hence use data from all nine experiments with different delays, resulting in estimated values $\hat{k} = 200$ and $\hat{c} = 2.12$ for the spring and damping constants, respectively. This yields parametric magnitude responses shown in Fig. 3.8 in dashed blue, closely matching the theoretical derivations with the exact parameter values. Phase responses for this parametric model will be used in the next section to identify input and measurement delays in the system.

As is evident from (3.47), a comparison of phase responses associated with HTF components for the models with and without time delay can be used to separately estimate input and measurement delays. These phase plots are illustrated in Fig. 3.9, with the undelayed parametric model and the delayed input–output estimates are shown in solid blue and dotted black, respectively. The phase error between these two responses can be expressed using (3.47) with respect to input and measurement delays as

$$\angle G_{err} = \angle G_m(j\omega) - \angle \hat{G}_m(j\omega) + [(\omega - m\omega_p)\tau_u + \omega\tau_y]. \quad (3.49)$$

Based on this expression, we formulate a minimization problem as a function of unknown input and measurement delays, taking the form

$$(\tau_u^*, \tau_y^*) = \underset{(\tau_u, \tau_y)}{\operatorname{argmin}} \sqrt{\int_0^{40} (|\hat{G}_m(j\omega)| \angle G_{err})^2 \, d\omega} \,. \quad (3.50)$$

Green dashed plots in Fig. 3.9 show phase responses of the identified system compensated with the delay estimates resulting from this minimization problem.

More systematically, Table 3.1 shows estimation results for input and measurement delays by only using phase responses from $\hat{G}_0$, $\hat{G}_{-1}$ and $\hat{G}_1$ with respect to parametrically identified harmonic transfer functions. As predicted by (3.47), $\hat{G}_0$ alone can not separate the effects of input and measurement delays. Consequently, we evaluate the estimation performance of the total delay in the system for this case. Table 3.1 shows that individual estimates of both types of delay stay below 5% for all nine experiments with differing amounts of actual delays.

### 3.6.5 Discussion

In this section, we presented a system identification strategy to estimate input and measurement delays for a hybrid, vertical, spring–mass–damper system. We first show how rhythmic locomotor systems can be identified in frequency domain using data-driven system identification techniques. To this end, we linearize the system dynamics around an asymptotically stable limit cycle and approximate the hybrid transitions between different states as a time-periodic behavior, so that we obtain a linear time periodic system representation for our simple hybrid model.

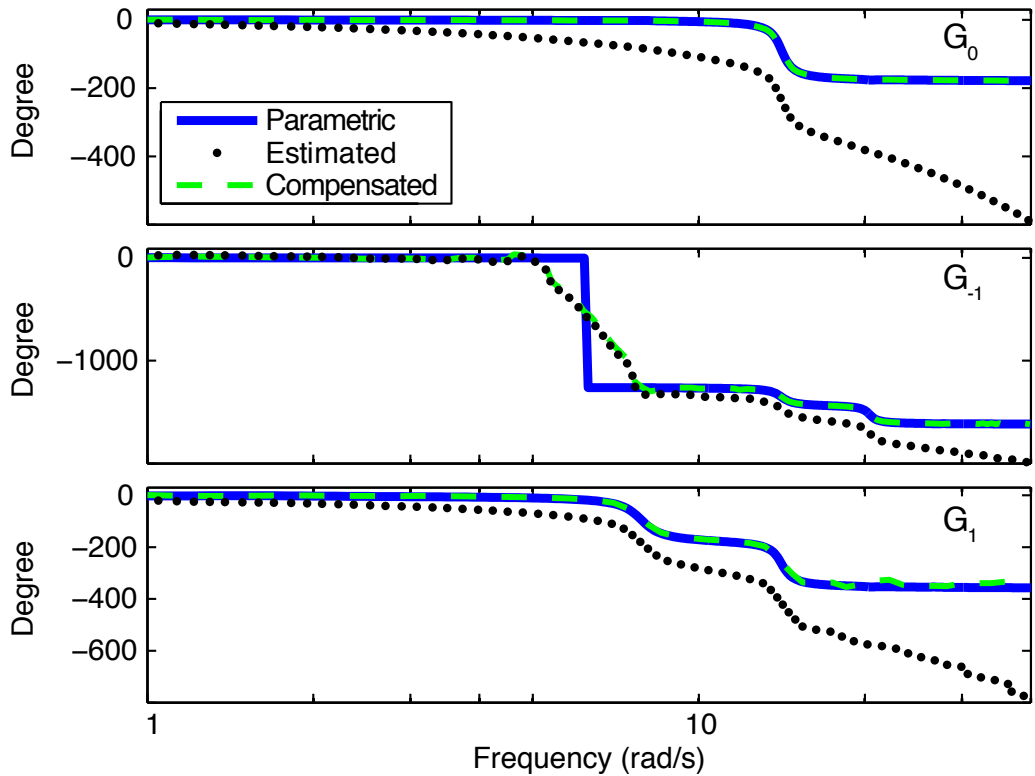

Figure 3.9: Phase responses of HTF components for the parametrically fitted (dashed blue) and estimated (dotted black) models. The dashed green plot shows the phase responses of the estimated model compensated with the identified input and measurement delays.

Our system identification process requires perturbing the limit cycle with chirp signals and recording deviations from the limit cycle as measured in the system output. However, our goal is to perform system identification under input and measurement delay and estimate the delay in the system using our transfer function estimates. To accomplish this, we performed nine experiments with different input and measurement delays in the system and estimated harmonic transfer functions corresponding to input–output characteristics of the system.

As for LTI transfer functions, we show that input and measurements delay on harmonic transfer functions do not effect the magnitudes of harmonic transfer functions. Therefore, we perform parametric identification based on the estimated harmonic transfer functions by only using the magnitude responses. The key point in our theoretical analysis is that the use of harmonic transfer functions allows us to independently estimate input and measurement delay in the system when the higher order harmonics are considered in the estimation process. We compare the phase response of the estimated and parametrically identified harmonic transfer functions for this purpose and estimate input and measurement delay in the system for the nine different experiments performed in this study.

## 3.7 Identification of VHOP

In this section, we apply the harmonic transfer functions (HTFs) based system identification method described in Section 3.3.3 to a more realistic vertical hopping robot model (VHOP) that captures some crucial aspects of the well-known Spring-Loaded Inverted Pendulum (SLIP) model of running. We provide simulation results to illustrate our approach and to characterize the performance of the system identification method applied to the VHOP system. Note that all simulations including the estimation of HTFs are implemented in Matlab environment using standard ordinary differential equation solvers and built-in matrix operators.

### 3.7.1 The Vertical Hopper (VHOP) Model

The Vertical Hopper (VHOP) model, illustrated in Fig. 3.10 consists of a vertically constrained point mass attached to a compliant leg with viscous damping. A linear actuator in parallel with the leg spring is incorporated to compensate for energy losses due to damping, and to implement behavioral controllers. The model also incorporates a very small mass at the toe which also allows both better correspondence to physical robot platforms, as well as the ability to apply inputs to the system during flight.

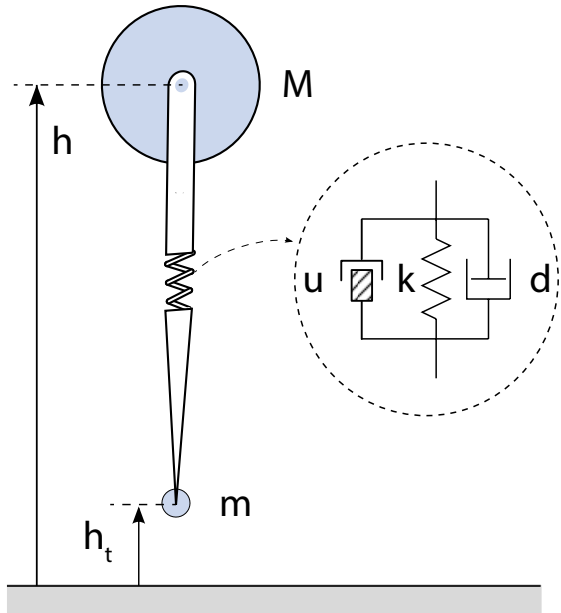

Figure 3.10: The Vertical Hopper (VHOP) model with leg compliance, damping and a parallel linear actuator, where $h$ and $h_t$ represents the height of the body and toe mass, respectively.

Our use of a linear actuator as an input to the system is inspired by similarly structured robot implementations in the literature [25, 80, 81]. There are two significant roles for the linear actuator in the VHOP model. First, it allows us to obtain a clock-driven structure, facilitating the construction of asymptotically stable limit cycles. Second, it can also be used to inject additional "external" input signals to support our system identification approach. This clock-driven structure with an additive input allows us to avoid phase resetting [71] and the complications associated with estimating phase [73]. Previously, [35] also adopted clock driven models to perform LTP type analysis on nonlinear systems.

Our ultimate goal is to apply the techniques in this study to more complex models such as the SLIP and its many variants. As a first step, however, we use the relatively simple VHOP model to keep our focus on LTP system identification of hybrid dynamics. Nonlinearities and control systems challenges associated with more complex models are hence left out of scope for the present study. Nevertheless, despite its simplicity, the VHOP model still possesses some of the critical features of locomotor dynamics, including hybrid dynamics with flight and stance phases, discontinuities in the state due to collisions, as well as periodic behavior in the form of limit cycles.

The VHOP model alternates between stance and flight phases and hence can be modeled as a hybrid dynamical system consisting of a set of smooth flows with discrete transitions between them [74]. As usual, the stance phase corresponds to states where the foot is in contact with the ground, while the flight phase corresponds to states when the robot is off the ground. Transitions from and to stance called liftoff and touchdown events, respectively. A simple diagram of VHOP phases and transitions is shown in Fig. 3.11. The system state for both phases is defined as

$$\mathbf{x} := \begin{bmatrix} h & h_t & \dot{h} & \dot{h}_t \end{bmatrix}^T. \quad (3.51)$$

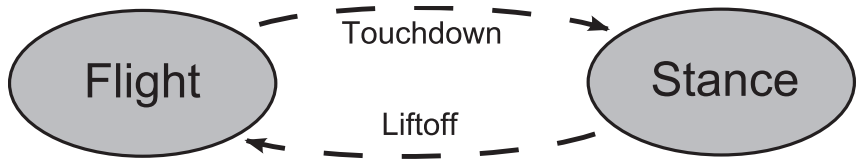

Figure 3.11: The two phases of locomotion with the VHOP model and associated transition events. Each phase has its own smooth flow.

### 3.7.2 VHOP System Dynamics

Structural differences between the stance and flight phases require their modeling through separate differential equations, leading to the hybrid nature of the VHOP model. During flight, the toe has no contact with the ground and is hence free to move vertically, leading to the equations of motion

$$M\ddot{h} = -Mg - d(\dot{h} - \dot{h}_t) + k(\ell_0 - (h - h_t)) + u(t) \tag{3.52}$$
$$m\ddot{h}_t = -mg + d(\dot{h} - \dot{h}_t) - k(\ell_0 - (h - h_t)) - u(t),$$

where the viscoelastic leg has damping coefficient, $d$, spring constant, $k$, and rest length $\ell_0$. Even though this formulation leaves the actuator input $u(t)$ unspecified, we will impose a clock-driven structure on this input signal in the form $u(t) = a\cos(\omega_p t) + u_c(t)$, incorporating a periodic open-loop forcing term to achieve a limit cycle, and an additive $u_c(t)$ for additional control affordance. The signal $u_c(t)$ will be the input used for system identification.

During stance, the toe is assumed to remain stationary on the ground until liftoff and the body mass experiences the spring, damper and actuator forces. The corresponding equations of motion hence take the form

$$M\ddot{h} = -Mg - d\dot{h} + k(\ell_0 - h) + u(t) \tag{3.53}$$
$$m\ddot{h}_t = 0, \tag{3.54}$$

where the same actuator action is used and initial height and velocity for the toe mass is also zero. Kinematic and dynamic parameters for both phases are detailed in Table 3.2, chosen to be consistent with the physical monopod platform in our laboratory [82] for future extensions to real-world applications.

The final remaining component in the hybrid dynamics are the threshold and transition functions. The liftoff event occurs during stance when the net vertical force on the toe mass crosses zero, beyond which the toe lifts off the ground and the flight phase starts. The corresponding boundary condition is defined as

$$f_{lo}(\mathbf{x}) := -d\dot{h} + k(\ell_0 - h) + a\cos(\omega_p t) + u_c(t) = 0. \tag{3.55}$$

Table 3.2: VHOP Model Parameters.

| Parameter | Description | Value | Unit |
|---|---|---|---|
| $M$ | Body mass | 2.7 | kg |
| $m$ | Toe mass | 0.05 | kg |
| $k$ | Compliance | 6500 | N/m |
| $d$ | Damping | 12 | Ns/m |
| $a$ | Pumping magnitude | 75 | N |
| $\omega_p$ | Pumping frequency | $2\pi / 0.33$ | rad / s |
| g | Gravity | 9.81 | m/s$^2$ |
| $\ell_0$ | Rest length | 0.2 | m |

In contrast, the touchdown event, defining the transition from flight to stance, occurs when the toe touches the ground, captured through the boundary condition $f_{td}(\mathbf{x}) := h_t = 0$.

System trajectories remain continuous through the liftoff event, but the touchdown event induces discontinuous state trajectories due to the associated collision, modeled through a transition function ensuring that $\dot{h}_t(t_{td}^+) = 0$. An important consequence of this discrete change is that the derivation of a closed-form, time-varying state space model is not feasible with available methods. However, [69] explicitly showed that a hybrid dynamical system with discrete jumps in system states and even system dimension can be modeled (locally) with a discrete-time impulse response function by choosing a set of Poincaré sections and considering the mapping between those sections. Motivated by this result, we conjecture that the discrete jumps and hybrid transitions can be embedded into a continuous-time, time-periodic impulse response function model. This is more general than a state space model and hence allows us to utilize the HTF structure. We leave attempts to prove this conjecture as a future work. Note, however, that analytic derivations of time-periodic impulse response functions are not straightforward even for very simple LTP systems, which further motivates our use of a data-driven system identification approach.

### 3.7.3 Non-parametric System Identification for the VHOP Model

The VHOP dynamics of (3.52) and (3.53) clearly do not correspond to a Linear Time-Periodic (LTP) system. Linearization of these dynamics around an isolated point is also not useful since the expected behavior takes the form of periodic trajectories that never stabilize around a single point in the state space. Consequently, our approach is to linearize the system *around its periodic behavior*.

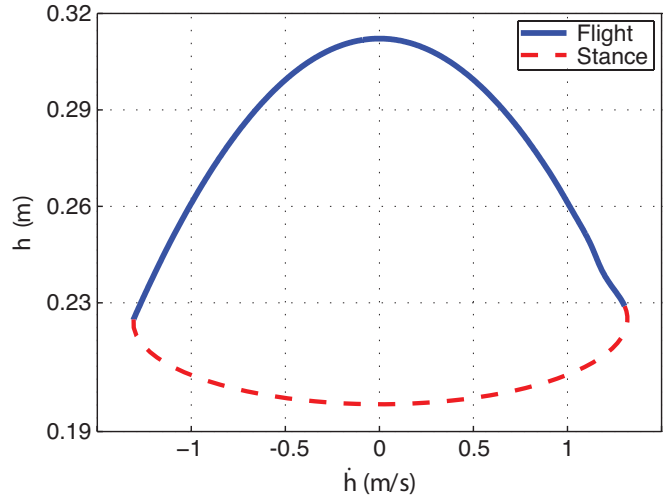

Figure 3.12: A cross section of an example VHOP limit cycle obtained with the periodic excitation $u(t) = 75\cos(2\pi t/0.33)$. Red and blue sections represent stance and flight phases on the limit cycle, respectively.

We start by assuming that the system has an asymptotically stable limit cycle $\mathbf{x}_{lc}(t) := [h_{lc}(t), h_{t,lc}(t), \dot{h}_{lc}(t), \dot{h}_{t,lc}(t)]^T$ with period $T$ when $u_c(t) = 0$. Such a limit cycle can be obtained through suitable choices of the periodic excitation component $a\cos(\omega_p t)$ within $u(t)$. For example, choosing $a = 75$ and $\omega_p = 2\pi/0.33$, results in the limit cycle illustrated in Fig. 3.12 that appears to be asymptotically stable according to numerical simulations. In this study, we selected the height of the upper mass, $h(t)$, as our output measurement to construct input–output model. We treat the output relative to the nominal behavior on the limit–cycle, i.e. $y(t) := h(t) - h_{lc}(t)$, where $h_{lc}(t)$ is the robot height on the limit cycle, assumed to be known (e.g., recorded during operation with $u_c(t) = 0$). Then, the resulting LTP system can be modeled via its impulse response:

$$y(t) = \int_0^t H(t,\tau) u_c(\tau) d\tau, \tag{3.56}$$

where the impulse response function is periodic, $H(t,\tau) = H(t-T, \tau-T)$, with period $T = 2\pi/w_p$. We apply the system identification method described in Section 3.3.3 on these coordinates. To this end, we use $u_c(t)$ to perturb the system away from the limit cycle and analyze the effects on system trajectories. As noted in Section 3.3.3, chirp signals are a good choice for these perturbations, exciting as many modes and components in the system dynamics as possible.

In particular, we use an input signal formed by the concatenation of 21 phase-shifted instances of the chirp signal illustrated in Fig. 3.3, whose frequency changes linearly in the range of zero to 5 Hz in 20 s. Each instance of this chirp signal is shifted in time by $T/21$ relative to the previous signal. The magnitude of these chirp signals was chosen through manual tuning in such a way that the perturbations are sufficient, but do not appear to excite significant nonlinearities. Each chirp signal $u_c^i(t_i)$ with $i \in \{1, 2, ..., 21\}$, is hence defined as

$$u_c^i(t_i) = 0.1 \sin(0.5\pi t_i^2) . \tag{3.57}$$

where $t_i := t - ((i-1)D - (i-1)T/21) \in [0, 20)$ and $D = 20s$ is the duration of each chirp application. This yields the final form of our perturbation input (partially shown in Fig. 3.13) as

$$u_c(t) = \begin{cases} u_c^1(t_1), & \text{if } 0 \leq t < D \\ \vdots & \vdots \\ u_c^{21}(t_{21}), & \text{if } 20D \leq t < 21D \\ 0, & \text{if } 21D \leq t . \end{cases} \tag{3.58}$$

Finally, we compute the Fourier transforms of the input and output signals as $U_c(jw)$ and $Y(jw)$, respectively. The identification methods described in Section 3.3.3 compare the actual output to the predicted output $\hat{Y}(jw)$, resulting in the desired estimates of the three harmonic transfer functions $\hat{\mathbf{G}}_0$, $\hat{\mathbf{G}}_{-1}$ and $\hat{\mathbf{G}}_1$ for the VHOP system.

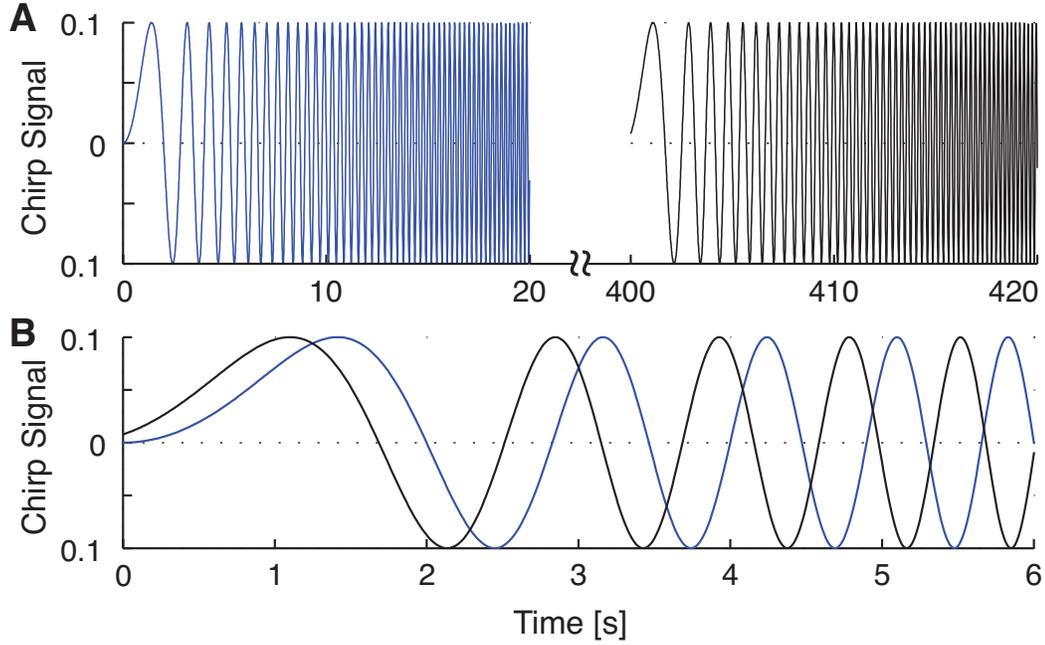

Figure 3.13: The perturbation input $u_c(t)$ used for the system identification process. (A) Phase-shifted repetitions of the original chirp signal, concatenated sequentially (only the $1^{st}$ and $21^{st}$ are shown for better illustration). (B) $1^{st}$ and $21^{st}$ chirp signals superimposed on top of each other for better visualization of phase difference between them.

We evaluate the accuracy of the system identification by comparing the output from VHOP dynamics to the inverse Fourier transform of $\hat{Y}(jw)$ obtained using the estimated harmonic transfer functions in $\hat{G}$. Fig. 3.14(A) shows the output $y(t)$ for the VHOP response to the chirp signal defined by (3.26), whereas Fig. 3.14(B) shows the discrepancy between the actual and predicted system outputs as a function of time.

As seen in Fig. 3.14, the estimated harmonic transfer functions can successfully predict system response. This result, however, is for the response of the system to the chirp input used for system identification itself and hence is not suitable for a fair evaluation of the prediction performance. A good predictor should be able to estimate system outputs for input signals that might differ substantially from those used for system identification. In the following section, we will present a more systematic characterization of the predictive accuracy obtained through the estimated harmonic transfer functions using sinusoid and step inputs.

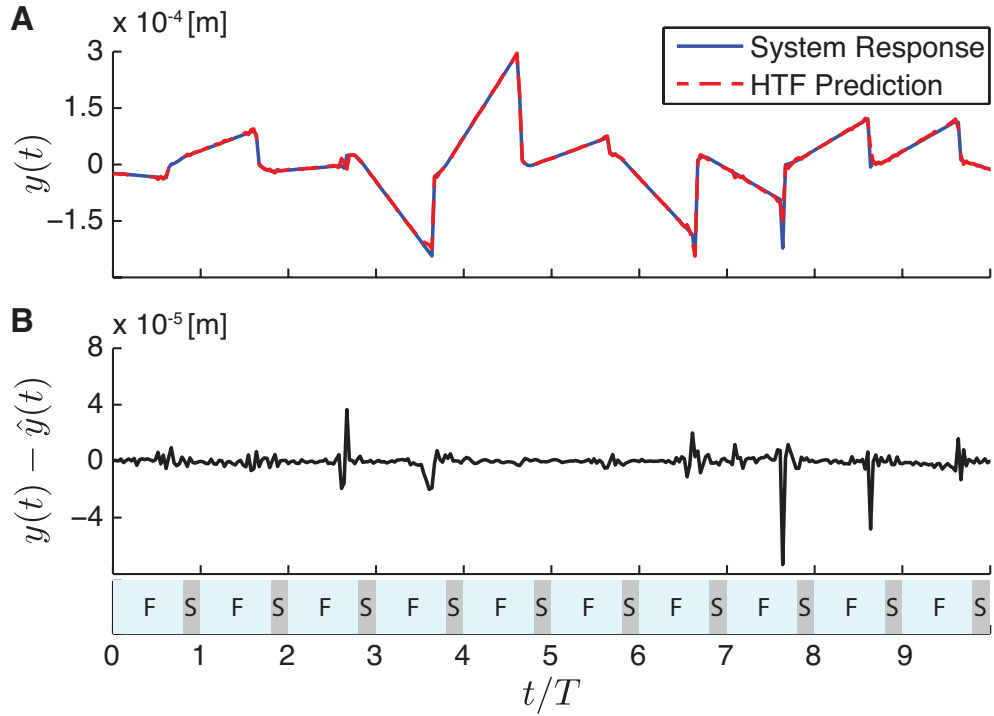

Figure 3.14: Prediction performance of harmonic transfer functions with the chirp input training signal. (A) The VHOP system output, (B) Discrepancy between the actual and predicted system outputs. The time axis is normalized with the hopping period. Stance and flight phases are indicated at the bottom by the letters S and F, respectively.

### 3.7.4 Prediction Accuracy of HTF Responses to Sinusoid and Step Inputs

To evaluate the prediction performance of the harmonic transfer functions, we validate the input–output model identified using chirp excitation on sinusoidal and step input waveforms. We calculate the percentage error based on the difference between the actual and predicted system responses to a particular input:

$$E_{\text{rms}} := 100 \frac{\sqrt{\frac{1}{T_{\text{rms}}} \int_0^{T_{\text{rms}}} (y(t) - \hat{y}(t))^2 dt}}{\sqrt{\frac{1}{T_{\text{rms}}} \int_0^{T_{\text{rms}}} y(t)^2 dt}}, \qquad (3.59)$$

where $T_{\text{rms}}$ is the duration of the sinusoidal input.

We simulate the VHOP dynamics of Section 3.7.2 using sinusoidal inputs with amplitude 0.1 and frequencies in the range $f \in [0.1, 20]\,\text{Hz}$ to find the "actual" outputs of the system. We then compute the output predictions by plugging the previously estimated harmonic transfer functions into the output equation with $n_h$ harmonics for comparison.

Fig. 3.15(A) illustrates $E_{\text{rms}}$ for each input sinusoid frequency when the system identification was performed with a chirp signal covering frequencies from 0 to 5 Hz, whereas Fig. 3.15(B) illustrates the same quantity when system identification was performed with a wider chirp signal covering frequencies from 0 to 10 Hz. Our results show that increasing the frequency coverage of the chirp signal used for training increases the accuracy of the resulting harmonic transfer function representation for higher frequencies. This is, of course, expected since exciting the system with a wider frequency range allows the harmonic transfer functions to be properly trained to also match system response for these higher frequencies. Both the smoothness condition imposed by the system identification process, as well as the nature of LTP systems wherein an input with a single frequency component yields many other frequencies in the output, result in improvements in prediction accuracy for even higher frequency ranges when the chirp spectrum is increased. These results show that the choice of the training input has substantial impact on the accuracy of the resulting HTF estimates.

As noted, practical implementation of this system identification method requires truncating the HTF beyond a certain order, but this threshold cannot be determined beforehand, in general. Consequently, we have explored the effect of incorporating different numbers of HTFs on prediction performance. The four plots in each graph of Fig. 3.15 correspond to system identification with 11, 15, 21 and 31 harmonic transfer functions taken into account. Theoretically, we should expect prediction accuracy to increase when more harmonics are included in the representation, which we verify empirically as illustrated in Fig. 3.15. Beyond a certain number of harmonics (21 in our case), improvements no longer seem substantial, suggesting a reasonable threshold for our system. Including additional harmonics only increases computational complexity without any further significant improvements in accuracy.

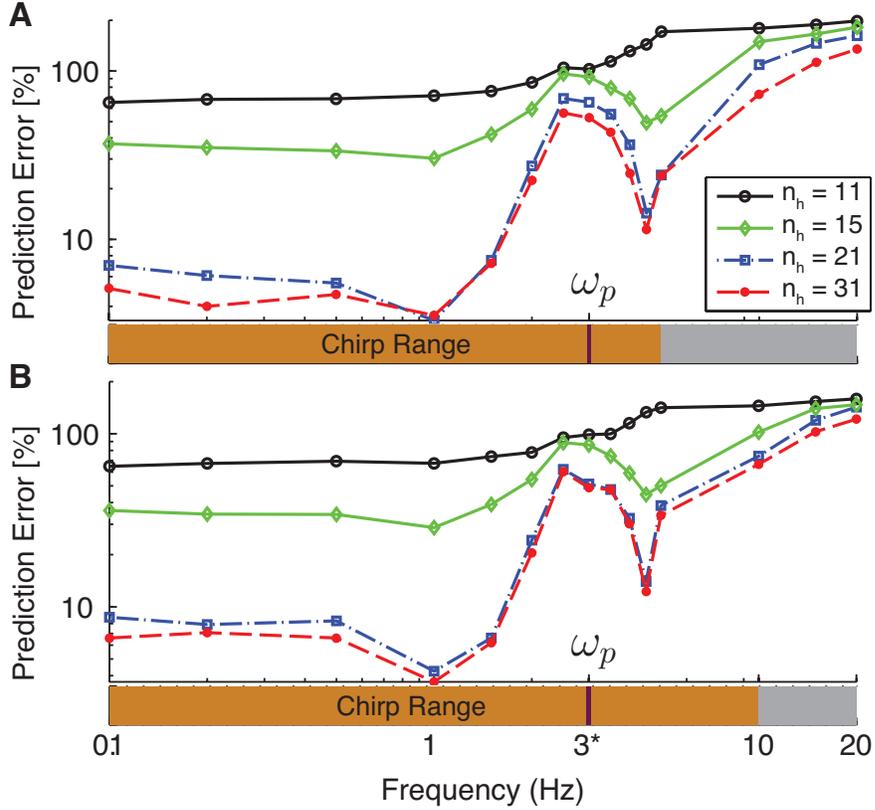

Figure 3.15: Percentage output prediction errors $E_{rms}$ for the HTF representation of the VHOP system in response to single sinusoid excitations at different frequencies in the range $[0, 20]$ Hz. (A) The HTF representations obtained through training with a chirp signal spanning frequencies in the range 0 to 5 Hz. (B) HTF representations obtained using a chirp input from 0 to 10 Hz.

A final comment is required on the performance results of Fig. 3.15. The predictive performance of the HTF degrades significantly around 3 Hz, which corresponds to the VHOP limit cycle frequency. As detailed before, LTP systems generate output signals at the input frequency plus harmonics of the pumping frequency. When the input and system frequencies are same ($\omega_p = 3\,\text{Hz}$ in our case), different harmonics at multiples of $\omega_p$ may also effect the response at this frequency. [69] showed that the identification of HTF should not be performed at frequencies $k\frac{\omega_p}{2}$, $k \in \mathbb{Z}$ when using sum-of-sines input stimuli for identification. A similar phenomena may be contributing to the errors in our result. However, it is also possible that our linearity assumption is being violated near the pumping frequency. Addressing this problem further is left to future work.

In addition to sinusoid inputs at specific frequencies, we have also investigated HTF prediction performance under a step input with magnitude 0.01. Note, once again, that system identification is still performed with a chirp signal and the step input is only applied for characterizing the prediction performance of the resulting HTF representation. The step input was applied to the system after the $200^{th}$ cycle, making sure that it has reached steady-state. In this case, the estimated system output was computed using the estimated harmonic transfer functions with $n_h$ harmonics identified using the chirp signal. Then Inverse Fourier transform was used to compute the time domain step response prediction of the HTF system.

Fig. 3.16 illustrates a comparison of measured and predicted responses of the HTF system for a single sinusoid at frequency $1\,\text{Hz}$ in panel (A) and the step input in panel (B). As expected, the HTF representation accurately predicts system response as shown in Fig. 3.16(A), consistent with the results shown in Fig. 3.15 ($E_{\text{rms}} = 3.3\%$). On the other hand, the prediction results for the step inputs exhibit significant errors as shown in Fig. 3.16(B). Even though the HTF response captures the qualitative behavior of the step response, including the spiked response around touchdown, noticeable errors remain with a percentage root mean square error $E_{rms} = 68\%$. This is somewhat similar to the large prediction errors observed around the system periodic frequency and may be due to the harmonics of the step input in the pumping frequency.

One possible reason for these errors may be the relatively small range of frequencies covered by the chirp signal for the identification process. Unfortunately, increasing the frequency range covered by the chirp signal requires substantially longer durations for the input signal. This dramatically increases both the duration of the simulation for data collection, as well as the computational complexity of the identification process. Even though increasing the frequency range, and hence the chirp duration, may increase accuracy, it would impair the feasibility of the system identification method on physical robot platforms due to the resulting need for substantially longer experiments with sustained locomotion.

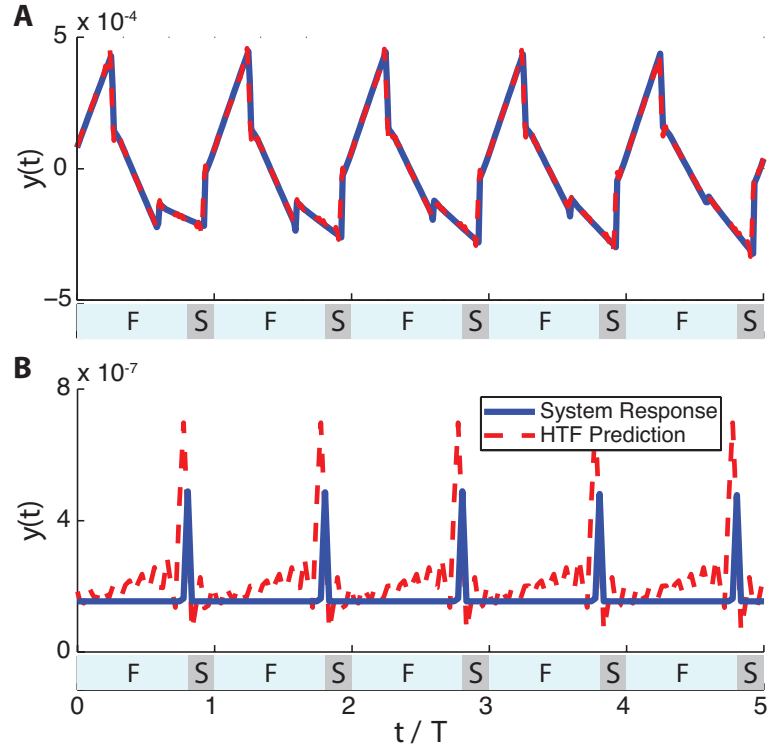

Figure 3.16: Prediction performance of harmonic transfer functions with the test input signals. Stance and flight phases indicated by S and F, respectively. Comparison between measured and predicted system responses for two types of inputs (not used for training): (A) 1 Hz sinusoidal input and (B) step input. In both cases, steady state response is shown.

The nonlinearities and the hybrid nature of the VHOP model may also be contributing to these errors. Note that both the system identification and evaluation tests are performed around a limit cycle of the VHOP model. Our method assumes that trajectories around the limit cycle exhibit hybrid transitions that are consistently timed with the limit cycle. However, especially for trajectories obtained with a step input, these transitions may occur much earlier than the limit cycle's corresponding transition. This violates one of the underlying assumptions in representing system behavior around the limit cycle as an LTP system. Moreover, the unidirectional forcing of the step input results in a larger deviation from the limit cycle than the symmetric sinusoid input, which may lead to a further violation of the linearity assumption behind the LTP representation. These effects can be more significant around the harmonics of the pumping frequency, which also includes the DC frequency.

Despite these prediction errors, however, an important feature observed in Fig. 3.16(B), is the ability of the HTF representation to predict the hybrid transitions in the system output resulting from the collision at touchdown. The effects of such impact collisions were investigated in a number of studies [22, 51], which are rather difficult to model within explicitly constructed models. Consequently, the ability of a data-driven system identification approach to model such hybrid features of a system is promising since they are an inevitable part of any locomotor system.

### 3.7.5 Prediction Accuracy of HTF Responses Under Uncorrelated Input and Output Noise

In Section 3.7.4, we investigated the prediction performance of the estimated harmonic transfer functions for different input waveforms assuming perfect measurement of input and output signals. However, our goal is to develop a data-driven system identification framework applicable to physical legged robot platforms. Therefore, we contaminate the input–output data used for system identification with zero-mean Gaussian noise in order to simulate its performance in more realistic settings.

In order to accomplish this, we use the noise modeling approach for HTFs adopted by [34]. Fig. 3.17 illustrates a block diagram representation of how noise affects the system identification data. Measured input and output data are corrupted by uncorrelated noise with zero mean and standard deviations; $\sigma_U$ and $\sigma_Y$, respectively. In other words, the measured input, $\tilde{u}_c(t)$, and output, $\tilde{y}(t)$, are represented with

$$\tilde{u}_c(t) = u_c(t) + n(t) \tag{3.60}$$
$$\tilde{y}(t) = y(t) + v(t), \tag{3.61}$$

where $n(t)$ and $v(t)$ are zero-mean noise signals affecting the input and output data, respectively.

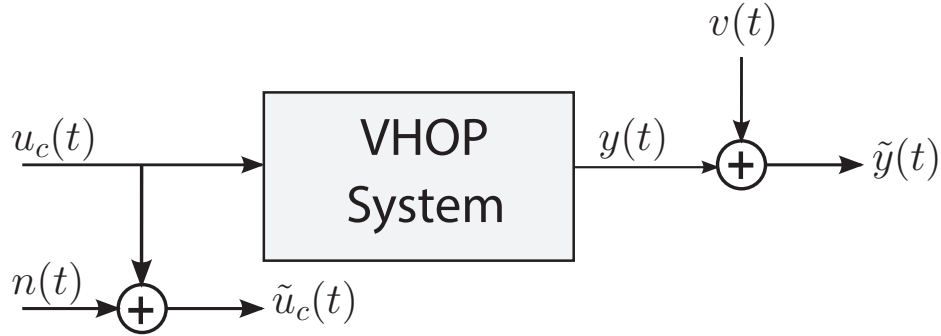

Figure 3.17: Block diagram representation of the VHOP system incorporating measurement noise on the input and output signals.

Fig. 3.18 illustrates the prediction performance of the estimated harmonic transfer functions for different cases of how the noise applied to sinusoid input tests with different frequencies. Note that addition of noise deteriorates the prediction performance of the estimated harmonic transfer functions as expected. The single sine excitation results show that our identification strategy is robust to noise corruption in input and output data up to SNR values of 12.5 (where $SNR := (A_{signal}/A_{noise})^2$, with $A$ denoting root mean square (RMS) amplitude) as in the case of [34]. However, further increasing the standard deviations drastically reduces the prediction performance, since noise starts to dominate the information necessary for system identification.

### 3.7.6 Summary of Approach

In this section, we proposed a data-driven system identification strategy to represent a simple, vertical hopping robot (VHOP) model. Most existing work on models of legged locomotion is based on explicit mathematical models. Even though these models are sufficiently accurate to describe various aspects of locomotory behaviors, their correspondence to physical behavior degrades in the presence of unmodeled components in actual hardware platforms. Our strategy is to use data-driven system identification techniques to describe legged locomotion systems.

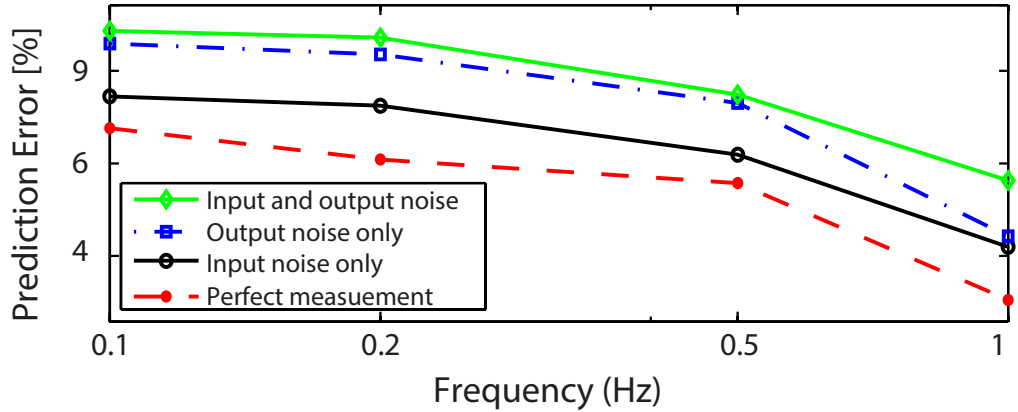

Figure 3.18: Percentage prediction errors $E_{rms}$ for the HTF representation of the VHOP system with input and measurement noise (with SNR values of 12.5) in response to single sinusoid excitations at different frequencies in the range $[0, 1]$ Hz.

The identification strategy we use relies on the perturbation of locomotion behaviors with small chirp signals, with the resulting system response used as output data. LTP system identification techniques are then applied to this input–output data. As a specific example, we estimate Harmonic Transfer Functions for our VHOP model around a periodic, stable hopping trajectory. We then compare the output prediction of the estimated harmonic transfer functions with actual output data obtained from VHOP simulations. Our results show that harmonic transfer functions can be used as predictors of simple locomotion models on training data.

Specifically, we performed prediction tests with numerically identified harmonic transfer functions on step inputs. Our results showed that the predictive performance of the HTF representation on step inputs is not as good as its performance on sinusoid inputs. Nevertheless, our results revealed that harmonic transfer functions are still capable of capturing the qualitative effects of hybrid transitions associated with touchdown collisions on system output. This type of phenomena is usually difficult to incorporate into explicit mathematical models. We have shown that harmonic transfer functions can capture and predict such discrete jumps in system dynamics, which is a promising result that highlight possible advantages of using data-driven techniques for the identification of legged locomotion models.

### 3.7.7  Limitations and Possible Future Extensions

The hybrid nature of the VHOP model does not allow us to obtain theoretical harmonic transfer functions due to discontinuities in the system state associated with collisions. A comparison of the system identification results to an ultimate, theoretical HTF representation was hence not possible. Consequently, we performed systematic simulation tests to characterize the adequacy of the numerically identified harmonic transfer functions in representing system behavior for simple legged locomotion models. To this end, we presented single sinusoid input tests to evaluate the prediction performance of harmonic transfer functions. In addition, we corrupted the input–output data with uncorrelated noise to investigate its accuracy towards experimental inquiries. Our simulation studies showed that LTP system identification techniques yield promising results on the identification of simple locomotion models when sufficient number of harmonics are considered during the identification process. However, an important next step will be to more formally address hybrid transitions in a continuous HTF framework.

Our approach is based on modeling legged locomotion systems as linear time periodic (LTP) systems around their limit cycles. Legged locomotion models exhibit hybrid dynamics during their locomotion, with discrete transitions between different dynamics. For trajectories in close proximity to the limit-cycle, phase resetting does not occur and these transitions can be approximated as time-dependent for stable clock-driven systems. This enables us to use an LTP structure for local system behavior around limit cycles. We are planning to extend our methods to systems which are not clock-driven by adopting methods developed by [69, 71].

## 3.8 Conclusions

In this chapter, we presented a system identification strategy to estimate input–output transfer functions for a simple vertically constrained vertical hopper robot model towards data-driven models of legged locomotion. We first showed that a class of hybrid locomotion models can be approximated as a hybrid Linear Time Periodic (LTP) system in close proximity to their asymptotically stable limit cycle.

We used the concept of Harmonic Transfer Functions (HTFs) to analyze the frequency response characteristics of the LTP system. We identify the HTFs using input–output data of the system in order to estimate the input–output models for legged locomotion models. Our results showed that HTFs can also be used to investigate stability characteristics of the legged locomotor systems via data-driven system identification.

Motivated by the successful identification results of this chapter, we plan to extend our work to the identification of the Spring-Loaded Inverted Pendulum (SLIP) model [21] and its extensions, widely used as models of locomotory behaviors in the literature. Our future goal is to apply our system identification methods to our physical monopod robot platform and to compare the identification performances with our previously verified analytical model [22].

# Chapter 4

# Towards State Space Models of Legged Locomotion

This chapter presents our studies towards developing state space models of legged locomotion. To accomplish this, we first present a state space system technique for a class of hybrid LTP systems, formulated in the frequency domain based on frequency domain input–output data. Our goal in this study is to develop a technique for estimating time-periodic system and input matrices for a hybrid LTP system, assuming full state measurement. The identification problem is formulated in a linear regression framework using Fourier transformations in order to estimate Fourier series coefficients of the time-periodic system and input matrices via a least-squares solution. In addition, we propose a new subspace-based state space identification method for a more general class of LTP systems without any assumption on full state measurement. Our method is based on the fact that LTP systems can be transformed into equivalent discrete-time LTI systems using bilinear transform and lifting. To this end, we use frequency domain subspace identification methods to estimate an equivalent discrete-time LTI system from the input–output data of the original LTP system. Subsequently, we use optimization methods to obtain an LTP realization by exploiting the specific parametric structure of Fourier series coefficients in our lifting method. The work presented in this chapter has also been reported and appeared in [91, 92].

## 4.1 Parametric Identification of Hybrid Linear Time Periodic Systems

Our main focus in this part is the identification of nonlinear, hybrid dynamical systems that operate near their periodic solutions. A wide variety of dynamical phenomena in biology and engineering include *oscillatory* and *hybrid* characteristics [93–95]. Thus, such dynamical behaviors are commonly modeled as nonlinear hybrid dynamical systems that operate near some isolated periodic orbits (a.k.a. limit-cycle). Though, there are remarkably fewer studies focusing on the problem of system identification for hybrid dynamical systems operating around limit-cycles than system identification studies on dynamical systems that operate near their point equilibria (e.g. LTI systems).

In the broadest sense, a hybrid dynamical system is one that both flows smoothly (defined by a set of differential equations) and jumps discretely (defined by a set of transition maps) [96]. These discrete jumps are often accompanied by a switch between different smooth flows, punctuating system trajectories with discontinuous jumps, sometimes even changing the dimension of the underlying state space [16]. Despite the generality of this definition, we limit our scope to hybrid systems for which state trajectories are continuous, but possibly non-differentiable. In other words, we exclude systems that undergo discrete jumps in states as well as changes in the state dimensions.

Under certain assumptions, the linearization of smooth nonlinear systems around their periodic solutions (orbit), yields linear time-periodic (LTP) systems [96], whereas the linearization of the class of nonlinear hybrid systems we consider around their periodic orbits yields hybrid LTP systems [97]. Since we exclude hybrid transitions with discrete jumps in system state and dimension, the class of induced hybrid LTP systems that we study exhibit continuous but only piece-wise differentiable vector fields [61, 62]. In Section 4.2.1, we formally define the general form of LTP systems that we focus on. Our main contribution in this study is a parametric system identification method for hybrid LTP systems that we consider, using frequency domain representations of input-output data.

Unlike the literature on LTP and/or hybrid system identification, the identification of LTI systems is a relatively mature field. There is a wide range of techniques for the identification of LTI systems, appropriate for widely differing needs of engineers and scientists [36]. For example, subspace system identification methods provide a powerful and popular parametric tool for estimating state space representations of LTI systems from input–output data [98].

There are a number of methods that extend the LTI identification techniques to the identification of LTP systems. For example [99] utilizes the subspace system identification method of [98] to estimate physical parameters of smooth linear time-varying systems, whereas [100] developed a different subspace system identification technique for discrete time periodically time-varying systems. In the context of piece-wise smooth system identification, [101] introduced a formulation to estimate state space models for piecewise LTI systems, which may be considered as a special case of our formulation in Section 4.2.1, when the switching time between the subsystems is known. Similarly, [102] utilizes a data-driven input–output system identification method to estimate piecewise affine models for approximating dynamics of a hexapedal robot. However, none of these methods completely cover our class of LTP systems and they all perform identification based on time domain input-output data.

In our formulation, we assume that switching times between different continuous LTP vector fields are known. This information is used to separately identify individual contributions from each LTP subsystem to the overall periodic system. In our approach, we obtain Fourier series coefficients for the state and input matrices and then formulate the problem in a linear regression framework. Different than [103], where the authors also use a linear regression approach to formulate state space subspace system identification problem for LTI systems, we consider LTP systems and the variables to be estimated are the Fourier series coefficients of the periodic system and input matrices. After estimating system matrices using a least squares solution, we use Fourier synthesis equations to construct time-periodic system and input matrices.

### 4.1.1 Problem Formulation

In this study, we focus on linear time-periodic systems, whose state evaluation equation can be written as

$$\dot{\mathbf{x}}(t) = A(t)\mathbf{x}(t) + B(t)\mathbf{u}(t) , \qquad (4.1)$$

where $A(t) \in \mathbb{R}^{n \times n}$, $B(t) \in \mathbb{R}^n$ and both system matrices are periodic with a fixed, known period $T > 0$ such that $A(t) = A(t + nT)$ and $B(t) = B(t + nT)$, $\forall n \in \mathbb{Z}$. In this study, we further assume that we can measure all system states.

Existing literature on frequency domain system identification of LTP systems concentrates mainly on non-parametric estimation of the harmonic transfer functions (HTFs) [33, 34, 62, 104], Even though a number of previous studies perform parametric identification by fitting parameterized transfer function models to the non-parametrically identified HTFs [61, 63, 69], the present section focuses on a direct state-space parametric identification method for the hybrid LTP system without dealing with computational details of HTFs.

In this formulation, the steady-state response of the system can be represented as

$$X(j\omega) = \sum_{n=-\infty}^{\infty} H_n(j\omega - jn\omega_p)U(j\omega - jn\omega_p) , \qquad (4.2)$$

where $H_n(s)$ can be theoretically derived for certain special cases when the state space representation of the system is available, such as for systems with finite harmonic expansions or constant system matrices [32, 78]. In the general case, these harmonic transfer functions can only be approximated by truncating the infinite dimensional Toeplitz matrices that arise during the derivation phase Section 3.3.1.

In this section, we will work with systems in the form of (4.1), assumed to be driven by an observable input, $u(t)$, with measurements provided for all of its states. Moreover, we also require the following assumptions to hold.

**Assumption 1** *Our models of interest consist of $M$ alternating "unknown" LTP sub-dynamics, $A^1(t)$, $A^2(t)$, $\cdots$, $A^M(t)$, whose activations are triggered by $M$ complementary "known" rectangular switching functions, $s^1(t)$, $s^2(t)$, $\cdots$, $s^M(t)$, during each cycle of the system. Both $A^i(t)$ and $s^i(t)$ are $T$-periodic functions, with the switching functions taking the form*

$$s^i(t) = \begin{cases} 1, & \text{if } \quad t_i + nT \leq t < t_{i+1} + nT, \quad \forall n \in \mathbb{Z} \\ 0, & \text{otherwise,} \end{cases} \quad (4.3)$$

*where $t_i$'s denote the known switching times and satisfy the conditions $t_1 = 0$, $t_{M+1} = T$, and $t_i < t_{i+1}, \forall i \in 1, \cdots, M$. The state and input matrices of (4.1) can hence be written as*

$$A(t) = \sum_{i=1}^{M} A^i(t) s^i(t), \qquad B(t) = \sum_{i=1}^{M} B^i(t) s^i(t). \quad (4.4)$$

□

**Assumption 2** *We assume that the system period $T$ as well as the transition times between different sub-system dynamics can be measured and are known. This information is sufficient to construct the switching functions, $s^1(t)$, $s^2(t)$, $\cdots$, $s^M(t)$, that trigger the activation of alternating sub-systems.* □

Based on the LTP framework and our assumptions listed above, the problem we are interested can be defined as: **Given**

- a number of single-sine (or sums-of-sines) input measurements applied at different frequencies, $u(t)$,

- corresponding state measurements, $x(t)$,

- the system period, $T$, and the switching times between successive subsystems, $s^1(t)$, $s^2(t)$, $\cdots$, $s^M(t)$,

**Estimate** piecewise smooth, linear time-periodic state and input matrices, $A^1(t)$, $A^2(t)$, $\cdots$, $A^M(t)$ and $B^1(t)$, $B^2(t)$, $\cdots$, $B^M(t)$.

## 4.1.2 Estimation of LTP System Matrices

Our analysis begins with obtaining Fourier series expansions for the state and input matrices, $A(t)$ and $B(t)$ as

$$A(t) = \sum_{n=-\infty}^{\infty} A_n e^{jn\omega_p t}, \qquad B(t) = \sum_{n=-\infty}^{\infty} B_n e^{jn\omega_p t}, \tag{4.5}$$

and transforming (4.1) into

$$\dot{\mathbf{x}}(t) = \sum_{n=-\infty}^{\infty} A_n e^{jn\omega_p t} \mathbf{x}(t) + \sum_{n=-\infty}^{\infty} B_n e^{jn\omega_p t} \mathbf{u}(t). \tag{4.6}$$

Assuming that the system is stable and that oscillations reach steady-state, we can then switch to the frequency domain through the Fourier transformation to yield

$$(j\omega) X(j\omega) = \sum_{n=-\infty}^{\infty} A_n X(j\omega - jn\omega_p) + \sum_{n=-\infty}^{\infty} B_n U(j\omega - jn\omega_p). \tag{4.7}$$

Fourier series coefficients of the multiplication of two periodic signals with the same period can be obtained as the convolution of the Fourier coefficients of the each individual signal. Considering the Fourier series coefficients for the rectangular switching functions as

$$s^i(t) = \sum_{n=-\infty}^{\infty} S_n^i e^{jn\omega_p t} \tag{4.8}$$

and using (4.4), Fourier series coefficients of $A(t)$ can then be obtained as

$$A_n = \sum_{k=-\infty}^{\infty} A_k^1 S_{n-k}^1 + \cdots + \sum_{k=-\infty}^{\infty} A_k^M S_{n-k}^M. \tag{4.9}$$

Substituting (4.9) and a similar expansion for $B_n$ into (4.7), we obtain

$$(j\omega)X(j\omega) = \sum_{n=-\infty}^{\infty}\left\{\sum_{k=-\infty}^{\infty} A_k^1 S_{n-k}^1\right\}X(j\omega - jn\omega_p) +$$
$$\vdots$$
$$\sum_{n=-\infty}^{\infty}\left\{\sum_{k=-\infty}^{\infty} A_k^M S_{n-k}^M\right\}X(j\omega - jn\omega_p) +$$
$$\sum_{n=-\infty}^{\infty}\left\{\sum_{k=-\infty}^{\infty} B_k^1 S_{n-k}^1\right\}U(j\omega - jn\omega_p) +$$
$$\vdots$$
$$\sum_{n=-\infty}^{\infty}\left\{\sum_{k=-\infty}^{\infty} B_k^M S_{n-k}^M\right\}U(j\omega - jn\omega_p). \qquad (4.10)$$

After reorganizing the terms, we obtain

$$(j\omega)X(j\omega) = \sum_{k=-\infty}^{\infty} A_k^1 \underbrace{\left\{\sum_{n=-\infty}^{\infty} S_{n-k}^1 X(j\omega - jn\omega_p)\right\}}_{X_k^1(j\omega)} +$$
$$\vdots$$
$$\sum_{k=-\infty}^{\infty} A_k^M \underbrace{\left\{\sum_{n=-\infty}^{\infty} S_{n-k}^M X(j\omega - jn\omega_p)\right\}}_{X_k^M(j\omega)} +$$
$$\sum_{k=-\infty}^{\infty} B_k^1 \underbrace{\left\{\sum_{n=-\infty}^{\infty} S_{n-k}^1 U(j\omega - jn\omega_p)\right\}}_{U_k^1(j\omega)} +$$
$$\vdots$$
$$\sum_{k=-\infty}^{\infty} B_k^M \underbrace{\left\{\sum_{n=-\infty}^{\infty} S_{n-k}^M U(j\omega - jn\omega_p)\right\}}_{U_k^M(j\omega)}. \qquad (4.11)$$

Here, $X_k^i(j\omega)$ and $U_k^j(j\omega)$ correspond to the convolution of the Fourier coefficients for the switching functions with the Fourier coefficients of the state and input functions, respectively. Now, truncating the infinite Fourier series to only $K$ components in either direction and converting (4.11) into matrix form yields

$$(j\omega)X(j\omega) = \underbrace{\begin{bmatrix} A^1_{-K} & \cdots & A^1_0 & \cdots & A^1_K \end{bmatrix}}_{\mathcal{A}^1} \underbrace{\begin{bmatrix} X^1_{-K}(j\omega) \\ \vdots \\ X^1_0(j\omega) \\ \vdots \\ X^1_K(j\omega) \end{bmatrix}}_{\mathcal{X}^1(j\omega)} +$$

$$\vdots$$

$$\underbrace{\begin{bmatrix} A^M_{-K} & \cdots & A^M_0 & \cdots & A^M_K \end{bmatrix}}_{\mathcal{A}^M} \underbrace{\begin{bmatrix} X^M_{-K}(j\omega) \\ \vdots \\ X^M_0(j\omega) \\ \vdots \\ X^M_K(j\omega) \end{bmatrix}}_{\mathcal{X}^M(j\omega)} +$$

$$\underbrace{\begin{bmatrix} B^1_{-K} & \cdots & B^1_0 & \cdots & B^1_K \end{bmatrix}}_{\mathcal{B}^1} \underbrace{\begin{bmatrix} U^1_{-K}(j\omega) \\ \vdots \\ U^1_0(j\omega) \\ \vdots \\ U^1_K(j\omega) \end{bmatrix}}_{\mathcal{U}^1(j\omega)} +$$

$$\vdots$$

$$\underbrace{\begin{bmatrix} B^M_{-K} & \cdots & B^M_0 & \cdots & B^M_K \end{bmatrix}}_{\mathcal{B}^M} \underbrace{\begin{bmatrix} U^M_{-K}(j\omega) \\ \vdots \\ U^M_0(j\omega) \\ \vdots \\ U^M_K(j\omega) \end{bmatrix}}_{\mathcal{U}^M(j\omega)} . \qquad (4.12)$$

**Remark 2** *In the identification of real systems and even for most simulated nonlinear systems, prior information on the proper choice of $K$, the limit on the number of Fourier series to be estimated, is unavailable. Moreover, the "true" value of $K$ can be even infinity. Since we currently focus on the identification of deterministic systems, an ad-hoc, yet acceptable solution is to choose a sufficiently big value for $K$ and disregard Fourier series coefficients that are less than a certain threshold.* □

**Remark 3** *Note that our choice of $K$ does not require truncating infinite summations for computing $X_k^1(j\omega), \cdots, X_k^M(j\omega), U_k^1(j\omega), \cdots, U_k^M(j\omega)$ in (4.11). On the other hand, computing infinite summations in computerized environments is of course not generally possible and hence another, possibly larger truncation $n = N$ can be used for these summations involving known quantities.* □

Before proceeding with a least-squares solution, we add an additional constraint to (4.12) to capture the requirements that system matrices, states and inputs are real valued. Let

$$A_K^i = A_{K,Re}^i + jA_{K,Im}^i, \qquad (4.13)$$

where $A_{K,Re}^i$ and $A_{K,Im}^i$ denote real and imaginary parts of the $K^{th}$ Fourier coefficient of the $i^{th}$ system matrix. We must then have

$$A_{-K}^i = A_{K,Re}^i - jA_{K,Im}^i \qquad (4.14)$$

to ensure that the system matrix is real-valued in the time domain. This yields

$$\mathcal{A}^i = \begin{bmatrix} A_{K,Re}^i - jA_{K,Im}^i & \cdots & A_0^i & \cdots & A_{K,Re}^i + jA_{K,Im}^i \end{bmatrix}$$

In order to simplify the formulation of our least-squares solution, we re-organize the terms in $\mathcal{A}^i$ to eliminate repetitions. More formally, we define

$$\bar{\mathcal{A}}^i := \begin{bmatrix} A_{K,Re}^i & \cdots & A_0^i & \cdots & A_{K,Im}^i \end{bmatrix} \qquad (4.15)$$

and

$$P := \begin{bmatrix} \mathcal{I} & & & & & & \mathcal{I} \\ & \ddots & & & & \iddots & \\ & & \mathcal{I} & 0 & \mathcal{I} & & \\ & & 0 & \mathcal{I} & 0 & & \\ & & -j\mathcal{I} & 0 & j\mathcal{I} & & \\ & \iddots & & & & \ddots & \\ -j\mathcal{I} & & & & & & j\mathcal{I} \end{bmatrix}, \qquad (4.16)$$

which are specifically constructed to satisfy

$$\mathcal{A}^i = \bar{\mathcal{A}}^i P \,. \qquad (4.17)$$

Using the decomposition above, (4.12) can be simplified by reorganizing terms and grouping known and unknown quantities in separate matrices as

$$\underbrace{(j\omega)X(j\omega)}_{\mathbf{y}^T(j\omega)} = \underbrace{\begin{bmatrix} \bar{\mathcal{A}}^1 & \cdots & \bar{\mathcal{A}}^M & \bar{\mathcal{B}}^1 & \cdots & \bar{\mathcal{B}}^M \end{bmatrix}}_{\mathbf{v}^T} \underbrace{\begin{bmatrix} P\bar{\mathcal{X}}^1(j\omega) \\ \vdots \\ P\bar{\mathcal{X}}^M(j\omega) \\ P\bar{\mathcal{U}}^1(j\omega) \\ \vdots \\ P\bar{\mathcal{U}}^M(j\omega) \end{bmatrix}}_{\mathbf{n}^T(j\omega)}.$$

Transposing both sides yields a linear equation as

$$\mathbf{n}(j\omega)\,\mathbf{v} = \mathbf{y}(j\omega) \qquad (4.18)$$

As explained in Section 3.3, LTP system outputs contain components not only in the input frequency but also at frequencies shifted by the harmonics of the pumping frequency. Consequently, we will evaluate (4.7) both at the input frequency $\omega$ as well as the shifted harmonics $\omega \pm h\omega_p, h \in \mathbb{Z}$ in order to capture

time-periodic system as

$$\underbrace{\begin{bmatrix} \mathbf{n}(j\omega + h\omega_p) \\ \vdots \\ \mathbf{n}(j\omega) \\ \vdots \\ \mathbf{n}(j\omega - h\omega_p) \end{bmatrix}}_{N(j\omega)} \mathbf{v} = \underbrace{\begin{bmatrix} \mathbf{y}(j\omega + h\omega_p) \\ \vdots \\ \mathbf{y}(j\omega) \\ \vdots \\ \mathbf{y}(j\omega - h\omega_p) \end{bmatrix}}_{Y(j\omega)}. \quad (4.19)$$

In order to ensure that the solutions are real-valued, we separate the real and imaginary parts of the possibly complex-valued components computed from out test data as

$$\underbrace{\begin{bmatrix} Re\{N(j\omega)\} \\ Im\{N(j\omega)\} \end{bmatrix}}_{N_w} \mathbf{v} = \underbrace{\begin{bmatrix} Re\{Y(j\omega)\} \\ Im\{Y(j\omega)\} \end{bmatrix}}_{Y_w}. \quad (4.20)$$

**Remark 4** *Separating real and imaginary components of complex-valued components computed from data doubles the number of tests used for the least squares solution.* □

Subsequently, collecting together multiple measurements from different frequencies yields

$$\underbrace{\begin{bmatrix} \vdots \\ N_w \\ \vdots \end{bmatrix}}_{\mathcal{N}} \mathbf{v} = \underbrace{\begin{bmatrix} \vdots \\ Y_w \\ \vdots \end{bmatrix}}_{\mathcal{Y}} \quad (4.21)$$

Now, assuming that $\mathcal{N}$ has full rank, the least squares error solution can be found as

$$\mathbf{v} = (\mathcal{N}^H \mathcal{N})^{-1} \mathcal{N}^H \mathcal{Y}. \quad (4.22)$$

We can extract Fourier series coefficient matrices from $\mathbf{v}$ and then construct $A^1(t)$, $A^2(t), \cdots, A^M(t)$ and $B^1(t), B^2(t), \cdots, B^M(t)$ using Fourier series synthesis as

$$A^i(t) = \sum_{n=-K}^{K} A_n^i e^{jn\omega_p t}, \quad B^i(t) = \sum_{n=-K}^{K} B_n^i e^{jn\omega_p t}. \quad (4.23)$$

### 4.1.3 Application: Switching Damped Mathieu Function

In this section, we present an example system, a piecewise smooth linear time-periodic function, and evaluate the performance of the proposed algorithm on this example. The piecewise smooth LTP system we consider in this example consists of two switching damped Mathieu function with the form

$$\ddot{x}(t) + \underbrace{2\zeta\omega_n}_{c}\dot{x}(t) + \underbrace{(1 + 2\beta \cos \omega_p t)\omega_n^2}_{\kappa(t)} x(t) = u(t) \tag{4.24}$$

where $c$ represents piecewise constant damping term, while $\kappa(t)$ represents piecewise smooth time-periodic compliance term in the Mathieu function. Piecewise smooth LTP system dynamics can now be written as

$$\ddot{x}(t) = \begin{cases} u(t) - c_1 \dot{x}(t) - \kappa_1(t)x(t), & \text{if } Tn \leq t \leq Tn + T/2 \\ u(t) - c_2 \dot{x}(t) - \kappa_2(t)x(t), & \text{otherwise.} \end{cases} \tag{4.25}$$

By using (4.25), state and input matrices can be obtained as

$$A^1(t) = \begin{bmatrix} 0 & 1 \\ -(1 + 2\beta_1 \cos \omega_p t)\omega_n^2 & -2\zeta_1 \omega_n \end{bmatrix} \quad B = \begin{bmatrix} 0 \\ 1 \end{bmatrix}$$

$$A^2(t) = \begin{bmatrix} 0 & 1 \\ -(1 + 2\beta_2 \cos \omega_p t)\omega_n^2 & -2\zeta_2 \omega_n \end{bmatrix}$$

where the input matrix is time-invariant for this example. By using the parameters specified in Table 4.1, actual values of the Fourier series coefficients to be estimated can be found as

$$A_0^1 = \begin{bmatrix} 0 & 1 \\ -39.4784 & -3.7699 \end{bmatrix}, \quad A_1^1 = \begin{bmatrix} 0 & 0 \\ -3.9478 & 0 \end{bmatrix}$$

$$A_0^2 = \begin{bmatrix} 0 & 1 \\ -39.4784 & -1.2566 \end{bmatrix}, \quad A_1^2 = \begin{bmatrix} 0 & 0 \\ -7.8957 & 0 \end{bmatrix} \tag{4.26}$$

Notice that $A_{-1}^1 = A_1^1$ and $A_{-1}^2 = A_1^2$, since they have real values.

In order to begin our estimation process, we first simulate the system in order to collect necessary data for the parametric identification process. To accomplish

Table 4.1: Mathieu Function Parameters

| $T$ | $\omega_p$ | $\omega_n$ | $\zeta_1$ | $\beta_1$ | $\zeta_2$ | $\beta_2$ |
|---|---|---|---|---|---|---|
| 0.5 | $4\pi$ | $2\pi$ | 0.3 | 0.1 | 0.1 | 0.2 |

this goal, we simulate the piecewise smooth LTP dynamics of (4.25) by applying single sine inputs as

$$u(t) = \sin\left(2\pi(0.2 + 0.4k)t\right) \tag{4.27}$$

where $k$ is chosen in the range, $k \in [0, 19]$, so that we generate 20 input stimuli for our example and record the state measurements. Our input signals are 10 $s$. long and all data are sampled at 100 $Hz$. Note that we currently use single sine inputs in our tests but our modeling framework also supports sums-of-sines methodology, which decreases the number of tests required for system identification

Note that previously, [34] showed that in order to estimate HTF components uniquely using single sinusoidal signals (i.e. one experimental data per each frequency), the input signal must not be equal to the harmonics of the half of the pumping frequency, i.e. $\omega \neq k\omega_p/2$, $k \in \mathbb{Z}$. Since we are attempting to compute the parametric LTP matrices from input-output data, we no longer need to satisfy this constraint for the case of pure piece-wise smooth LTP systems. However, [105] also showed that for the identification of non-linear systems that operate around a limit-cycle, input frequencies that are equal to the harmonics of the pumping frequency should also be avoided in order to isolate the frequency components of the limit-cycle and response around the limit-cycle. For this reason we also do not include the harmonics of the pumping frequency in our input signals.

Once we applied the input stimuli and collected our state measurements, we need to make three implementation choices before building our least squares estimation matrices. First of all, we consider $N = 20$, yielding 41 Fourier series coefficients for the computation of $X_k^i(j\omega)$ and $U_k^i(j\omega)$ in (4.11), since our bound here originates from the signal length (see Remark 3). Secondly, we choose $K = 1$, yielding the number of Fourier series coefficients to be estimated as 3, so that we can capture the time-periodic behavior in our estimations. Finally, we consider $h = 2$ harmonics of the LTP system for (4.19), so that we evaluate each input at 5 different frequencies.

Based on our implementation choices explained above, Fourier series coefficients are estimated as

$$\hat{A}_0^1 = \begin{bmatrix} 0.0002 & 0.9999 \\ -39.4404 & -3.7786 \end{bmatrix} \quad (4.28)$$

$$\hat{A}_1^1 = \begin{bmatrix} -0.0001 - j0.0003 & -0.0001 - j0.0000 \\ -3.9253 + j0.0095 & 0.0090 - j0.0012 \end{bmatrix}$$

$$\hat{A}_0^2 = \begin{bmatrix} 0.0009 & 1.0003 \\ -40.0038 & -1.2824 \end{bmatrix}$$

$$\hat{A}_1^2 = \begin{bmatrix} 0.0002 - j0.0008 & -0.0001 - j0.0002 \\ -7.9860 + j0.0242 & 0.0017 + j0.0084 \end{bmatrix}$$

with $A_{-1}^1 = A_1^{1*}$ and $A_{-1}^2 = A_1^{2*}$. We then reconstruct $\hat{A}^1(t)$ and $\hat{A}^2(t)$ using (4.23). Similarly, $B$ is estimated as $\hat{B} = \begin{bmatrix} 0 \\ 0.9965 \end{bmatrix}$.

In order to better express our estimation results, we plot time domain graphs of $\kappa(t)$ and $c(t)$ defined in (4.24) both using the actual and estimated system matrices as illustrated in Fig. 4.1. $\kappa(t)$ represents the piecewise time-periodic compliance behavior in our Mathieu function, while $c(t)$ represents piecewise time-invariant damping loss. Both of these variables switch to another parameter at the half of the period, which brings a time-periodic nature to both functions.

As illustrated in Fig. 4.1, our estimations with $K = 1$ fits well to the actual system parameters, since we didn't contaminate our simulation data with noise for experiments. Apart from that, $K = 1$ is the exact number of Fourier series coefficients for our system as seen in (4.26). Actually, it gives an exact fit for $\kappa(t)$ but it is an overfit for $c(t)$, since $K = 0$ would be sufficient to represent its piecewise LTI nature. However, overfitting does not cause any problem for $c(t)$, since higher order Fourier series coefficients are estimated too small as in (4.28).

In order to investigate the effects of under-fitting, we repeated our estimations for $K = 0$ and re-estimated our system parameters as illustrated in Fig. 4.1. In this case, it can be observed that $\kappa(t)$ can not be estimated accurately, although we can still obtain accurate estimations for $c(t)$.

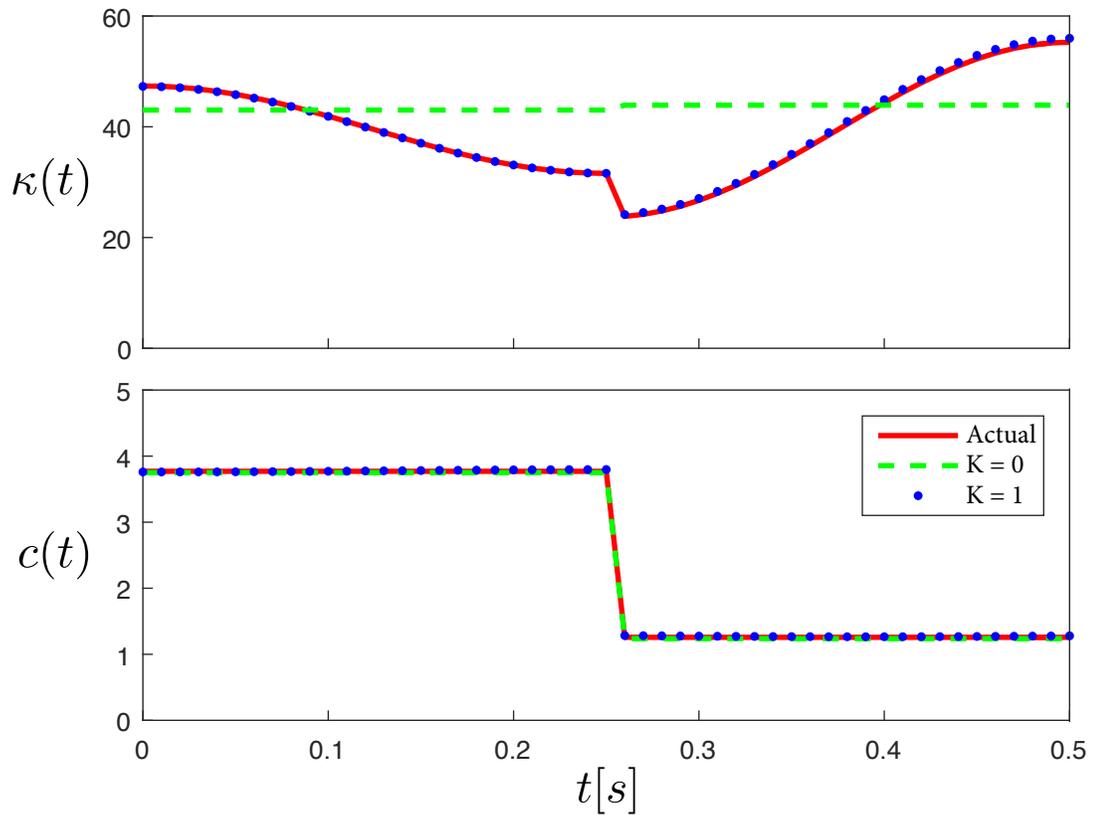

Figure 4.1: Estimation results for compliance and damping term for a single period. $K = 0$ corresponds to piecewise LTI case and $K = 1$ corresponds to piecewise LTP case.

### 4.1.4 Discussion

A huge class of physical physical dynamical systems exhibit quasi-periodic trajectories and hybrid characteristics. However, it is fair to assume that only a few of the system identification studies in literature concentrate on the identification of hybrid dynamic system that operate around some periodic orbits, which is the main goal of this study. Specifically, we limit our attention to the hybrid systems that has continuous state trajectories but potentially discontinuous vector fields. Under some assumptions, the local flow around the periodic orbit of such a system can be approximated with a hybrid LTP system.

Based on these motivations, we introduced a state space parametric identification framework for hybrid LTP systems for which the periodic switching times are assumed to be known. We formulated the problem in a linear regression framework in frequency domain, where we estimated Fourier series coefficients of the time-periodic system and input matrices. Then, we re-constructed the time domain system and input matrices using Fourier synthesis after a least squares solution. Currently, our formulations assume full state measurements which is the main limitation of our method. In the next section, we will attempt to improve our method such that we can relax this assumption also including process and measurement noise.

## 4.2 Frequency Domain Subspace Identification of Linear Time Periodic Systems

In this part, we introduce a frequency domain subspace-based state space identification method for linear time periodic (LTP) systems. Many problems in engineering and biology, such as wind turbines [106], rotor bearing systems [31], aircraft models [107], power distribution networks [108], and human walking [109] require the consideration of time periodic dynamics. As such, the analysis, identification, and control of LTP systems received considerable attention [32, 69, 77, 110].

The frequency domain analysis method for LTP systems introduced by Wereley [32, 111], wherein time periodic system matrices and signals in the LTP state space formulation were expanded into their Fourier series coefficients, is among important first contributions in this area. In this work, the principle of harmonic balance was used to obtain the concept of harmonic transfer functions (HTFs) for LTP systems. The initial formulation was developed for continuous-time LTP systems as infinite-dimensional operators, which were subsequently extended to handle discrete-time LTP systems in the form of finite-dimensional transfer functions operators [34].

Most existing literature on LTP system identification focuses on non-parametrically identifying HTFs as an input–output characterization of LTP systems [31, 34, 69, 112], including our previous work on identification of legged locomotion behaviors around stable periodic orbits [61–63]. Even though there are many representations for a given finite-dimensional linear dynamical system that would produce equivalent input–output characterizations, state space models are commonly used and widely accepted to be practical for multi-variable finite dimensional linear systems. Consequently, this study aims to develop a frequency domain state space identification method for LTP systems.

A well-established solution to state space identification for linear time invariant (LTI) systems is the subspace identification framework [98], wherein system matrices are estimated using oblique projections and singular value decomposition. These are non-iterative algorithms that produce "optimal" state space estimates with a very low time complexity. More importantly, the subspace identification framework supports both time domain [98] and frequency domain [113] data to obtain state space estimates for LTI systems. There are also studies on extending this framework to time-varying systems. Verhaegen et al. developed a subspace identification method for estimating successive discrete state transition matrices from time domain data for periodically and arbitrarily time varying systems [100]. A similar time domain subspace identification method for linear time varying (LTV) systems has been proposed by Shi et al. in [99]. There are also some studies which assume known/chosen scheduling functions for identifying linear parameter varying (LPV) systems [91, 114]. Similarly, a continuous-time subspace identification method for periodically time varying state space models with known/chosen scheduling functions is presented in [115].

Nevertheless, to the best of our knowledge, there are currently no general methods that can address the frequency domain subspace identification problem for LTP systems without any prior assumptions on scheduling functions. In this context, our aim in this study is to develop a new general subspace identification methodology for estimating state space models from frequency response data for LTP systems.

LTP systems can generally be transformed to equivalent discrete-time LTI systems via bilinear transform [116] and lifting [117]. Based on this observation, we first use the frequency domain subspace identification method of [113] to estimate a discrete-time LTI state space equivalent model from the input–output data of the original LTP system. A key property of our frequency domain lifting method is the specific parametric structure of Fourier series coefficients associated with the original LTP system. In order to identify these coefficients in the estimated system, we propose to use an optimization framework based on particle swarm optimization (PSO) methods. An important contribution of this study is hence the method we propose to obtain an LTP state-space realization using the estimated lifted LTI system. Our identification–realization algorithm also allows the realization of Floquet-transformed state space models for LTP systems with arbitrary time-periodic system matrices, whose analytic derivation are often very challenging and may even be impossible [118].

### 4.2.1 Problem Formulation and Solution Methodology

In this study, we consider multi-input/multi-output (MIMO), stable, linear time-periodic (LTP) systems represented by

$$\dot{\bar{x}}(t) = \bar{A}(t)\bar{x}(t) + \bar{B}(t)u(t)$$
$$y(t) = \bar{C}(t)\bar{x}(t) + \bar{D}(t)u(t) ,$$
(4.29)

where $u(t) \in \mathbb{R}^{n_u}$, $y(t) \in \mathbb{R}^{n_y}$ and $\bar{x}(t) \in \mathbb{R}^{n_p}$ represent input, output and state vectors, respectively. The system matrices are assumed to be periodic with a fixed, known common period $T > 0$, with $\bar{A}(t) = \bar{A}(t + nT)$, $\bar{B}(t) = \bar{B}(t + nT)$, $\bar{C}(t) = \bar{C}(t + nT)$ and $\bar{D}(t) = \bar{D}(t + nT)$, $\forall n \in \mathbb{Z}$. We now formulate the identification problem as follows:

**Given**

- a number of single-sine (or sum-of-sines [69]) input signals applied at different frequencies, $u(t)$,

- corresponding output measurements, $y(t)$,

- the system period, $T$,

**Estimate**

- linear, time-periodic system matrices $\bar{A}(t)$, $\bar{B}(t)$, $\bar{C}(t)$ and $\bar{D}(t)$ up to a similarity transformation.

In principle, the subspace identification method we propose and describe in this section is applicable to both continuous-time and discrete-time LTP systems for which the associated monodromy matrix is non-singular. However, it has been noted that the data (Hankel) matrices used for frequency domain subspace identification of continuous-time systems may become ill-conditioned with increasing system dimension [119]. Even though different methods have been proposed to address this issue [119, 120], we will find it more convenient to formulate our estimation method for discrete-time LTP systems, noting that continuous-time LTP systems can be transformed into discrete-time equivalents using bilinear (Tustin) transformations [117].

As stated earlier, for an LTP system, a complex exponential input with frequency $\omega$ produces output not only at the input frequency (which is the case for LTI systems), but also at different harmonics $\omega \pm k\omega_p$, $k \in \mathbb{Z}$ separated by the system frequency, $\omega_p = 2\pi/T$, with possibly different magnitudes and phases in steady state. In this context, the concept of Harmonic Transfer Functions (HTFs) was developed to represent each harmonic response of the LTP system with a distinct transfer function $G_k(w + k\omega_p)$ for $k \in \mathbb{Z}$ [32]. This approach represents an LTP system as the superposition of multiple modulated LTI systems. As such, HTFs can be used as a lifting technique to transform an LTP system to an LTI equivalent [121] for which subspace identification methods can now be expected to yield sufficiently accurate state space estimates [98, 113, 114, 122, 123]. Among available alternatives, we utilize the frequency domain subspace identification technique presented in [113].

To this end, we first introduce the transformations used for obtaining a discrete-time LTI equivalent to an LTP system. Naturally, the original state space form of (4.29) will not be available before system identification is performed. Therefore, the transformations explained in this section are not directly applied. Nevertheless, these derivations are necessary to transform input–output data (not the state space model) collected from the original unknown LTP system to the reciprocal input–output data of the equivalent discrete-time LTI system.

### 4.2.2 The Floquet Theory (Transform)

The Floquet theorem is a crucial tool used for the analysis and control of linear time periodic (LTP) systems. One of the main results of the Floquet theorem is the existence of a coordinate change by using which any LTP system in the form (4.29) can be represented with a time-invariant state update matrix with the other system matrices preserve their periodic nature.

In a mathematical sense, Let $\Phi(t,\tau)$ be the state transition matrix and $\Phi(T,0)$ be the fundamental monodromy matrix (state transition matrix over one period) of (4.29). Then, the main results of the Floquet theorem can be summarized as below.

**Theorem 1 (Floquet [124])** *If the monodromy matrix, $\Phi(T,0)$ for an LTP system in the form (4.29) is nonsingular, then the following results hold:*

1. **State transition matrix:** *The state transition matrix of (4.29) can always be represented in the form*

$$\Phi(t,\tau) = P(t)e^{\mathbf{A}(t-\tau)}P^{-1}(\tau) \tag{4.30}$$

   *where $P(t)$ is an invertible $T$-periodic matrix and $\mathbf{A}$ is a (possibly complex-valued) constant matrix.*

2. **Similarity transformation:** Then, the state transformation $P(t)$ transforms the system in (4.29) to a new form for which the state update matrix is time-invariant as

$$\begin{aligned}\dot{\mathbf{x}}(t) &= \mathbf{A}\mathbf{x}(t) + \mathbf{B}(t)u(t), \\ y(t) &= \mathbf{C}(t)\mathbf{x}(t) + \mathbf{D}(t)u(t).\end{aligned} \quad (4.31)$$

where

$$\begin{aligned}\mathbf{A} &= P^{-1}(t)\{\bar{A}(t)P(t) - \dot{P}(t)\} & (4.32) \\ \mathbf{B}(t) &= P^{-1}(t)\bar{B}(t) & (4.33) \\ \mathbf{C}(t) &= \bar{C}(t)P(t) & (4.34) \\ \mathbf{D}(t) &= \bar{D}(t) & (4.35)\end{aligned}$$

**Remark 5** *Note that although the initial definition of Floquet transformation does not guarantee a real-valued state update matrix $\mathbf{A}$, it is always possible to find a real-valued $\mathbf{A}$ by using a $2T$-periodic $P(t)$ [118].*

### 4.2.3 Discretization via Bilinear (Tustin) Transform

We now describe how the continuous-time state-space model in (4.31) is transformed to a discrete-time LTP equivalent. Motivated by the literature [116], we utilize the bilinear (Tustin) transform to map the complex values in $s$ domain to the $z$ domain with a trapezoidal rule approximation as

$$s = \frac{2(z-1)}{T_s(z+1)}, \quad (4.36)$$

where $T_s$ is the sampling period for the continuos-time system. Taking the Laplace transform of (4.31), we have

$$\begin{aligned}s\mathbf{X}(s) &= \mathbf{A}\mathbf{X}(s) + \mathbf{B}(s) * \mathbf{U}(s), \\ \mathbf{Y}(s) &= \mathbf{C}(s) * \mathbf{X}(s) + \mathbf{D}(s) * \mathbf{U}(s).\end{aligned} \quad (4.37)$$

where, $*$ represents the convolution operator. We then transform (4.37) into the $z$-domain by using (4.36) as

$$(z-1)\mathbf{X}(z) = \mathbf{A}\frac{T_s}{2}(z+1)\mathbf{X}(z)$$
$$+ \frac{T_s}{2}(z+1)(\mathbf{B}(s) * \mathbf{U}(s)), \quad (4.38)$$
$$\mathbf{Y}(z) = \mathbf{C}(z) * \mathbf{X}(z) + \mathbf{D}(z) * \mathbf{U}(z). \quad (4.39)$$

Now, after transforming (4.38) to time domain yields the state update equation as

$$\mathbf{x}((k+1)T_s) - \mathbf{x}(kT_s) = \mathbf{A}\frac{T_s}{2}\left[\mathbf{x}((k+1)T_s) + \mathbf{x}(kT_s)\right]$$
$$+ \frac{T_s}{2}\left[\mathbf{B}((k+1)T_s)u((k+1)T_s)\right]$$
$$+ \frac{T_s}{2}\left[\mathbf{B}(kT_s)u(kT_s)\right] \quad (4.40)$$

Separating the terms for time $k$ and $k+1$ to different sides yields

$$\mathbf{x}((k+1)T_s) - \mathbf{A}\frac{T_s}{2}\mathbf{x}((k+1)T_s)$$
$$- \frac{T_s}{2}\left[\mathbf{B}((k+1)T_s)u((k+1)T_s)\right] \quad (4.41)$$
$$= \mathbf{x}(kT_s) + \mathbf{A}\frac{T_s}{2}\mathbf{x}(kT_s) + \frac{T_s}{2}\left[\mathbf{B}(kT_s)u(kT_s)\right].$$

By using a change of variables as

$$\sqrt{T_s}\mathbf{x}_d[k+1] := \mathbf{x}(kT_s) + \mathbf{A}\frac{T_s}{2}\mathbf{x}(kT_s) + \frac{T_s}{2}\left[\mathbf{B}(kT_s)u(kT_s)\right] \quad (4.42)$$

and evaluating (4.41) at time $k$ yields

$$(I - \mathbf{A}\frac{T_s}{2})\mathbf{x}(kT_s) = \sqrt{T_s}\mathbf{x}_d[k] + \frac{T_s}{2}\left[\mathbf{B}(kT_s)u(kT_s)\right] \quad (4.43)$$

Hence, solution for $\mathbf{x}(kT_s)$ can be obtained as

$$\mathbf{x}(kT_s) = (I - \mathbf{A}\frac{T_s}{2})^{-1}\sqrt{T_s}\mathbf{x}_d[k]$$
$$+ (I - \mathbf{A}\frac{T_s}{2})^{-1}\frac{T_s}{2}\left[\mathbf{B}(kT_s)u(kT_s)\right] \quad (4.44)$$

Now, considering the right hand side of (4.41) as

$$\sqrt{T_s}\mathbf{x}_d[k+1] = (I + \mathbf{A}\frac{T_s}{2})\mathbf{x}(kT_s) + \frac{T_s}{2}\left[\mathbf{B}(kT_s)u(kT_s)\right], \qquad (4.45)$$

and plugging (4.44) into (4.45) and organizing the terms yield

$$\mathbf{x}_d[k+1] = \underbrace{(I + \mathbf{A}\frac{T_s}{2})(I - \mathbf{A}\frac{T_s}{2})^{-1}}_{\mathbf{A}_d}\mathbf{x}_d[k] \\ + \underbrace{\frac{2}{\sqrt{T_s}}(\frac{2}{T_s}I - \mathbf{A})^{-1}\mathbf{B}(kT_s)}_{\mathbf{B}_d[k]}\mathbf{u}_d[k], \qquad (4.46)$$

where $\mathbf{u}_d[k] := u(kT_s)$.

In order to obtain a discrete-time representation for the output equation, we first sample the output equation (4.31) as

$$y(kT_s) = \mathbf{C}(kT_s)\mathbf{x}(kT_s) + \mathbf{D}(kT_s)u(kT_s). \qquad (4.47)$$

Now, plugging (4.44) into (4.47) and organizing the terms yields

$$\mathbf{y}_d[k] = \underbrace{\frac{2}{\sqrt{T_s}}\mathbf{C}(kT_s)(\frac{2}{T_s}I - \mathbf{A})^{-1}}_{\mathbf{C}_d[k]}\mathbf{x}_d[k] \\ + \underbrace{\left(\mathbf{D}(kT_s) + \mathbf{C}(kT_s)(\frac{2}{T_s}I - \mathbf{A})^{-1}\right)}_{\mathbf{D}_d[k]}\mathbf{u}_d[k], \qquad (4.48)$$

where $\mathbf{y}_d[k] := y(kT_s)$. The transformed discrete-time LTP state space model of (4.31) can now be formulated as

$$\begin{aligned}\mathbf{x}_d[k+1] &= \mathbf{A}_d\mathbf{x}_d[k] + \mathbf{B}_d[k]\mathbf{u}_d[k], \\ \mathbf{y}_d[k] &= \mathbf{C}_d[k]\mathbf{x}_d[k] + \mathbf{D}_d[k]\mathbf{u}_d[k],\end{aligned} \qquad (4.49)$$

where

$$\mathbf{A}_d = (\frac{2}{T_s}I + \mathbf{A})(\frac{2}{T_s}I - \mathbf{A})^{-1}, \qquad (4.50)$$

$$\mathbf{B}_d[k] = \frac{2}{\sqrt{T_s}}(\frac{2}{T_s}I - \mathbf{A})^{-1}\mathbf{B}(kT_s), \qquad (4.51)$$

$$\mathbf{C}_d[k] = \frac{2}{\sqrt{T_s}}\mathbf{C}(kT_s)(\frac{2}{T_s}I - \mathbf{A})^{-1}, \qquad (4.52)$$

$$\mathbf{D}_d[k] = \mathbf{D}(kT_s) + \mathbf{C}(kT_s)(\frac{2}{T_s}I - \mathbf{A})^{-1}\mathbf{B}(kT_s). \qquad (4.53)$$

Note that (4.49) also forms a periodic system, where $\mathbf{B}_d[k] = \mathbf{B}_d[k+nN]$, $\forall n \in \mathbb{Z}$ (also valid for $\mathbf{C}_d[k]$ and $\mathbf{D}_d[k]$) and $N$ is the discrete-time system period defined as $N := T/T_s$. For the sake of simplicity, $N$ is assumed to be even.

### 4.2.4 Lifting to a Time-Invariant Reformulation

It has been shown that a majority of the LTP systems can be represented with time invariant formulations [117]. This motivates our use of the frequency domain lifting method based on the principle of harmonic balance to obtain an LTI equivalent state space model for the discrete-time LTP system of (4.49). Among possible alternatives (see [117] for a survey), we use the approach proposed in [34] due to the convenient structure of Fourier series coefficients for periodic system matrices in time invariant Toeplitz matrices.

Wereley showed that an LTP system maps a complex exponential input to another complex exponential signal, which is modulated by the complex Fourier series expansion of a periodic signal of $\omega_p$ [32]. These signals are called exponentially modulated periodic (EMP) signals and are defined as

$$u(t) = e^{st} \sum_{n=-\infty}^{\infty} U_n e^{jn\omega_p t} \tag{4.54}$$

where $t \geq 0$ and $s \in \mathbb{C}$. Sampling the continuous-time signal (4.54) with a period $T_s$ yields the discrete-time EMP signal

$$\mathbf{u}_d[k] := z^k \sum_{n=-N/2}^{N/2-1} U_n e^{j2\pi \frac{nk}{N}}. \tag{4.55}$$

where, $U_n$ are called modulated Fourier series coefficients for EMP signals and are defined as

$$U_n := \frac{1}{N} \sum_{k=0}^{N-1} (\mathbf{u}_d[k] z^{-k}) e^{-j2\pi \frac{nk}{N}}. \tag{4.56}$$

It has been shown that when an LTP system is given an EMP input, state and output signals are also EMP in steady-state [32]. Therefore, similar to (4.55), we obtain the state and output signals as

$$\mathbf{x}_d[k] = z^k \sum_{n=-\frac{N}{2}}^{\frac{N}{2}-1} X_n e^{j2\pi \frac{nk}{N}}, \quad (4.57)$$

$$\mathbf{y}_d[k] = z^k \sum_{n=-\frac{N}{2}}^{\frac{N}{2}-1} Y_n e^{j2\pi \frac{nk}{N}}. \quad (4.58)$$

In addition, discrete-time Fourier synthesis equation for the periodic system matrices are computed as

$$\mathbf{B}_d[k] = \sum_{n=-\frac{N}{2}}^{\frac{N}{2}-1} B_n e^{j2\pi \frac{nk}{N}}, \quad (4.59)$$

$$\mathbf{C}_d[k] = \sum_{n=-\frac{N}{2}}^{\frac{N}{2}-1} C_n e^{j2\pi \frac{nk}{N}}, \quad (4.60)$$

$$\mathbf{D}_d[k] = \sum_{n=-\frac{N}{2}}^{\frac{N}{2}-1} D_n e^{j2\pi \frac{nk}{N}}, \quad (4.61)$$

Note that since $\mathbf{A}_d$ is in time-invariant form due to Floquet transform, there is no need to compute Fourier series coefficients for $\mathbf{A}_d$.

Now, plugging in the Fourier synthesis equations of (4.59), (4.60) and (4.61) into (4.49) yields

$$z^{k+1} \sum_{n=-\frac{N}{2}}^{\frac{N}{2}-1} X_n e^{j2\pi \frac{n(k+1)}{N}} =$$

$$\mathbf{A}_d \left( z^k \sum_{n=-\frac{N}{2}}^{\frac{N}{2}-1} X_n e^{j2\pi \frac{nk}{N}} \right) \quad (4.62)$$

$$+ \left( \sum_{n=-\frac{N}{2}}^{\frac{N}{2}-1} B_n e^{j2\pi \frac{nk}{N}} \right) \left( z^k \sum_{m=-\frac{N}{2}}^{\frac{N}{2}-1} U_m e^{j2\pi \frac{mk}{N}} \right).$$

Organizing the terms in both sides yields

$$z^{k+1} \sum_{n=-\frac{N}{2}}^{\frac{N}{2}-1} X_n e^{j2\pi \frac{n}{N}} e^{j2\pi \frac{nk}{N}} = \\ z^k \sum_{n=-\frac{N}{2}}^{\frac{N}{2}-1} \left( \mathbf{A}_d X_n + \sum_{m=-\frac{N}{2}}^{\frac{N}{2}-1} B_{n-m} U_m \right) e^{j2\pi \frac{nk}{N}}. \quad (4.63)$$

Now, rearranging the terms in the right hand side yields

$$0 = z^k \sum_{n=-\frac{N}{2}}^{\frac{N}{2}-1} \left( z X_n e^{j2\pi \frac{n}{N}} - \mathbf{A}_d X_n - \sum_{m=-\frac{N}{2}}^{\frac{N}{2}-1} B_{n-m} U_m \right) e^{j2\pi \frac{nk}{N}} \quad (4.64)$$

The set of exponentials, $\{ e^{j2\pi \frac{nk}{N}} \mid n \in [-\frac{N}{2}, \frac{N}{2} - 1] \}$, constitute an orthonormal basis. Thus, by the principle of harmonic balance, each term enclosed by the brackets must individually equal zero to ensure that the overall sum is zero. Therefore, for all $n \in [-\frac{N}{2}, \frac{N}{2} - 1]$, we have

$$z e^{j2\pi \frac{n}{N}} X_n = \mathbf{A}_d X_n + \sum_{m=-N/2}^{N/2-1} B_{n-m} U_m. \quad (4.65)$$

Note that the above equation is valid since Fourier coefficients, $B_m$, are also periodic with $N$. Performing similar steps for the output equation yields

$$Y_n = \sum_{m=-N/2}^{N/2-1} C_{n-m} X_n + \sum_{m=-N/2}^{N/2-1} D_{n-m} U_m \quad (4.66)$$

for all $n \in [-N/2, N/2 - 1]$.

Similar to continuos-time systems, (4.65) and (4.66) can be represented with Toeplitz matrices towards obtaining an LTI state space model. To this end, we first redefine the state, input and output vectors as

$$\mathcal{X}_d := \begin{bmatrix} X_{-\frac{N}{2}} \\ \vdots \\ X_{-1} \\ X_0 \\ X_1 \\ \vdots \\ X_{\frac{N}{2}-1} \end{bmatrix} \quad \mathcal{U}_d := \begin{bmatrix} U_{-\frac{N}{2}} \\ \vdots \\ U_{-1} \\ U_0 \\ U_1 \\ \vdots \\ U_{\frac{N}{2}-1} \end{bmatrix} \quad \mathcal{Y}_d := \begin{bmatrix} Y_{-\frac{N}{2}} \\ \vdots \\ Y_{-1} \\ Y_0 \\ Y_1 \\ \vdots \\ Y_{\frac{N}{2}-1} \end{bmatrix}. \quad (4.67)$$

In addition, time-invariant reformulation of the originally $N$-periodic input matrix can be obtained as

$$\mathcal{B}_d := \begin{bmatrix} B_0 & B_{-1} & \ldots & B_{-\frac{N}{2}} & 0 & \ldots & 0 \\ B_1 & B_0 & B_{-1} & \ldots & B_{-\frac{N}{2}} & \ddots & \ldots \\ \ldots & \ddots & \ddots & \ddots & \ddots & \ddots & 0 \\ B_{\frac{N}{2}-1} & \ldots & B_1 & B_0 & B_{-1} & \ldots & B_{-\frac{N}{2}} \\ 0 & \ddots & \ddots & \ddots & \ddots & \ddots & \ldots \\ 0 & \ddots & B_{\frac{N}{2}-1} & \ldots & B_1 & B_0 & B_{-1} \\ 0 & \ldots & 0 & B_{\frac{N}{2}-1} & \ldots & B_1 & B_0 \end{bmatrix}. \quad (4.68)$$

Similarly, Toeplitz forms for $\mathcal{C}_d$ and $\mathcal{D}_d$ matrices can be obtained in terms of their Fourier series coefficients, $\{C_n \mid n \in [-\frac{N}{2}, \frac{N}{2}-1]\}$ and $\{D_n \mid n \in [-\frac{N}{2}, \frac{N}{2}-1]\}$, respectively. Note that, since $\mathbf{A}_d$ is time-invariant, its Toeplitz form, $\mathcal{A}_d$ includes only $\mathbf{A}_d$ in its diagonals

$$\mathcal{A}_d := blkdiag\{\mathbf{A}_d\} \mid \mathcal{A}_d \in \mathbb{R}^{Nn_p \times Nn_p} \quad (4.69)$$

where $blkdiag$ represents a block-diagonal matrix composed of $\mathbf{A}_d$ in its diagonals. As a final step, we define a modulation matrix, $\mathcal{N}_d$ to capture the exponential terms in (4.65) as

$$\mathcal{N}_d := blkdiag\{e^{j2\pi\frac{n}{N}} I_{n_p} \mid \forall n \in [-N/2, N/2 - 1]\}. \quad (4.70)$$

by using which we also define

$$\mathcal{A}_{dN} := \mathcal{N}_d^{-1} \mathcal{A}_d, \quad \mathcal{B}_{dN} := \mathcal{N}_d^{-1} \mathcal{B}_d. \quad (4.71)$$

Now, (4.65) and (4.66) is represented as an LTI system using the Toeplitz matrices as

$$\begin{aligned} z\mathcal{X}_d &= \mathcal{A}_{dN}\mathcal{X}_d + \mathcal{B}_{dN}\mathcal{U}_d \\ \mathcal{Y}_d &= \mathcal{C}_d\mathcal{X}_d + \mathcal{D}_d\mathcal{U}_d. \end{aligned} \quad (4.72)$$

and harmonic transfer functions of the system becomes

$$G_d(z) = \mathcal{C}_d(zI - \mathcal{A}_{dN})^{-1}\mathcal{B}_{dN} + \mathcal{D}_d. \quad (4.73)$$

**Remark 6** *Note that different than continous-time harmonic transfer functions (HTFs) defined in [32], the discrete-time HTFs has finite dimensions due to discrete-time Fourier series expansion. This feature allows us to obtain finite-dimensional LTI equivalents for LTP systems without needing a truncation.* □

**Remark 7** *Note that both the bilinear (Tustin) transformation in Section 4.2.3 and the lifting operation in Section 4.2.4 can be derived without needing a Floquet transformation. However, having a time-invariant state update matrix $\mathbf{A}$ allows proving the existence of the inverses in (4.50) and (4.73) with simple conditions based on the eigenvalues of $\mathbf{A}$.* □

### 4.2.5 Transforming to a Real-Valued State Space Model

An important problem for the LTI state-space model of (4.72) is that its system matrices may be complex-valued. However, the majority of state-space subspace identification techniques in the literature assume real-valued input–output data, estimating real-valued system matrices [113, 125–128]. To address this problem, we propose a similarity transformation to obtain a real-valued form for the complex-valued system structure. Note that the original LTP system given in (4.29) was assumed to be real-valued as indicated in Section 4.2.1. Therefore, the Fourier series coefficients for these time-periodic matrices are in complex conjugate form, which allows us to separate the real and imaginary parts via a simple similarity transformation as in [91, 121] without changing the system dimension. To this end, we define a complex-valued transformation matrix

$$\mathcal{T}_u := \begin{bmatrix} 0.5I_{N/2} & 0_{N/2} & 0.5I_{N/2} \\ 0_{N/2}^T & I & 0_{N/2}^T \\ -jI_{N/2} & 0_{N/2} & jI_{N/2} \end{bmatrix} \quad (4.74)$$

which is used to transform the input signal with $\mathcal{U} = \mathcal{T}_u \tilde{\mathcal{U}}_d$, with $I_{N/2}$ denoting a $N/2$ by $N/2$ identity matrix, $0_{N/2}$ a zero-column vector with length $N/2$ and $\mathcal{U}_d$ expanded as $\tilde{\mathcal{U}}_d = [\mathcal{U}_d^T \ U_{\frac{N}{2}}]^T$. Similar expansions can be used to obtain $\tilde{\mathcal{A}_{dN}}, \tilde{\mathcal{B}_{dN}}, \tilde{\mathcal{C}_d}, \tilde{\mathcal{D}_d}$ and $\tilde{\mathcal{Y}}$. Defining similarity transformation matrices for state and output as $\mathcal{T}_x$ and $\mathcal{T}_y$, respectively, the real-valued state space model can hence be obtained as

$$z\mathcal{X} = \underbrace{\mathcal{T}_x\tilde{\mathcal{A}}_{dN}\mathcal{T}_x^{-1}}_{A}\mathcal{X} + \underbrace{\mathcal{T}_x\tilde{\mathcal{B}}_{dN}\mathcal{T}_u^{-1}}_{B}\mathcal{U}$$
$$\tilde{\mathcal{Y}} = \underbrace{\mathcal{T}_y\tilde{\mathcal{C}}_d\mathcal{T}_x^{-1}}_{C}\mathcal{X} + \underbrace{\mathcal{T}_y\tilde{\mathcal{D}}_d\mathcal{T}_u^{-1}}_{D}\mathcal{U}. \qquad (4.75)$$

Note that different than a standard similarity transformation, $\mathcal{T}_x$, we use $\mathcal{T}_u$ and $\mathcal{T}_y$ to transform input–output data of the system. Thus, the newly defined real-valued system matrices and signals results in the LTI system

$$x[k+1] = Ax[k] + Bu[k]$$
$$y[k] = Cx[k] + Du[k], \qquad (4.76)$$

which is suitable for most LTI frequency domain subspace identification methods such as [113].

## 4.2.6 Subspace Identification via Frequency Response Data

Having showed that an LTP system can also be represented with an equivalent discrete-time LTI system, this section briefly summarizes the frequency domain subspace identification method of [113] to estimate a discrete-time LTI equivalent for the original LTP system.

Consider the discrete-time Fourier transform of the LTI system represented in (4.76)

$$e^{j\omega}X(\omega) = AX(\omega) + BU(\omega)$$
$$Y(\omega) = CX(\omega) + DU(\omega). \qquad (4.77)$$

Let $u[k] = e^{j\omega k}$, then $U(\omega) = [0 \ \ldots \ 0 \ 1 \ 0 \ \ldots \ 0]^T$ by computing (4.56) for each Fourier series coefficients, where the only non-zero term is the zeroth Fourier coefficient. Then, plugging in $U(\omega)$ into (4.77) yields

$$e^{j\omega}X(\omega) = AX(\omega) + \bar{B} \qquad (4.78)$$
$$G(e^{j\omega}) = CX(\omega) + \bar{D}, \qquad (4.79)$$

where $G(e^{j\omega})$ is the frequency response of the multi-output LTI system, $\bar{B}$ and $\bar{D}$ corresponds to $(\frac{N}{2}+1)^{th}$ column of $B$ and $D$, respectively.

In order to derive the data matrices for the subspace identification, we recursively multiply (4.79) with $e^{j\omega}$ and plug in (4.78) into (4.79), which yields

$$\underbrace{\begin{bmatrix} G(e^{j\omega}) \\ e^{j\omega}G(e^{j\omega}) \\ \dots \\ e^{j(q-1)\omega}G(e^{j\omega}) \end{bmatrix}}_{G_\omega} = QX(\omega) + \Gamma \underbrace{\begin{bmatrix} I \\ e^{j\omega}I \\ \dots \\ e^{j(q-1)\omega}I \end{bmatrix}}_{U_\omega}. \quad (4.80)$$

Here $Q$ is the extended observability matrix, which is calculated up to $q-1^{th}$ power of $A$ as

$$Q = \begin{bmatrix} C \\ CA \\ CA^2 \\ \dots \\ CA^{q-1} \end{bmatrix} \quad (4.81)$$

and $\Gamma$ is a lower triangular block Toeplitz matrix defined as

$$\Gamma = \begin{bmatrix} \bar{D} & 0 & \dots & \dots & 0 \\ C\bar{B} & \bar{D} & 0 & \dots & \dots \\ CA\bar{B} & C\bar{B} & \bar{D} & 0 & \dots \\ \dots & \ddots & \ddots & \ddots & 0 \\ CA^{q-2}\bar{B} & CA^{q-3}\bar{B} & \dots & \dots & \bar{D} \end{bmatrix}. \quad (4.82)$$

Note that $q$ is a chosen number that should be greater than the "dimension" of the LTI system represented in (4.76). Assuming that the dimension of the original LTP system to be $n_p$, dimension of the LTI system to be estimated becomes $2Nn_p$ when there is no dimension reduction.

One advantage of using frequency domain identification techniques is that multiple measurements can be simply combined by expanding the input, output

and state matrices as

$$\mathbf{U} = \begin{bmatrix} I & I & \ldots & I \\ e^{j\omega_1}I & e^{j\omega_2}I & \ldots & e^{j\omega_M}I \\ \ldots & \ldots & \ddots & \ldots \\ e^{j(q-1)\omega_1}I & e^{j(q-1)\omega_2}I & \ldots & e^{j(q-1)\omega_M}I \end{bmatrix}$$

$$\mathbf{G} = \begin{bmatrix} G_1 & G_2 & \ldots & G_M \\ e^{j\omega_1}G_1 & e^{j\omega_2}G_2 & \ldots & e^{j\omega_M}G_M \\ \ldots & \ldots & \ddots & \ldots \\ e^{j(q-1)\omega_1}G_1 & e^{j(q-1)\omega_2}G_2 & \ldots & e^{j(q-1)\omega_M}G_M \end{bmatrix}$$

$$\mathbf{X} = \begin{bmatrix} X(\omega_1) & X(\omega_2) & \ldots & X(\omega_M) \end{bmatrix}, \tag{4.83}$$

where $M$ is the number of tests at different frequencies. In this context, (4.80) can be expanded as

$$\mathbf{G} = Q\mathbf{X} + \Gamma\mathbf{U} \tag{4.84}$$

Since the original LTI system to be estimated (4.76) has real-valued state space representation, (4.84) are separated to their real and imaginary parts to force real-valued estimations

$$\underbrace{[Re\{\mathbf{G}\}\ Im\{\mathbf{G}\}]}_{\mathcal{G}} = Q\underbrace{[Re\{\mathbf{X}\}\ Im\{\mathbf{X}\}]}_{\mathbf{X}} + \Gamma\underbrace{[Re\{\mathbf{U}\}\ Im\{\mathbf{U}\}]}_{\mathbf{U}}. \tag{4.85}$$

which yields an equation of the form

$$\mathcal{G} = Q\mathbf{X} + \Gamma\mathbf{U}. \tag{4.86}$$

The estimation process starts by projecting (4.86) onto the null space of $\mathbf{U}$ using the projector

$$\mathbf{U}^\perp = I - \mathbf{U}^T(\mathbf{U}\mathbf{U}^T)^{-1}\mathbf{U} \tag{4.87}$$

and hence we obtain

$$\mathcal{G}\mathbf{U}^\perp = Q\mathbf{X}\mathbf{U}^\perp + \cancelto{0}{\Gamma\mathbf{U}\mathbf{U}^\perp}. \tag{4.88}$$

A numerically efficient implementation of this projection can be performed via QR decomposition [98]. We then compute the singular value decomposition for $\mathcal{G}\mathbf{U}^\perp$ as

$$\mathcal{G}\mathbf{U}^\perp = \begin{bmatrix} \hat{U}_n & \hat{U}_o \end{bmatrix} \begin{bmatrix} \hat{\Sigma}_n & 0 \\ 0 & \hat{\Sigma}_o \end{bmatrix} \begin{bmatrix} \hat{V}_n^T \\ \hat{V}_o^T \end{bmatrix} \tag{4.89}$$

where $n$ is the estimated system dimension, which is chosen based on LTP system properties mentioned in Remark 8.

**Remark 8** *In classical LTI subspace identification, the estimated system order, $n$, is chosen based on the drastical drops in singular values of (4.89). However, if one seeks to find a possible LTP realization for the estimated system, $n$ needs to be chosen to satisfy additional constraints. Let the eigenvalues of $\mathbf{A}_d$ be $S_d = \{\lambda_i^d\}_{i=1}^{n_p}$. Lifting to a time-invariant form as in (4.72) results in $\mathcal{A}_{dN}$ with the following eigenvalues*

$$S = \left\{\left\{\lambda_i^d e^{-j2\pi \frac{k}{N}}\right\}_{i=1}^{n_p} \middle| \forall k \in [-N/2, N/2 - 1]\right\}. \tag{4.90}$$

*It is quite possible that the user could limit the estimated number of harmonics, $K$, such that $K < N/2 - 1$. In this case (4.90) will take the form*

$$\hat{S} = \left\{\left\{\lambda_i^d e^{-j2\pi \frac{k}{N}}\right\}_{i=1}^{n_p} \middle| \forall k \in [-K, K]\right\}. \tag{4.91}$$

*Note that under these constraints $n$ would be equal to the cardinality of $\hat{S}$, i.e. $n = |S| = (2K + 1)n_p$.* □

One possible estimate for the extended observability matrix is found as (see [113])

$$Q = \hat{U}_n \hat{\Sigma}_n^{1/2}. \tag{4.92}$$

Hence, by definition (4.81), we can estimate $C$ and $A$ as

$$\hat{A} = (J_1 Q)^\dagger J_2 Q \tag{4.93}$$

$$\hat{C} = J_3 Q \tag{4.94}$$

where

$$J_1 = \begin{bmatrix} I_{(q-1)n_y \times (q-1)n_y} & 0_{(q-1)n_y \times n_y} \end{bmatrix} \tag{4.95}$$

$$J_2 = \begin{bmatrix} 0_{(q-1)n_y \times n_y} & I_{(q-1)n_y \times (q-1)n_y} \end{bmatrix} \tag{4.96}$$

$$J_3 = \begin{bmatrix} I_{n_y \times n_y} & 0_{n_y \times (q-1)n_y} \end{bmatrix}. \tag{4.97}$$

Having found $\hat{A}$ and $\hat{C}$, estimates for input and feedthrough matrices can be obtained via a least-squares problem as

$$(\hat{B}, \hat{D}) = \underset{(B,D)}{\arg\min} \sum_{k=0}^{M} ||G(z) - \hat{C}(zI - \hat{A})^{-1}B - D||_F^2. \quad (4.98)$$

where $||.||_F$ denotes the Frobenious norm.

At this point, our method provides a parametric dynamical system representation which can predict the output of the original system to any type of input using a compact representation of information compared to non-parametric representations such as time-periodic impulse response functions and HTFs. To the best of our knowledge such a method which yields a state space realization for a general class of LTP systems based on frequency domain input–output data does not appear in the literature.

The main drawback of this representation—lifted LTI— is that it is unintuitive for the general audience. In Section 4.2.7, we will introduce an optimization based transformation which maps the lifted LTI representation to a more intuitive classical state-space LTP representation in the Floquet transformed form. This optimization step adds extra computational burden, but helps to obtain an intuitive representation.

### 4.2.7 Reconstructing LTP State Space Estimates

The estimated system matrices $\hat{A}$, $\hat{B}$, $\hat{C}$ and $\hat{D}$ yield an equivalent discrete-time LTI state space model for the system formulated in (4.29) as

$$\begin{aligned}\hat{x}[k+1] &= \hat{A}\hat{x}[k] + \hat{B}u[k] \\ \hat{y}[k] &= \hat{C}\hat{x}[k] + \hat{D}u[k].\end{aligned} \quad (4.99)$$

Our goal in this section is to obtain an LTP realization for the estimated system (4.99). In Section 4.2.4, we showed that it is possible to find a discrete-time LTI equivalent for LTP systems having a specific parametric structure of the Fourier series coefficients. Unfortunately, subspace identification yields state space estimates up to a similarity transformation. Therefore, we need to find a similarity transformation matrix, $\mathcal{T}$, for (4.99) in order to identify Fourier series coefficients of the underlying LTP system towards obtaining a periodic realization.

Let the LTI system of (4.99) be represented in the form

$$\hat{\mathcal{P}} = \begin{bmatrix} \hat{A} & \hat{B} \\ \hat{C} & \hat{D} \end{bmatrix}. \tag{4.100}$$

Also let $\mathcal{P}$ denote the form satisfying specific parametric structure of the Fourier series coefficients explained in Section 4.2.4. Then, our goal is to find a similarity transformation matrix, $\mathcal{T}$ satisfying

$$\underbrace{\begin{bmatrix} \mathcal{T}^{-1} & 0 \\ 0 & I \end{bmatrix}}_{\mathcal{M}^{-1}} \hat{\mathcal{P}} \underbrace{\begin{bmatrix} \mathcal{T} & 0 \\ 0 & I \end{bmatrix}}_{\mathcal{M}} = \mathcal{P}. \tag{4.101}$$

Note that $\mathcal{P}$ is a parametric representation with unknown system parameters. Therefore, any solution satisfying $\mathcal{P}$ yields an LTP realization for the estimated system. Our goal in this section is to find one of these solutions to obtain an LTP state space realization for the estimated system. To accomplish this, we first define an error metric, $\mathbf{d}(\mathcal{M}^{-1}\hat{\mathcal{P}}\mathcal{M}, \mathcal{P})$, which measures the distance between transformed system, $\mathcal{M}^{-1}\hat{\mathcal{P}}\mathcal{M}$ and $\mathcal{P}$. Then, our problem is reduced to a nonlinear optimization process as

$$\mathcal{M}_{opt} = \arg\min_{\mathcal{M}} \quad \mathbf{d}(\mathcal{M}^{-1}\hat{\mathcal{P}}\mathcal{M}, \mathcal{P}) \tag{4.102}$$

such that the distance between transformed estimated system and the solution set is minimized. Details on the computation of $\mathbf{d}$ are given in Section 4.2.8.

In order to solve this nonlinear optimization problem, we use particle swarm optimization (PSO) followed by the Nelder-Mead algorithm. PSO starts with $n^2$ particles to find a similarity transformation matrix that minimizes (4.102). Note that PSO does not guarantee convergence to the global minima. Thus, even though it is always possible to estimate an LTI state space equivalent for the original LTP system, it is not always guaranteed to find an LTP realization.

Once $\mathcal{M}_{opt}$ is estimated, we can extract the Fourier series coefficients and construct LTP state space estimates by using (4.59) to (4.61) to obtain estimated discrete-time LTP system as

$$\begin{aligned}\hat{\mathbf{x}}_d[k+1] &= \hat{\mathbf{A}}_d\hat{\mathbf{x}}_d[k] + \hat{\mathbf{B}}_d[k]\mathbf{u}_d[k], \\ \hat{\mathbf{y}}_d[k] &= \hat{\mathbf{C}}_d[k]\hat{\mathbf{x}}_d[k] + \hat{\mathbf{D}}_d[k]\mathbf{u}_d[k].\end{aligned} \qquad (4.103)$$

**Remark 9** *Note that the solution structure we aim to derive in the LTP realization step has a constant (time-invariant) system matrix. Therefore, the estimated LTP system will be in Floquet-transformed form based on the original LTP system. Thus, this method can be also used for finding the Floquet transformation for a known LTP system for which analytical derivation is challenging.* □

Finally, inverse bilinear (Tustin) transformation can be applied for (4.103) by using a similar methodology as in Section 4.2.3 to obtain continuos-time state space model estimates as

$$\begin{aligned}\dot{\hat{\mathbf{x}}}(t) &= \hat{\mathbf{A}}\hat{\mathbf{x}}(t) + \hat{\mathbf{B}}(t)u(t), \\ \hat{y}(t) &= \hat{\mathbf{C}}(t)\hat{\mathbf{x}}(t) + \hat{\mathbf{D}}(t)u(t).\end{aligned} \qquad (4.104)$$

where

$$\begin{aligned}\hat{\mathbf{A}} &= \frac{2}{T_s}(\hat{\mathbf{A}}_d + I)^{-1}(\hat{\mathbf{A}}_d - I), \\ \hat{\mathbf{B}}(kT_s) &= \frac{2}{\sqrt{T_s}}(\hat{\mathbf{A}}_d + I)^{-1}\hat{\mathbf{B}}_d[k], \\ \hat{\mathbf{C}}(kT_s) &= \frac{2}{\sqrt{T_s}}\hat{\mathbf{C}}_d[k](\hat{\mathbf{A}}_d + I)^{-1}, \\ \hat{\mathbf{D}}(kT_s) &= \hat{\mathbf{D}}_d[k] - \hat{\mathbf{C}}_d[k](\hat{\mathbf{A}}_d + I)^{-1}\hat{\mathbf{B}}_d[k],\end{aligned}$$

and intersample behavior is obtained via zero-order-hold.

### 4.2.8 Computation of Distance Function

In this section, we give the details of computations for the distance function $\mathbf{d}(\mathcal{M}^{-1}\hat{\mathcal{P}}\mathcal{M}, \mathcal{P})$ used in (4.102). To this end, we define an error metric for each estimated system matrix, $\hat{A}$, $\hat{B}$ and $\hat{C}$ as $\mathbf{d}_A$, $\mathbf{d}_B$ and $\mathbf{d}_C$, respectively and define the total error as

$$\mathbf{d} := \mathbf{d}_A + \mathbf{d}_B + \mathbf{d}_C. \tag{4.105}$$

Let $\hat{A}$ be represented as $(2K+1) \times (2K+1)$ block matrices with $\hat{A}_{i,j} \in \mathbb{C}^{n_p \times n_p}$, where $(i,j)$ corresponds to block in $i^{th}$ row and $j^{th}$ column. The parametric form of $\mathcal{A}_{dN}$ suggests a block diagonal structure with the exponential modulation terms multiplying the block-diagonal matrices as in (4.71). Thus, all non-diagonal block matrices must contribute to error. Besides, the diagonal matrices must have the specific parametric structure of (4.71). Let the mid-diagonal term of the estimated system be $\hat{A}_d$. Then, the error due to diagonal terms can be computed as

$$||\hat{A}_d - \underbrace{\frac{1}{2K+1} \sum_{k=1}^{2K+1} \hat{A}_{k,k} e^{-j2k\pi/N}}_{\bar{A}_{K+1,K+1}}||_F^2 \tag{4.106}$$

Since we seek a real-valued $\hat{A}_d$, we pick $\hat{A}_d = Re\{\bar{A}_{K+1,K+1}\}$ to minimize the error in (4.106). Then, $\mathbf{d}_A$ is defined as

$$\mathbf{d}_A := ||Im\{\bar{A}_{K+1,K+1}\}||_F^2 + \sum_{\substack{i,j=1 \\ i \neq j}}^{2K+1} ||\hat{A}_{i,j}||_F^2. \tag{4.107}$$

Similarly, let $\hat{B}$ be represented as $(2K+1) \times 1$ block matrices with $\hat{B}_i \in \mathbb{C}^{n_p \times 1}$. We also define $\bar{B}$ as the transformed $\hat{B}$ satisfying the specific parametric structure of the Fourier series coefficients. Then, error due to input matrix, $\hat{B}$ can be defined as $||\hat{B} - \bar{B}||_F^2$. Similarly, we pick $\bar{B}_{K+1} = Re\{\hat{B}_{K+1}\}$ to minimize the error. Then, we choose $\bar{B}_n = \frac{1}{2}(\hat{B}_n + \hat{B}_{2K+2-n})$ for $n = 1, 2, \ldots, K$ and $\bar{B}_{2K+2-n} = \bar{B}_n^*$ for $n = 1, 2, \ldots, K$. Thus, $\mathbf{d}_B$ is defined as

$$\mathbf{d}_B := ||Im\{\bar{B}_{K+1}\}||_F^2 + \sum_{\substack{i=1 \\ i \neq K+1}}^{2K+1} ||\frac{1}{2}\hat{B}_i - \frac{1}{2}\hat{B}_{2K+2-i}^*||_F^2. \tag{4.108}$$

Finally, let $\hat{C}$ be represented as $(2K+1) \times (2K+1)$ block matrices with $\hat{C}_{i,j} \in \mathbb{C}^{1 \times n_p}$. We also define $\bar{C}$ as the transformed $\hat{C}$ satisfying the specific parametric structure of the Fourier series coefficients for the output matrix. Then, the error due to this term can be defined as $||\hat{C} - \bar{C}||_F^2$. Now, we seek to find $\bar{C}$ in order to compute the error.

To accomplish this, we first define $\bar{\bar{C}}_0$ as

$$\bar{\bar{C}}_0 := Re\left\{\frac{1}{2K+1} \sum_{i=1}^{2K+1} \hat{C}_{i,i}\right\} \tag{4.109}$$

as the zeroth Fourier series coefficient. Then, we define the candidate solutions for the other Fourier series coefficients as

$$\bar{\bar{C}}_n := \frac{1}{2(2K-n)}\left\{\sum_{i=n+1}^{2K+1} \hat{C}_{i,i-1} + \left[\sum_{i=n}^{2K} \hat{C}_{i,i+1}\right]^*\right\} \tag{4.110}$$

and $\bar{\bar{C}}_{-n} := \bar{\bar{C}}_n^*$ for all $n = 1, 2, \ldots, 2K$. Then,

$$\bar{C}_{i,j} = \bar{\bar{C}}_{i-j} \mid (i,j) = 1, 2, \ldots, 2K+1. \tag{4.111}$$

Having computed $\bar{C}$, error due to output matrix can be computed as

$$\mathbf{d}_C := ||\hat{C} - \bar{C}||_F^2 \tag{4.112}$$

### 4.2.9 Numerical Example

In this section, we provide an example system based on the well-known Mathieu function to evaluate the performance of the proposed method. Consider the system

$$\ddot{x}(t) = -\omega_n^2 x(t) - 2\zeta\omega_n \dot{x}(t) + (1 + 2\beta_b cos\omega_p t)u(t) \tag{4.113}$$

and the measurement function

$$y(t) = (1 + 2\beta_c cos\omega_p t)x(t). \tag{4.114}$$

where $\omega_p = 4\pi$, $\omega_n = 2\pi$, $\zeta = 0.3$, $\beta_b = 0.2$ and $\beta_c = 0.1$.

In state space, the LTP system can be represented as

$$\dot{\mathbf{x}}(t) = \begin{bmatrix} 0 & 1 \\ -\omega_n^2 & -2\zeta\omega_n \end{bmatrix} \mathbf{x}(t) + \begin{bmatrix} 0 \\ 1 + 2\beta_b cos\omega_p t \end{bmatrix} u(t), \quad (4.115)$$

$$y(t) = \begin{bmatrix} 1 + 2\beta_c cos\omega_p t & 0 \end{bmatrix} \mathbf{x}(t)$$

where the numerical representation of the system matrices can be obtained as

$$\mathbf{A} = \begin{bmatrix} 0 & 1 \\ -39.4784 & -3.7699 \end{bmatrix}$$

$$\mathbf{B}(t) = \begin{bmatrix} 0 \\ 1 + 0.4\cos(\omega_p t) \end{bmatrix} \quad (4.116)$$

$$\mathbf{C}(t) = \begin{bmatrix} 1 + 0.2\cos(\omega_p t) & 0 \end{bmatrix}.$$

In order to start the identification process, we simulate the LTP system in (4.115) by applying single cosine inputs as

$$u(t) = cos(2\pi(0.15 + 0.1k)t) \quad (4.117)$$

where $k \in \{1, 2, \ldots, 400\}$, so that we generate 400 input stimuli and record the output measurements as in (4.114). The input signals, and hence the simulation durations, were chosen as 200 $s$ and all data were sampled with $f_s = 100\ Hz$. Note that our solution method also supports sum-of-cosines tests, which would drastically reduce the number of tests required for system identification [69].

Once we obtain the input–output data from the "unknown" system, we apply the proposed subspace identification method to estimate an LTP realization for the original system. We obtain the following LTP system as an equivalent representation of the system given in (4.115) as

$$\dot{\hat{\mathbf{x}}}(t) = \hat{\mathbf{A}}\hat{\mathbf{x}}(t) + \hat{\mathbf{B}}(t)u(t),$$
$$\hat{y}(t) = \hat{\mathbf{C}}(t)\hat{\mathbf{x}}(t) + \hat{\mathbf{D}}(t)u(t). \quad (4.118)$$

where

$$\hat{\mathbf{A}} = \begin{bmatrix} 5.8078 & 4.5134 \\ -21.0736 & -9.5772 \end{bmatrix} \quad (4.119)$$

$$\hat{\mathbf{B}}(t) = \begin{bmatrix} -0.0238 - 0.0097\cos(\omega_p t + 0.0026) \\ 0.0552 + 0.0224\cos(\omega_p t + 0.0012) \end{bmatrix} \quad (4.120)$$

$$\hat{\mathbf{C}}(t) = \begin{bmatrix} 10.0552 + 2.1630\cos(\omega_p t + 0.0003) \\ 4.3252 + 0.8815\cos(\omega_p t + 0.0084) \end{bmatrix}^T. \quad (4.121)$$

Note that subspace identification yields a state space estimate up to a similarity transformation. Therefore, we transform the estimated system to controllable canonical form (similar to the original system) for comparison and obtain

$$\hat{\mathbf{A}} = \begin{bmatrix} 0 & 1 \\ -39.4911 & -3.7694 \end{bmatrix} \quad (4.122)$$

$$\hat{\mathbf{B}}(t) = \begin{bmatrix} -0.0004\cos(\omega_p t + 0.3337) \\ 1 + 0.4071\cos(\omega_p t + 0.0023) \end{bmatrix} \quad (4.123)$$

$$\hat{\mathbf{C}}(t) = \begin{bmatrix} 0.9980 + 0.2055\cos(\omega_p t + 0.0063) \\ -0.0029 - 0.0027\cos(\omega_p t - 0.1409) \end{bmatrix}^T. \quad (4.124)$$

The proposed subspace identification methodology produces accurate parametric estimations of the state space structure of the original system. We also computed normalized matrix norm of the error between the predicted and actual system matrices as well as normalized signal norms of error signals between the time periodic input and output vectors:

$$||\mathbf{A} - \hat{\mathbf{A}}||_2/||\mathbf{A}||_2 = 0.0003$$
$$\Delta\mathbf{B}(t) = \mathbf{B}(t) - \hat{\mathbf{B}}(t) \rightarrow ||\Delta\mathbf{B}||_2/||\mathbf{B}||_2 = 0.0049 \quad (4.125)$$
$$\Delta\mathbf{C}(t) = \mathbf{C}(t) - \hat{\mathbf{C}}(t) \rightarrow ||\Delta\mathbf{C}^T||_2/||\mathbf{C}^T||_2 = 0.0056.$$

It can be seen that normalized quantitive errors between the actual and predicted data are very small. To evaluate the prediction performance, we illustrate the HTFs of the continuos-time LTP systems (actual and estimated) in Fig. 4.2. We observe negligible errors in magnitude plots and some minor errors in phase plots. In order to investigate the effect of these differences on input–output data, we simulated the actual and estimated system with a square wave input with period $\pi$ and magnitude varying between $-1$ and $1$. Fig. 4.3 illustrates the response of both the actual and the estimated system. As seen in Fig. 4.3, the estimated LTP system can accurately predict the response to a square wave function.

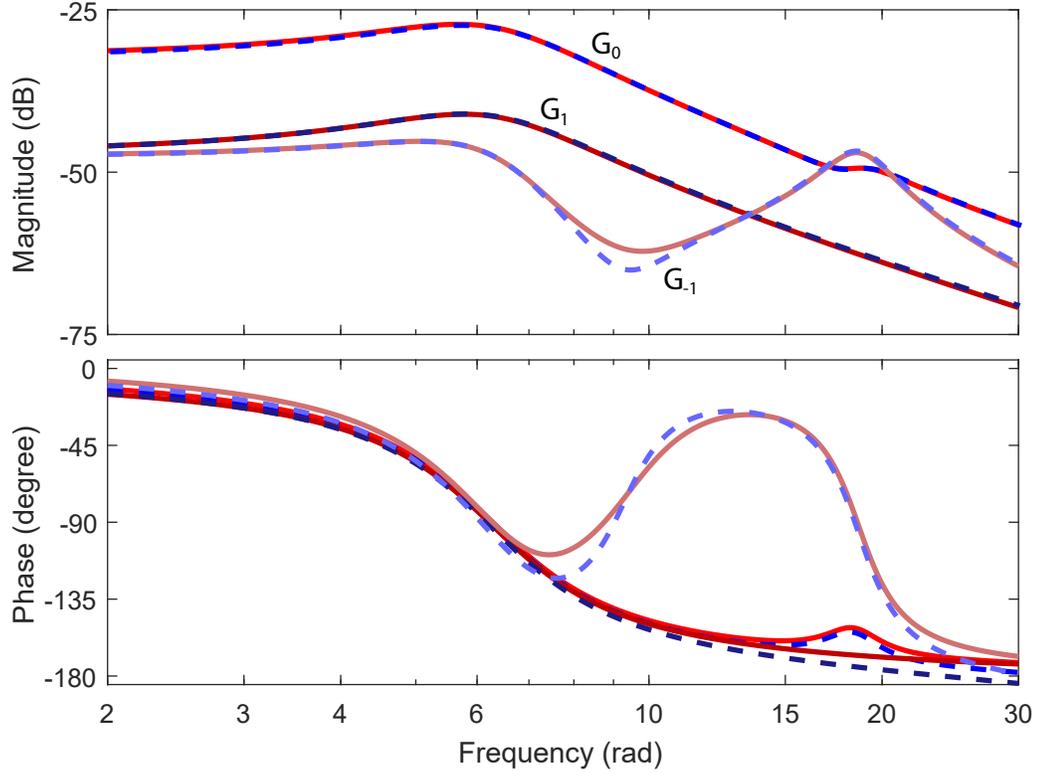

Figure 4.2: Harmonic transfer functions of the original (red-solid) and estimated (blue-dashed) systems. Magnitude and phase plots for $G_0$ are illustrated with standard red-solid and blue-dashed lines, respectively. $G_1$ and $G_{-1}$ are illustrated with dark and light tones of the corresponding colors.

Table 4.2: N-RMSE errors for different test signals

|         | Single Sine | Square Wave | Periodic Ramp |
|---------|-------------|-------------|---------------|
| N-RMSE  | 0.0030      | 0.0042      | 0.0040        |

In addition, we used normalized root mean squared error (N-RMSE) to quantify the prediction performance of the estimated LTP system for different test signals. Table 4.2 reports N-RMSE errors for the square wave example illustrated in Fig. 4.3, a periodic ramp input with period, $\pi$ (varying between $-1$ and 1) and a sinusoidal input with period, $\pi$. Note that the square wave and periodic ramp signal predictions produce small N-RMSE errors when we consider the N-RMSE error of the sinusoidal input signal, which was used as a training data for the system identification process.

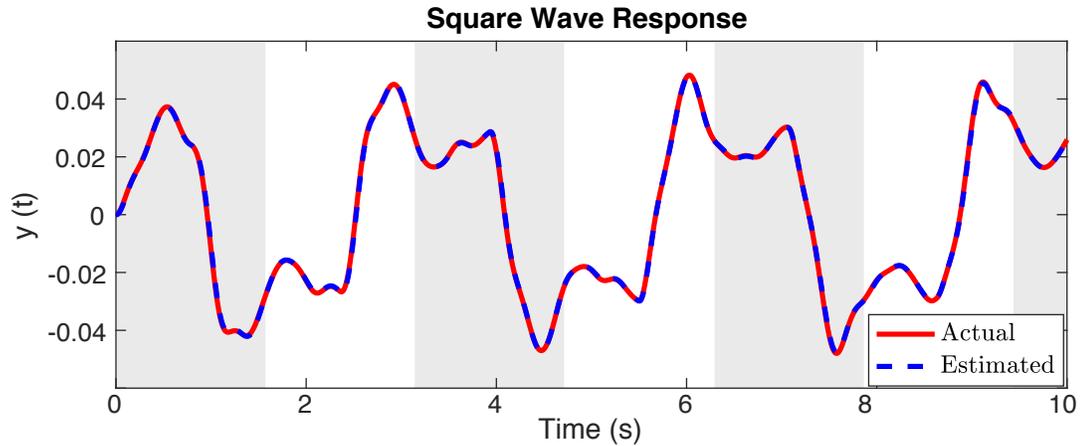

Figure 4.3: Response of the actual and estimated system to a square wave input signal. Shaded and white regions represent the $+1$ and $-1$ regions of the square wave, respectively.

## 4.3 Conclusions

In this chapter, we proposed two state space identification methods for linear time periodic (LTP) systems towards developing state space models of legged locomotor behaviors. In the first part, we limited our attention to the hybrid LTP systems that has continuous state trajectories but potentially discontinuous vector fields. We introduced a state space parametric system identification framework for these systems assuming full state measurement and switching time between successive subsystems are known.

In the second part, we proposed a new method for subspace-based state space identification of linear time periodic (LTP) systems using frequency response data. Our solution methodology is based on the fact that LTP systems can be transformed into equivalent discrete-time linear time-invariant (LTI) systems. To accomplish this, we utilize bilinear (Tustin) transformation and a frequency domain lifting method available in the literature. Then, we estimate an LTI system representation that can predict the input–output data of the original system.

We then introduced an optimization based transformation which helps us to identify the Fourier series coefficients in the lifted LTI representation. Then, we obtain an LTP realization by using Fourier synthesis equations. Note that the optimization process adds an extra computational burden to standard subspace identification as a cost of obtaining a more intuitive realization. Finally, the estimated LTP system has a time invariant state matrix in Floquet-transformed form. Therefore, our method allows finding Floquet transforms for known LTP systems via system identification.

# Chapter 5

# Conclusion and Future Works

Inspired from animals in nature, the legged locomotion is one of the trending areas in robotics to design legged robots that can move like their animal counterparts do in nature. To accomplish this, we first need to understand the functions and concepts in nature, so that we can engineer them to develop better robotics applications. To this end, novel system identification methods are required to mathemetically represent legged locomotion and understand the physics behind the animal locomotion.

In the first part of this thesis, we focused on mechanics-based mathematical modeling of legged locomotion. Especially, we aimed to experimentally validate the prediction performance of a recently proposed approximate analytical solution to Spring-Loaded Inverted Pendulum (SLIP) model dynamics. To this end, we first designed and built a one-legged hopping robot platform based on the SLIP template and then performed parametric system identification to identify model parameters. Then, we assessed the prediction performance of the approximate analytical solution on single-stride locomotion experiments that are collected with different leg springs and various initial conditions. Our results showed that even the approximate analytical solutions to SLIP dynamics provides sufficiently accurate predictions for a one-legged hopping robot platform.

In the second part, we focused on developing input–output models of legged locomotion without explicitly modeling the physical system dynamics. Our aim was to estimate transfer functions corresponding to input–output behavior of legged locomotor systems. To achieve this, we modeled the legged locomotor dynamics as a Linear Time Periodic (LTP) system by linearizing around a stable periodic orbit. We then utilized the frequency domain analysis and identification methods for LTP systems towards identification of input–output models for legged locomotion problems.

Finally, we worked on estimating state space models of legged locomotion to obtain a more intuitive system representation. To accomplish this, we proposed two state space identification methods to estimate time periodic system representations for LTP systems. Firstly, we considered a hybrid, LTP system with full state measurement assumption. We showed that frequency domain input–output data yields sufficiently accurate state space estimations for the state and input matrices for hybrid LTP systems. We then released our full state measurement assumption and formulated the identification problem for a general class of LTP systems. To this end, we utilized frequency domain subspace identification methods to estimate LTP state space realizations for unknown stable LTP systems.

As a natural future work, experimental validation of the data-driven system identification techniques is fundamental towards application of our methods to biological locomotor systems. Besides, extending the data-driven system identification methods presented in this thesis to closed-loop systems is also crucial for analysis and identification of biological systems. As a continuation of this study, we plan to utilize these methods for separate identification of plant and the controller in a biological locomotor system in order to understand the control principles of animal in nature.